\documentclass[prl,twocolumn,amsmath,amssymb,floatfix,longbibliography]{revtex4-2}
\usepackage{graphicx}
\usepackage{dcolumn}
\newcolumntype{d}[1]{D{.}{.}{#1}}
\usepackage{bm}
\usepackage{longtable}
\usepackage{mathtools}

\usepackage{xspace}
\usepackage{xpunctuate}
\DeclareRobustCommand\etal{\xperiodafter{\emph{et al}}}
\DeclareRobustCommand\ie{\xperiodafter{\emph{i.e}}}

\usepackage{epstopdf}

\usepackage{float}

\begin{document}

\title{Possible structural quantum criticality tuned by rare-earth ion substitution in infinite-layer nickelates}

\author{Alaska Subedi} 

\affiliation{CPHT, CNRS, Ecole Polytechnique, IP Paris, F-91128
  Palaiseau, France} 

\date{\today}

\begin{abstract}

Materials with competing phases that have similar ground state
energies often exhibit a complex phase diagram.  Cuprates are a
paradigmatic example of such a system that show competition between
charge, magnetic, and superconducting orders. The infinite-layer
nickelates have recently been revealed to feature similar
characteristics.  In this paper, I show that these nickelates are
additionally near a structural quantum critical point by mapping the
energetics of their structural instabilities using first priniciples
calculations.  I first confirm previous results that show a phonon
instability in the $P4/mmm$ phase leading to the $I4/mcm$ structure
for $R$NiO$_2$ with $R$ = Sm--Lu.  I then study the
non-spin-polarized phonon dispersions of the $I4/mcm$ phase and find
that they exhibit rare-earth size dependent instabilities at the $X$
and $M$ points for materials with $R$ = Eu--Lu.  Group-theoretical
analysis was used to enumerate all the isotropy subroups due to these
instablities, and the distorted structures corresponding to their
order parameters were generated using the eigenvectors of the unstable
phonons.  These structures were then fully relaxed by minimizing both
the atomic forces and lattice stresses. I was able to stabilize only
five out of the twelve possible distortions.  The $Pbcn$ isotropy
subgroup with the $M_5^+(a,a)$ order parameter shows noticeable energy
gain relative to other distortions for the compounds with late
rare-earth ions.  However, the order parameter of the lowest-energy
phase switches first to $X_2^- (0,a) + M_5^+ (b,0)$ and then to $X_2^-
(0,a)$ as the size of the rare-earth ion is progressively increased.
Additionally, several distorted structures lie close in energy for the
early members of this series.  These features of the structural
energetics persist even when antiferromagnetism is allowed. Such a
competition between different order parameters that can be tuned by
rare-earth ion substitution suggests that any structural transition
that could arise from the phonon instabilities present in these
materials can be suppressed to 0 K.
\end{abstract}


\maketitle

\section{Introduction}

The recent report of superconductivity in thin films of
Nd$_{0.8}$Sr$_{0.2}$NiO$_2$ by Li \etal has rekindled interest in the
infinite-layer nickelates because the nominal valence of Ni$^{1+}$ in
these materials has the same $3d^9$ electronic configuration with $s =
\frac{1}{2}$ that is found in the Cu$^{2+}$ ions of the cuprates
\cite{li2019}. Subsequently, signatures of superconductivity have also
been found in thin films of doped PrNiO$_2$ and LaNiO$_2$
\cite{osada2020nl,osada2021,zeng2022sa}.
A large amount of experimental
\cite{hepting2020,li2020absence,fu2019,lee2020ap,li2020,zeng2020prl,goodge2021,wang2020synthesis,gu2020,wang2021,xiang2021,osada2020prm,rossi2021orbital,
  cui2021nmr,he2021,ortiz2021prb,
  gao2021,zeng2022nc,zhou2021,zhao2021,lin2022,hsu2021,lu2021,li2021,puphal2021,chen2022,wang2021pressure,zhou2022,ortiz2021,rossi2021,
  krieger2021,tam2021,goodge2022,chow2022,fowlie2022,harvey2022,ding2022}
and theoretical
\cite{botana2020,sakakibara2020,jiang2020,wu2020,nomura2019,ryee2020,zhang2020prr,zhang2020,zhang2020type,jiang2019,werner2020,hu2019,zhou2020spin,
  bernardini2020,gu2020cp,choi2020prb,lechermann2020prb,chang2020hund,liu2020,karp2020,kitatani2020,been2021,leonov2020,lechermann2020,choi2020,
  wang2020,kang2020,kapeghian2020,kang2021, zhang2021magnetic,
  lechermann2021,katukuri2020,malyi2022,plienbumrung2021,klett2022,xia2021,bernardini2021,carrasco2021,zhang2022phase,karp2022,jiang2022}
effort has already been expended in understanding the electronic and
magnetic properties of these materials.  However, the microscopic
mechanism underlying the observed superconductivity has not been fully
clarified.

One reason for the lack of progress in this direction is the absence
of full chemical and structural information of the samples that
exhibit superconductivity, which stems from the difficulty in
performing diffraction experiments on thin films.  It is also not well
understood how the sample growth conditions affect the chemical and
structural factors that give rise to superconductivity in these
materials.  For example, superconductivity has not yet been observed
in polycrystalline powders with the same chemical composition as that
of the superconducting thin films
\cite{li2020absence,wang2020synthesis}.  Furthermore,
superconductivity seems to be observed in thin films of only around 10
nm in thickness \cite{li2019,gu2020,zeng2022sa,gao2021,zhou2021}.
Interestingly, density functional theory (DFT) calculations show that
the stoichiometric compound is thermodynamically unstable, with
hydrogenation reducing this instability \cite{si2020,malyi2022}.

In any case, experimental probes have so far been oblivious to any
presence of hydrogen in the superconducting thin films, and the
two-dimensional layers formed by NiO$_4$ squares have been thought to
be the key structural ingredient behind the superconductivity in these
materials.  DFT calculations show that the parent compound NdNiO$_2$
hosting these layers is stable, but the calculated electron-phonon
coupling is too small to account for the observed $T_c$
\cite{nomura2019}.  Calculations that utilize the DFT electronic
structure and take into account the many-body effects find a
$d_{x^2-y^2}$-wave superconducting instability, which has strengthened
the case for an unconventional nature of superconductivity in these
materials \cite{wu2020,sakakibara2020,kitatani2020}.  Such an order
parameter would exhibit nodes in the superconducting order parameter.
Currently, there is experimental support for both nodal and nodeless
superconducting gap \cite{chow2022,harvey2022}.

There is also experimental evidence for structural instability in
these materials.  Polycrystalline powders and single crystals of these
materials have been refined to the tetragonal $P4/mmm$ structure
\cite{crespin1983,levitz1983,hayward1999,hayward2003,puphal2021}. Recent
resonant x-ray scattering experiments, however, find a charge order
near the wave vector $(\frac{1}{3},0)$ in lightly-doped LaNiO$_2$ and
NdNiO$_2$ thin films \cite{rossi2021,krieger2021,tam2021}.  DFT-based
phonon dispersions calculations do not find any structural
instabilities in these two materials, but they do show the $P4/mmm$
structure to be dynamically unstable for the infinite-layer nickelates
with smaller rare-earth ions
\cite{xia2021,bernardini2021,carrasco2021,zhang2022phase}.

In this paper, I extend previous studies of the structural
instabilities in the infinite-layer nickelates presented in
Refs.~\cite{xia2021,bernardini2021,carrasco2021,zhang2022phase} by
investigating the instabilities that were not considered in these
works.  I first calculate the non-spin-polarized phonon dispersions of
all $P4/mmm$ $R$NiO$_2$ compounds with $R$ = La--Lu (including Y also
as a rare-earth element) and reproduce previous results that find a
phonon instability at the $A$ $(\frac{1}{2},\frac{1}{2},\frac{1}{2})$
point for $R$ = Sm--Lu.  This instability leads to an $I4/mcm$
structure that exhibits in-plane rotation of the NiO$_4$ squares that
is out-of-phase in the out-of-plane direction.

I then go beyond the previous studies by calculating the
non-spin-polarized phonon dispersions of this phase for the compounds
with $R$ = Sm--Lu.  I find only SmNiO$_2$ to be stable in the $I4/mcm$
structure.  The $I4/mcm$ phase of GdNiO$_2$ exhibits an instability at
the $X$ $(\frac{1}{2},\frac{1}{2},0)$ point, while the compounds with
$R$ = Gd--Lu exhibit an additional instability at the $M$ $(0,0,1)$
point. I used group-theoretical analysis to identify all the twelve
different distortions that are possible due to these instabilities.
Only five of these distorted structures could be stabilized after full
structural relaxations minimizing both the atomic forces and lattices
stresses.  While the $Pbcn$ structure with the order parameter
$M_5^+(a,a)$ exhibits significant energy gain relative to other
distortions for the compounds with late rare-earth elements, several
distorted structures lie close in energy in the early members of this
series.  Moreover, among the nearly degenerate structures, the order
parameter of the lowest-energy distortion changes first to $X_2^-
(0,a) + M_5^+ (b,0)$ and then to $X_2^-(0,a)$ as the size of the
rare-earth ion is progressively increased.  These aspects of the
structural energetics remain even in the antiferromagnetic state.
This presence of competing structural phases arising out of phonon
instabilities that can be tuned by rare-earth ion substitution
suggests that the infinite-layer nickelates lie in the vicinity of a
structural quantum critical point.

\section{Computational Approach}

The phonon calculations presented here were obtained using density
functional perturbation theory as implemented in the {\sc quantum
  espresso} package \cite{dfpt,qe}.  This is a pseudopotential-based
code, and I used the pseudopotentials generated by Dal Corso
\cite{pslib}.  The calculations were performed within the generalized
gradient approximation of Perdew, Burke and Ernzerhof (PBE) \cite{pbe}
using planewave cutoffs of 60 and 600 Ry, respectively, for the
basis-set and charge density expansions.  The Brillouin zone
integration was performed using $16\times16\times16$,
$12\times12\times12$, and $12\times12\times8$ $k$-point grids for the
$P4/mmm$, $I4/mcm$, and $Pbcn$ phases, respectively.  The dynamical
matrices were obtained on $8\times8\times8$, $4\times4\times4$, and
$6\times6\times4$ $q$-point grids for the $P4/mmm$, $I4/mcm$, $Pbcn$
phases, respectively.  A 0.008 Ry Marzari-Vanderbilt smearing was used
in all the calculations.

I used the {\sc isotropy} package to determine all the order
parameters that are possible due to the unstable phonon modes
\cite{isotropy}.  The distortions corresponding to these order
parameters were then generated by utilizing the eigenvectors of the
unstable phonon modes on 64-atom $2\times2\times2$ supercells of the
primitive $I4/mcm$ unit cells.  These structures were then fully
relaxed by minimizing both the atomic forces and lattice stresses using
the {\sc vasp} package, which is a planewave implementation of the
projector-augmented-wave method \cite{vasp}. A planewave cutoff of 600
eV, $k$-point grid of $8\times8\times8$, and Methfessel-Paxton
smearing of 0.1 eV were used in these calculations.  A magnetic
solution was not allowed.  However, the fully relaxed structures were
utilized to construct the conventional unit cells that allowed for a
nearest-neighbor (G-type) antiferromagnetic ordering. These structures
were again relaxed allowing for such an antiferromagnetic solution
using equivalent or denser $k$-point grids.  Some non-spin-polarized
and magnetic structural relaxations were also performed on the $Pbnm$
$a^-a^-c^+$ distortion of the $P4/mmm$ structure generated on 32-atom
$2\times2\times2$ supercells.  A planewave cutoff of 650 eV and
$k$-point grid of $8\times8\times8$ were used in these calculations.
Each component of the force is less than 1 meV/\AA\ in the relaxed
structures.

The spin-orbit interaction was neglected in all the calculations. The
pseudopotentials of the rare-earth ions have the $4f$ electrons in the
core such that the ions have a formal valence of $3+$.  I made
extensive use of the {\sc findsym} \cite{findsym}, {\sc amplimodes}, and {\sc spglib}
\cite{spglib} packages in the symmetry analysis of the relaxed
structures.

\section{Results and Discussion}

\subsection{Dynamical instability in the parent \bm{$P4/mmm$} phase}

\begin{figure*}
  \includegraphics[width=\textwidth]{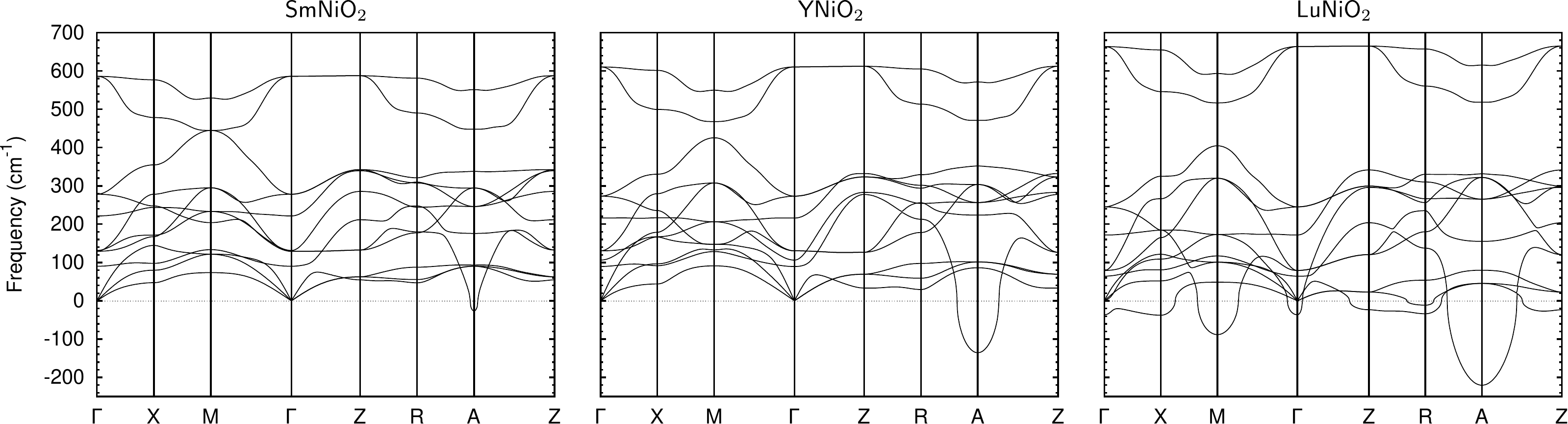}
  \caption{Calculated non-spin-polarized phonon dispersions of
    fully-relaxed SmNiO$_2$, YNiO$_2$, and LuNiO$_2$ in the $P4/mmm$
    phase. The high-symmetry points are $\Gamma$ $(0,0,0)$, $X$
    $(0,\frac{1}{2},0)$, $M$ $(\frac{1}{2}, \frac{1}{2}, 0)$, $Z$
    $(0,0,\frac{1}{2})$, $R$ $(0,\frac{1}{2},\frac{1}{2})$, and $A$
    $(\frac{1}{2},\frac{1}{2},\frac{1}{2})$ in terms of the reciprocal
    lattice vectors. Imaginary frequencies are indicated by negative
    values.}
  \label{fig:parentph}
\end{figure*}

\begin{figure}
  \includegraphics[width=\columnwidth]{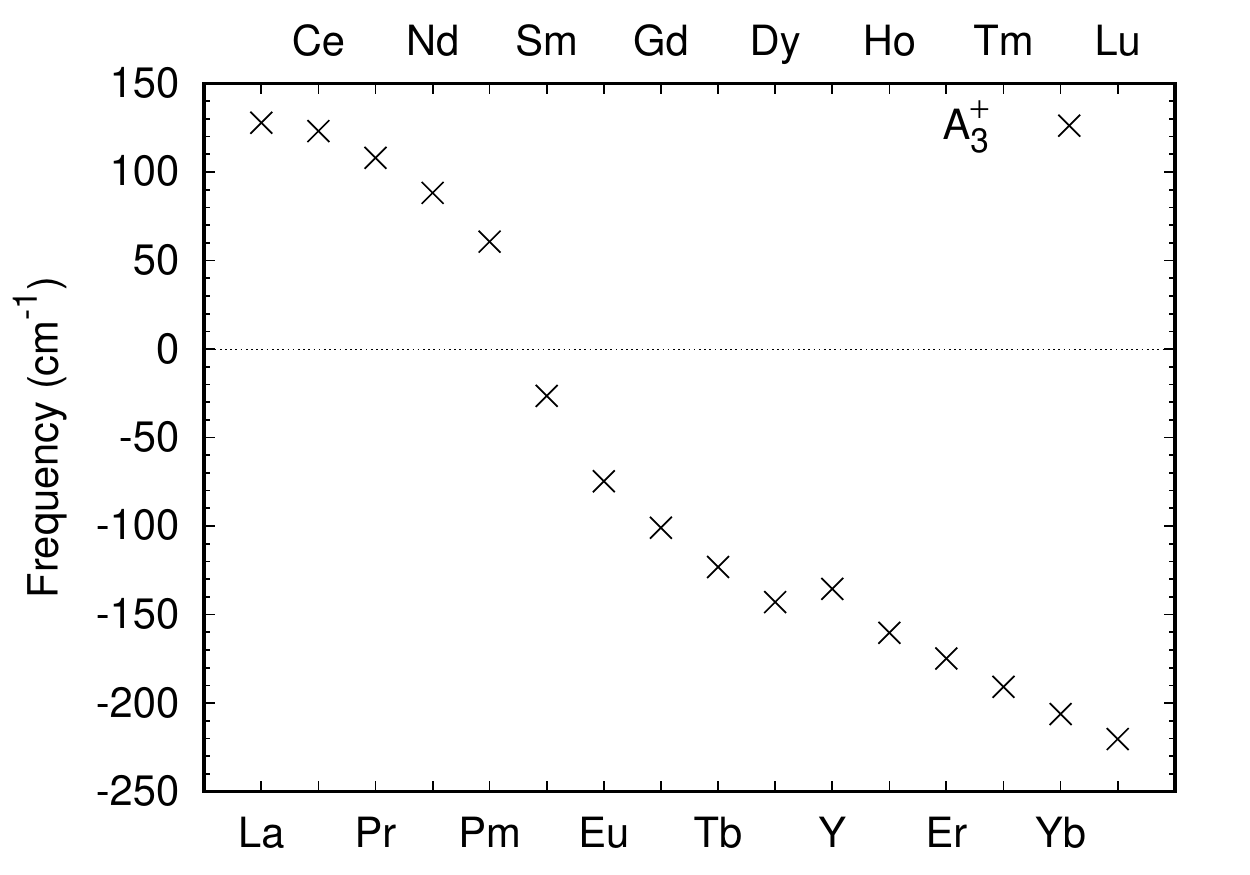}
  \caption{Calculated phonon frequency of the $A_3^+$ mode of
    $R$NiO$_2$ ($R$ = rare-earth) compounds in the non-spin-polarized
    $P4/mmm$ phase as a function of the rare-earth ions, which are sorted
    according to their ionic radii.  Imaginary frequencies are
    indicated by negative values.}
  \label{fig:omegaA}
\end{figure}


The calculated non-spin-polarized phonon dispersions of $R$NiO$_2$ in
the $P4/mmm$ phase are given in Fig.~\ref{fig:parentph} for $R =$ Sm,
Y, and Lu, which broadly agree with previous results
\cite{xia2021,bernardini2021,zhang2022phase}. The dispersions exhibit
a nondegenerate phonon branch that is unstable at $A$
$(\frac{1}{2},\frac{1}{2},\frac{1}{2})$. As noted by Xia \textit{et
  al.}, this instability has a strong dependence on the rare-earth ion size
\cite{xia2021}.  The calculated frequency of the unstable mode at $A$
as a function of the rare-earth ion is shown in Fig.~\ref{fig:omegaA}.
This mode is stable for compounds with large rare-earth ions La--Pm
but becomes unstable for those with rare-earth ions Sm--Lu that have
smaller ionic radii.  The instability is quite weak for SmNiO$_2$,
with a calculated value of 26$i$ cm$^{-1}$ for the imaginary frequency
of the unstable mode.  For LuNiO$_2$, which has the smallest
rare-earth ion, the instability is relatively strong with a calculated
imaginary frequency of 220$i$ cm$^{-1}$.

The instability at $A$ causes an in-plane
rotation of the NiO$_4$ squares that is out-of-phase along the $c$
axis \cite{xia2021,bernardini2021}. With the convention that the Ni
ion is at the origin, this mode has the irreducible representation
(irrep) $A_3^+$ \cite{bernardini2021}.  Since this mode
is nondegernate and $A$ has only one element in its star, this
instability leads to only one distorted structure.  Group-theoretical
analysis shows that this low-symmetry structure has the space group
$I4/mcm$ \cite{xia2021,bernardini2021,carrasco2021,zhang2022phase}.

I used the eigenvector of the unstable $A_3^+$ mode to generate the
$I4/mcm$ structure for all the $R$NiO$_2$ compounds and fully relaxed
the structure by minimizing both the atomic forces and lattice
stresses.  As expected, the $I4/mcm$ structure is lower in energy than
the corresponding $P4/mmm$ structure only for those compounds that
exhibit an instability of the $A_3^+$ mode in the $P4/mmm$ phase (\ie
$R$ = Sm--Lu), which is consistent with previous studies
\cite{xia2021,zhang2022phase}.  The calculated energy gain range from
relatively tiny 2.0 meV/atom for SmNiO$_2$ to noticeably large 74.3
meV/atom for LuNiO$_2$.
%

\subsection{Is the \bm{$I4/mcm$} phase stable?}

\begin{figure*}
  \includegraphics[width=\textwidth]{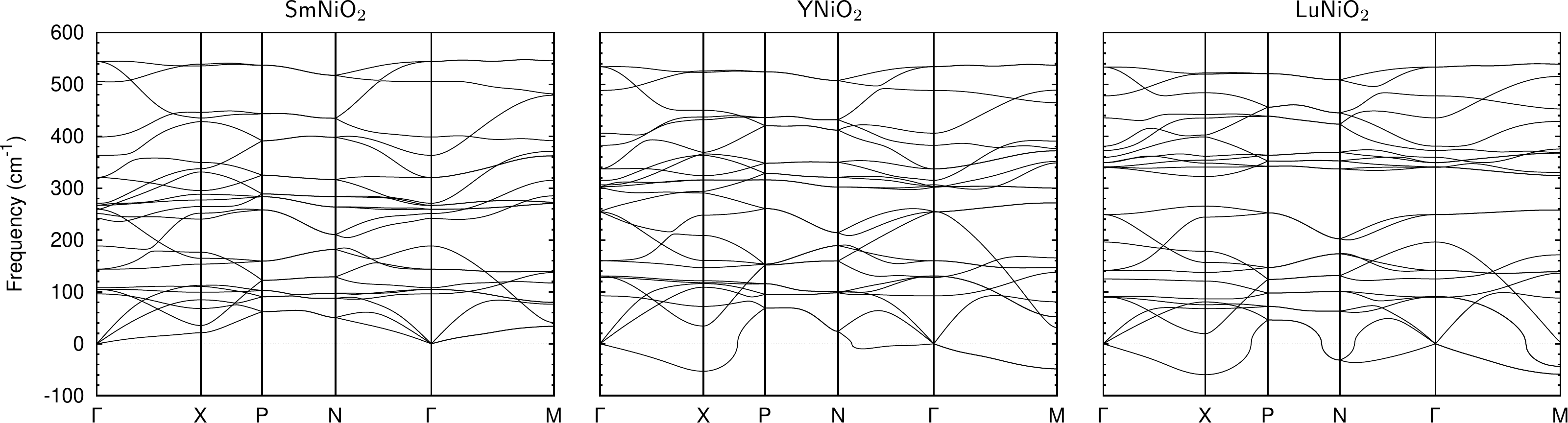}
  \caption{Calculated non-spin-polarized phonon dispersions of
    fully-relaxed SmNiO$_2$, YNiO$_2$, and LuNiO$_2$ in the
    high-symmetry $I4/mcm$ phase.  The high-symmetry points are
    $\Gamma$ $(0,0,0)$, $X$ $(\frac{1}{2},\frac{1}{2},0)$, $P$
    $(\frac{1}{2},\frac{1}{2},\frac{1}{2})$, $N$
    $(\frac{1}{2},0,\frac{1}{2})$, and $M$ $(0,0,1)$ in terms of the
    reciprocal lattice vectors of the conventional unit
    cell. Imaginary frequencies are indicated by negative values.}
  \label{fig:phoneforty}
\end{figure*}

\begin{figure}
  \includegraphics[width=\columnwidth]{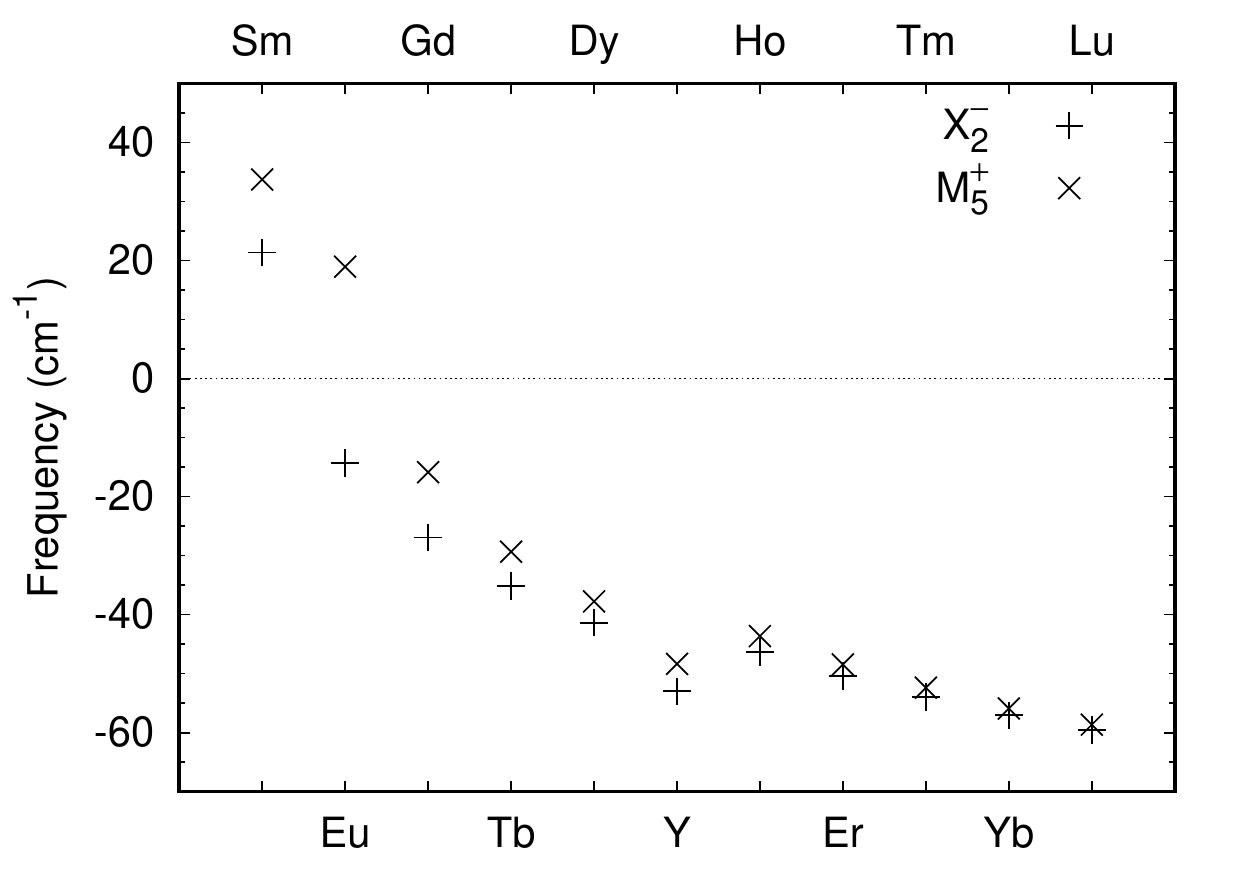}
  \caption{Calculated phonon frequency of the $X_2^-$ and $M_5^+$
    modes of $R$NiO$_2$ ($R$ = Sm--Lu and Y) compounds in the
    non-spin-polarized $I4/mcm$ phase as a function of the rare-earth
    ions.  Imaginary frequencies are indicated by negative values.}
  \label{fig:omegaXM}
\end{figure}

I studied the structural stability of the $I4/mcm$ phase of the
infinite layer rare-earth nickelates obtained above for $R$ = Sm--Lu
by calculating their non-spin-polarized phonon dispersions, which are
shown in Fig.~\ref{fig:phoneforty} for SmNiO$_2$, YNiO$_2$, and
LuNiO$_2$.  Fig.~\ref{fig:omegaXM} shows the frequencies of the
lowest-energy phonon modes at the $X$ $(\frac{1}{2}, \frac{1}{2},0)$
and $M$ $(0,0,1)$ points in their Brillouin zone for the compounds
with $R$ = Sm--Lu.  I find that only SmNiO$_2$ is dynamically stable
in the $I4/mcm$ phase.  All other compounds exhibit phonon
instabilities that are again strongly dependent on the rare-earth ion
size.  EuNiO$_2$ exhibits an instability only at the $X$
point. Compounds with smaller rare-earth ions exhibit instabilities at
both $X$ and $M$. There is only one unstable mode at $M$ for $R$ =
Gd--Er, while an additional unstable mode at $M$ appears for smaller
rare-earth ions Tm--Lu.  Moreover, yet another instability at $N$
appears for compounds with rare-earth ions from Ho to Lu.  It is also
noteworthy that the highly-dispersive soft branch between
$\Gamma$--$M$ shows a relatively small rare-earth size dependence.
Interestingly, the lowest-frequency unstable modes at $X$, $M$, and
$N$ are all connected to unstable acoustic branches at $\Gamma$, which
should manifest as a softening of the corresponding elastic moduli in
experiments.

As can be seen in Fig.~\ref{fig:omegaXM}, the instability at $X$ is
larger than at $M$ in all the $I4/mcm$ phases that exhibit these
instabilities.  The calculated difference of the imaginary frequencies
between these modes is 9$i$ cm$^{-1}$ for GdNiO$_2$, which is the first
member of this series that exhibits both these instabilities.  As the
size of the rare-earth ion is decreased, this difference decreases,
and the instabilities at $X$ and $M$ become almost degenerate.  For
LuNiO$_2$, the instabilities at $X$ and $M$ have imaginary frequencies
of 59.5$i$ and 58.6$i$ cm$^{-1}$, respectively.

\begin{figure}
  \includegraphics[width=0.75\columnwidth]{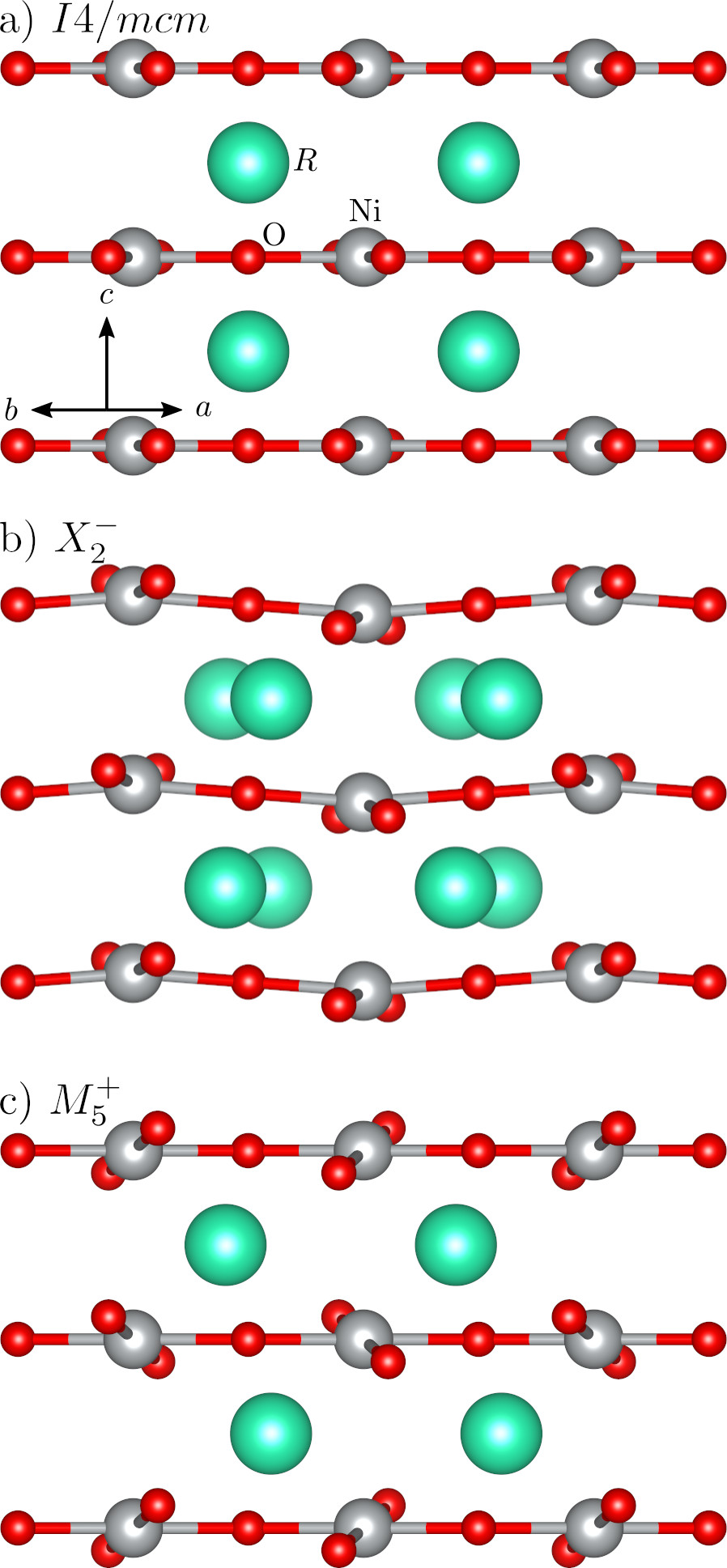}
  \caption{a) $I4/mcm$ structure of the infinite-layer $R$NiO$_2$
    compounds.  Atomic displacements due to b) $X_2^-$ and c) $M_5^+$
    phonon instabilities of the $I4/mcm$ phase. The structures are
    projected along the in-plane diagonal direction.}
  \label{fig:phonon-disp}
\end{figure}

The unstable phonon mode at $X$ in the $I4/mcm$ phase has the irrep
$X_2^-$ when the convention that one of the Ni ion is at the origin is
used.  This mode is nondegenerate, but the order parameter described
by this instability is two dimensional because the star of $X$ has two
elements $\{(\frac{1}{2}, \frac{1}{2}, 0), (\frac{1}{2}, -\frac{1}{2},
0)\}$.  The atomic displacements resulting from this instability are
shown in Fig.~\ref{fig:phonon-disp}(b).  A comparison with the
undistorted $I4/mcm$ structure in Fig.~\ref{fig:phonon-disp}(a) shows
that the $X_2^-$ mode causes an in-plane shear distortion of the $R$
squares that is out-of-phase along the $c$
axis. Additionally, the O-Ni-O bonds that are perpendicular to the
direction of the displacement the of $R$ ions shift and buckle in the
out-of-plane direction.  The shift and buckling are out-of-phase along the
displacement direction of the $R$ ions, but in-phase along the
$c$ axis.

The lowest-frequency unstable mode at $M$ has the irrep $M_5^+$.  This
mode is doubly degenerate and the star of $M$ has only one element.
So the order parameter due to this instability is also two
dimensional.  Fig.~\ref{fig:phonon-disp}(c) shows the atomic
displacements caused by this mode, which involves a sliding of the $R$
planes that is out-of-phase along the $c$ axis.  There is also an
out-of-plane tilting of the NiO$_4$ squares that is out-of-phase along
the sliding direction of the $R$ planes and in-phase along the $c$
axis.

\subsection{Which low-symmetry structures derive from the phonon
  instabilities of the \bm{$I4/mcm$} phase ?}

\begin{figure}
  \includegraphics[width=\columnwidth]{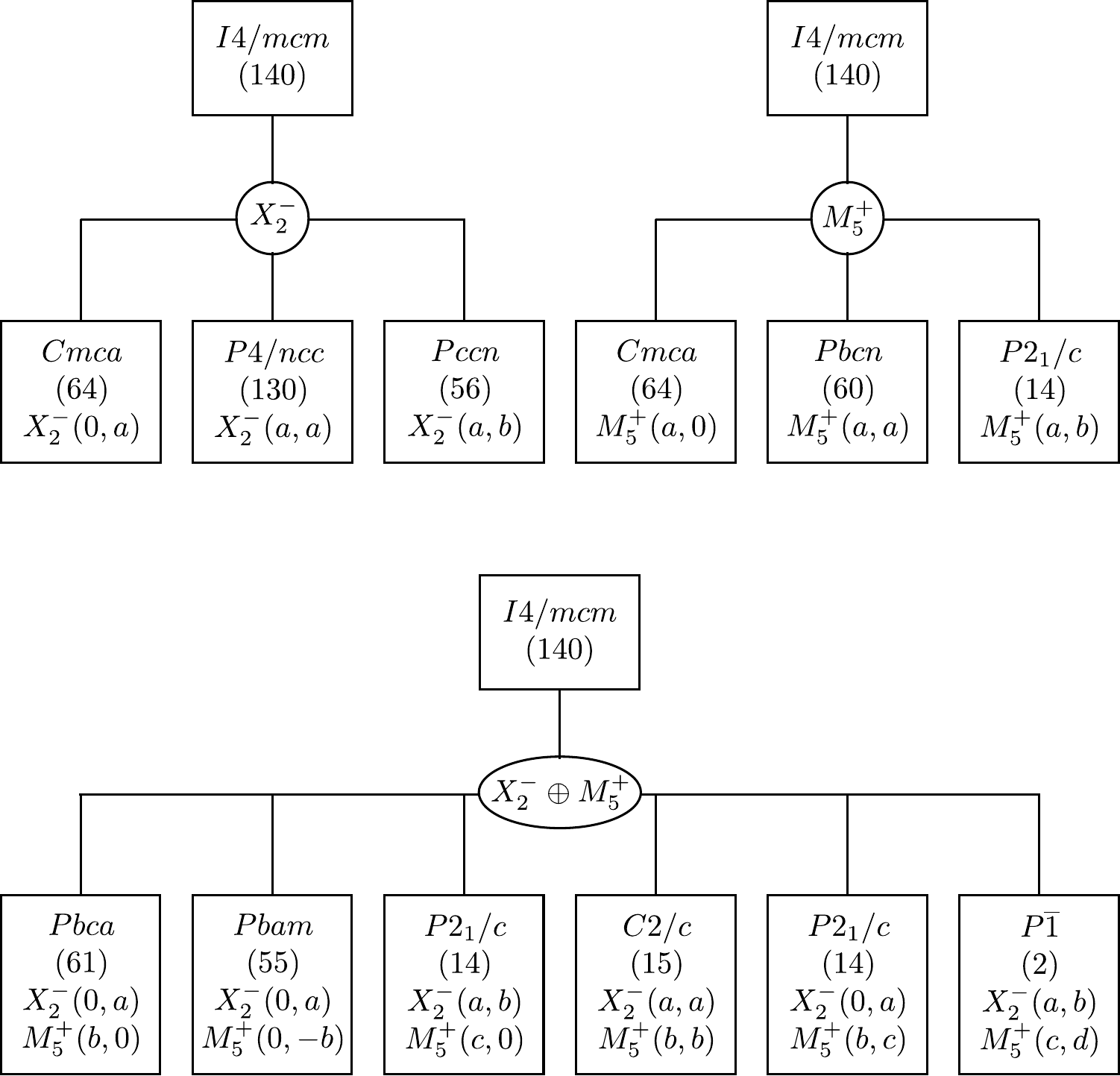}
  \caption{Isotropy subgroups and the corresponding order parameters
    of the $X_2^-$ and $M_5^+$ irreps of the $I4/mcm$ space group. The
    space group numbers are given in the parenthesis.}
  \label{fig:iso}
\end{figure}

I used group-theoretical analysis to identify all the distortions that
can arise due to the $X_2^-$ and $M_5^+$ instabilities.  These irreps
each have three isotropy subgroups.  In addition, there are six
subgroups due to the coupling of the $X_2^-$ and $M_5^+$ order
parameters. The order parameters and space groups of all the
low-symmetry structures that can arise due to these two instabilities
are shown in Fig.~\ref{fig:iso}.  I used the eigenvectors of these
unstable phonon modes to generate the twelve distortions of the
$I4/mcm$ phase in its 64-atom $2\times2\times2$ supercell for the
compounds with $R$ = Eu--Lu.  These were then fully relaxed by
minimizing both the atomic forces and lattice stresses without
allowing for a magnetic solution. Out of twelve possible distortions,
only five could be stabilized.  They are $Cmca$ $X_2^-(0,a)$, $P4/ncc$ 
$X_2^-(a,a)$, $Cmca$ $M_5^+(a,0)$, $Pbcn$ $M_5^+(a,a)$, and $Pbca$
$X_2^-(0,a) + M_5^+(b,0)$. The calculated total energies of these five
phases relative to their $I4/mcm$ phase is given in Table
\ref{tab:enensp}.

\begin{table}
  \caption{\label{tab:enensp} Total energies of the five isotropy
    subgroups of the $X_2^-$ and $M_5^+$ instabilities present in the
    $I4/mcm$ phase of the $R$NiO$_2$ compounds that could be
    stabilized using non-spin-polarized DFT calculations.  The
    energies are given relative to the respective $I4/mcm$ phase in
    the units of meV/atom. Note that the order parameters refer to
    that of the initial structures.  Symmetry-allowed distortions
    appear for the final relaxed structures. Only two distortions could
    be stabilized for EuNiO$_2$.}
  \begin{ruledtabular}  
    \begin{tabular}{l d{5.2} d{5.2} d{5.2} d{5.2} d{5.2}}
      & \multicolumn{1}{c}{$Cmca$} & 
      \multicolumn{1}{c}{$P4/ncc$} & 
      \multicolumn{1}{c}{$Cmca$} & 
      \multicolumn{1}{c}{$Pbcn$} & 
      \multicolumn{1}{c}{$Pbca$} \\
      & & & & & \multicolumn{1}{c}{$X_2^-(0,a)$}\\
      $R$NiO$_2$ & \multicolumn{1}{c}{$X_2^-(0,a)$} &
      \multicolumn{1}{c}{$X_2^-(a,a)$} &
      \multicolumn{1}{c}{$M_5^+(a,0)$} &
      \multicolumn{1}{c}{$M_5^+(a,a)$} &
      \multicolumn{1}{c}{+}\\
      & & & & & \multicolumn{1}{c}{$M_5^+(b,0)$} \\
      \hline
      Eu  & <-0.1 & <-0.1 &       &       &        \\
      Gd  &  -2.8 &  -2.1 & <-0.1 & <-0.1 &  -2.5  \\
      Tb  &  -2.3 &  -6.1 &  -1.4 & -10.3 & -10.4  \\
      Dy  &  -8.9 & -10.6 &  -5.1 & -17.5 & -16.7  \\ 
      Y   &  -9.0 & -11.4 &  -5.1 & -17.9 & -17.2  \\
      Ho  & -11.9 & -15.1 &  -7.9 & -24.5 & -22.8  \\
      Er  & -15.1 & -20.1 & -11.1 & -31.7 & -29.2  \\
      Tm  & -17.9 & -24.5 & -14.1 & -38.4 & -35.1  \\
      Yb  & -20.0 & -28.1 & -16.6 & -43.5 & -39.4  \\
      Lu  & -23.1 & -32.4 & -20.6 & -49.6 & -44.7  \\
      \end{tabular}
    \end{ruledtabular}
\end{table}

For EuNiO$_2$, which has only the $X_2^-$ instability, just the
$X_2^-(0,a)$ and $X_2^-(a,a)$ isotropy subgroups could be stabilized.
They are degenerate within the numerical accuracy of my calculations.
Correspondingly, the distortions from the $I4/mcm$ structures are also
small.  For example, the $X_2^-(0,a)$ phase exhibits a shear
distortion of the Eu squares of 0.16$^{\circ}$ and a buckling of the
O-Ni-O angles within the Ni-O squares of 0.11$^{\circ}$.

Compounds with rare-earth ions smaller than Eu exhibit additional
$M_5^+$ instability, and distortions due this mode could be stabilized
for these compounds.  In GdNiO$_3$, the $M_5^+(a,0)$ and $M_5^+(a,a)$
phases are degenerate with the $I4/mcm$ phase within the numerical
accuracy of the calculations.  The $X_2^-(0,a)$ phase has the lowest
energy with a relative energy of $-2.8$ meV/atom.  This is close to
that of the $X_2^-(a,a)$ and $X_2^-(0,a) + M_5^+(b,0)$ phases, which
have relative energies of $-2.1$ and $-2.5$ meV/atom, respectively.
Such a near degeneracy in the calculated total energies indicates that
there will be fluctuations among these structures at temperatures
higher than corresponding to the energy scale of $0.7$ meV/atom
($\approx$ 8 K).

TbNiO$_2$ exhibits a qualitative change in the energetics of the
structural distortions, with its $X_2^-(0,a) + M_5^+(b,0)$ phase now
having the lowest energy with a relative value of $-10.4$ meV/atom.
Its $M_5^+(a,a)$ distortion has a slightly larger relative energy of
$-10.3$ meV/atom.  The larger energy gain is accompanied by a
substantial increase of structural distortions.  For example, the Tb
squares distort by 9.6$^{\circ}$ and 4.5$^{\circ}$ in the two phases,
respectively.  This is a large distinction between two structures that
are so close in energy.

There is yet another qualitative change in the energetics of the
structural distortions as the size of the rare-earth ion is decreased
further.  For the compounds with $R$ = Dy--Lu, the $M_5^+(a,a)$ phase
has the lowest energy, with the $X_2^-(0,a) + M_5^+(b,0)$ phase lying
above it in relative energy.  The energy difference between the
$M_5^+(a,a)$ and $X_2^-(0,a) + M_5^+(b,0)$ phases is only 0.1 meV/atom
for TbNiO$_2$, and the difference remains below 2.0 meV/atom up to
HoNiO$_2$.  The energy gap between these phases progressively
increases to a value of
4.9 meV/atom for LuNiO$_2$.

Hence, the two defining characteristics of the energetics of the
structural distortions of the rare-earth infinite-layer nickelates is
the presence of competing phases with similar energies and tunability of
the order parameter of the lowest-energy phase as a function of the 
rare-earth ion size.  Both these features imply that any structural
transition that is possible due to the phonon instabilities present in
these materials can be driven to 0 K.

Interestingly, calculations show that a related material LaNiO$_3$
exhibits similar structural degeneracies and is also in the vicinity
of a structural quantum critical point, whereas YNiO$_3$ was proposed
not to host any structural fluctuations
\cite{subedi2018lno3}. Structural distortions in these materials occur
due to the $M_3^+$ and $R_4^+$ phonon instabilities present in the
parent cubic phase. It was argued that competing structural phases
occur in LaNiO$_3$ due to a large difference in the imaginary
frequencies between the unstable $M_3^+$ and $R_4^+$ modes, while a
similar value of the imaginary frequency of these modes resulted in a
large energy gain for the $Pnma$ structure with the $a^+b^-b^-$ tilt
pattern in YNiO$_3$.  Therefore, a large difference between the calculated
imaginary frequencies of two distinct phonon modes seems to lead to
energetically competing structural distortions.

\subsection{Does magnetic ordering change the energetics of the
  structural distortions?} 

\begin{table}
  \caption{\label{tab:enesp} Total energies of the five distortions
    due to the $X_2^-$ and $M_5^+$ instabilities of the $I4/mcm$ phase
    in the presence of the G-type antiferromagnetic order. The
    energies are given relative to the respective antiferromagnetic
    $I4/mcm$ phase in the units of meV/atom. Note that the order
    parameters refer to that of the initial structures.
    Symmetry-allowed distortions appear for the final relaxed
    structures. }
  \begin{ruledtabular}  
    \begin{tabular}{l d{5.2} d{5.2} d{5.2} d{5.2} d{5.2}}
      & \multicolumn{1}{c}{$Cmca$} & 
      \multicolumn{1}{c}{$P4/ncc$} & 
      \multicolumn{1}{c}{$Cmca$} & 
      \multicolumn{1}{c}{$Pbcn$} & 
      \multicolumn{1}{c}{$Pbca$} \\
      & & & & & \multicolumn{1}{c}{$X_2^-(0,a)$}\\
      $R$NiO$_2$ & \multicolumn{1}{c}{$X_2^-(0,a)$} &
      \multicolumn{1}{c}{$X_2^-(a,a)$} &
      \multicolumn{1}{c}{$M_5^+(a,0)$} &
      \multicolumn{1}{c}{$M_5^+(a,a)$} &
      \multicolumn{1}{c}{+}\\
      & & & & & \multicolumn{1}{c}{$M_5^+(b,0)$} \\
      \hline
      Eu  &  -0.3 &  -0.3 & <-0.1 & <-0.1 &  -0.4  \\
      Gd  &  -2.8 &  -4.1 &  -1.7 & -10.9 &  -9.1  \\
      Tb  &  -5.3 &  -8.3 &  -4.2 & -18.0 & -15.3  \\
      Dy  &  -7.9 & -12.4 &  -6.9 & -24.8 & -21.2  \\ 
      Y   &  -8.1 & -12.8 &  -7.1 & -24.5 & -21.0  \\
      Ho  & -10.6 & -16.5 &  -9.9 & -31.2 & -26.8  \\
      Er  & -13.7 & -21.0 & -13.5 & -37.8 & -32.7  \\
      Tm  & -16.8 & -24.9 & -17.0 & -43.8 & -37.9  \\
      Yb  & -19.4 & -27.7 & -20.2 & -48.0 & -41.5  \\
      Lu  & -23.1 & -31.3 & -24.6 & -53.2 & -46.1  \\
      \end{tabular}
    \end{ruledtabular}
\end{table}

Although magnetic ordering has not been experimentally observed in any
of the infinite-layer rare-earth nickelates, DFT calculations find a
small gain in energy when an antiferromagnetic solution is allowed,
suggesting a weak tendency toward antiferromagnetism in these
materials \cite{lee2004,botana2020,liu2020,been2021}.  Zhang \etal
have shown that the phonon instability at the $A$ point in the
$P4/mmm$ phase is sensitive to the presence of antiferromagnetic order
\cite{zhang2022phase}.  In particular, they find that the $A_3^+$ mode
in NdNiO$_2$, which is calculated to be stable in the
non-spin-polarized phase, becomes unstable when the G-type
antiferromagnetic solution is allowed.  They are then able to
stabilize the $I4/mcm$ phase for NdNiO$_2$, as well as for compounds
with Pr, Sm and smaller rare-earth ions.  I was able to reproduce this
result.  In fact, I find the $I4/mcm$ phase to be lower in energy than
the $P4/mmm$ phase for all $R$NiO$_2$ compounds except LaNiO$_2$ (\ie
$R$ = Ce--Lu) when the G-type order is imposed. [Similar structural
  instability emerges also when the C-type order is allowed because the
  calculated energy difference between two antiferromagnetic states is
  less than 1.0 meV/Ni.]  The calculated energy gain of the G-type
$I4/mcm$ phase relative to the respective non-spin-polarized $P4/mmm$
phase ranges from 17.1 eV/atom for CeNiO$_2$ to 111.9 meV/atom for
LuNiO$_2$.

Zhang \etal's results suggest that the inclusion of antiferromagnetic
order can qualitatively change the energetics of the structural
instabilities present in this family of compounds.  I generated the
minimal conventional unit cells that allow the G-type
antiferromagnetic ordering for the five isotropy subgroups of the
$I4/mcm$ phase stabilized above.  I then fully relaxed these
structures by minimizing the atomic forces and lattice stress in the
presence of the antiferromagnetic order.  As in the case of
non-spin-polarized calculations discussed above, the compounds with
$R$ = Ce--Sm relaxed back to the $I4/mcm$ phase also when the G-type
antiferromagnetic solution was allowed.  The energies of the rest of
the compounds that could be stabilized in the antiferromagnetically
ordered lower-symmetry phases are given in Table \ref{tab:enesp}
relative to that of the respective antiferromagnetically ordered
$I4/mcm$ phase.

I find the inclusion of the antiferromagnetic order marginally lowers
(by few meV/atom or less) the relative energies of the five distorted
structures.  For EuNiO$_2$, the $M_5^+ (a,0)$, $M_5^+ (a,a)$, and
$X_2^-(0,a) + M_5^+(b,0)$ phases become more stable than the $I4/mcm$
phase, which was not the case when a magnetic solution was not
allowed.  Otherwise, the structural energetics with antiferromagnetism
is remarkably similar to that of the non-spin-polarized case.  In
particular, it still exhibits near-degenerate structures and a
change in the order parameter of the lowest-lying phase as a function
of the rare-earth ion.  If these late rare-earth compounds also do not
exhibit magnetic order like the early ones, this suggests that a
structural quantum criticality occurs in the presence of magnetic
fluctuations.

\subsection{Are there other distortions with even lower energy?}

\begin{figure}
  \includegraphics[width=\columnwidth]{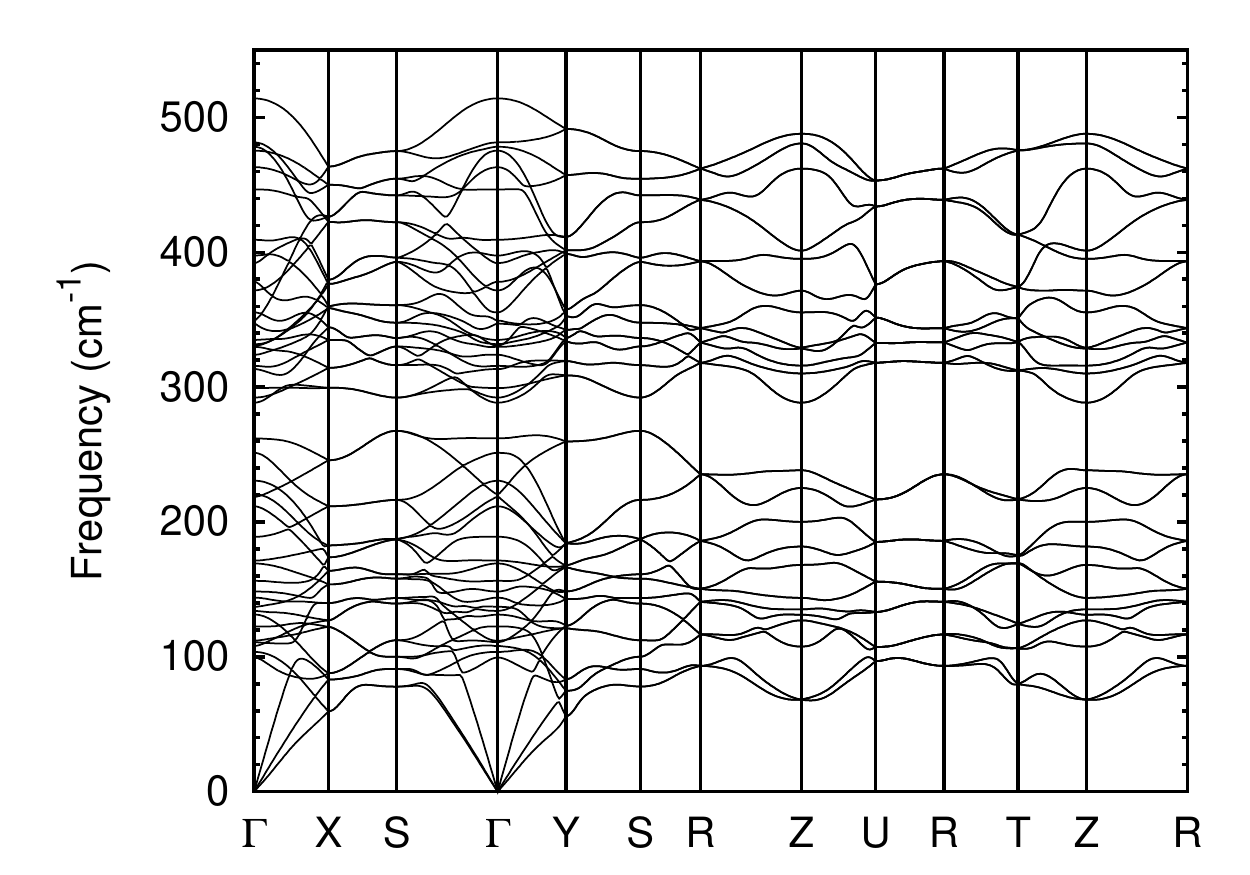}
  \caption{Calculated non-spin-polarized phonon dispersions of
    fully-relaxed YNiO$_2$ in the $Pbcn$ phase corresponding to the
    $M_5^+$ $(a,a)$ distortion of the $I4/mcm$ phase. The
    high-symmetry points are $\Gamma$ $(0,0,0)$, $X$
    $(\frac{1}{2},0,0)$, $S$ $(\frac{1}{2}, \frac{1}{2}, 0)$, $Y$
    $(0,\frac{1}{2},0)$, $R$ $(\frac{1}{2},\frac{1}{2},\frac{1}{2})$
    $Z$ $(0,0,\frac{1}{2})$, $U$ $(\frac{1}{2},0,\frac{1}{2})$, and
    $T$ $(0,\frac{1}{2},\frac{1}{2})$ in terms of the reciprocal
    lattice vectors. }
  \label{fig:phsixty}
\end{figure}

In principle, the five low-symmetry structures that could be stabilized
due to the $X_2^-$ and $M_5^+$ unstable modes of the $I4/mcm$ phase
may yet exhibit structural instabilities.  I did not have the
computational resources to calculate the phonon dispersions of all the
low-symmetry phases of the ten compounds with $R$ = Eu--Lu for which
these structures could be stabilized.  Nevertheless, I calculated the
phonon dispersions of YNiO$_2$ in its lowest-energy $Pbcn$ $M_5^+$
$(a,a)$ phase, which is shown in Fig.~\ref{fig:phsixty}.  This phase
does not exhibit any unstable phonon modes, implying that it is
dynamically stable.  Since $Pbcn$ $M_5^+ (a,a)$ phase is the
lowest-energy structure for the compounds with $R$=Dy--Lu, there are
unlikely to be other lower-energy distortions that can be continuously
connected to the parent $P4/mmm$ phase via phonon instabilities.

Carrasco \'Alvarez \etal have argued using first principles
calculations that the $Pbnm$ structure with a $a^-a^-c^+$ tilt pattern
of the NiO$_4$ square is the ground state phase of the infinite-layer
nickelates with smaller rare-earth ions \cite{carrasco2021}.  I tried
to able to stabilize this phase for YNiO$_2$ on a $2\times2\times2$
supercell of the $P4/mmm$ structure and was able to do so only when
antiferromagnetism was allowed.  This is again suggestive of the
strong coupling between magnetism and structural instabilities in
these materials \cite{zhang2022phase}.  The fully relaxed $Pbnm$
$a^-a^-c^+$ structure also exhibits the antipolar motion of the $R$
ions, as found by Carrasco \'Alvarez \etal.
I obtain a value
of 65.5 meV/atom for the calculated energy gain of the $Pbnm$
$a^-a^-c^+$ phase relative to the $P4/mmm$ phase when the C-type
antiferromagnetic order is imposed in both phases.
For comparison, the energy gain of the $M_5^+$ $(a,a)$
distortion of the $I4/mcm$ phase relative to the $P4/mmm$ phase for the
C-type antiferromagnetic state is 74.2 meV/atom.  Therefore, the
$Pbnm$ $a^-a^-c^+$ structure is likely not the ground state phase of
the infinite-layer nickelates with small rare-earth ions.

\section{Summary and Conclusions}

In summary, I have investigated the energetics of the structural
distortions of the infinite-layer rare-earth nickelates and find that
these materials may lie near a structural quantum critical point.  I
first confirmed the previous results showing the parent $P4/mmm$ phase
of the $R$ = Sm--Lu compounds to be unstable towards the $I4/mcm$
phase due to an $A_3^+$ phonon instability
\cite{xia2021,bernardini2021,zhang2022phase}.  I then calculated the
non-spin-polarised phonon dispersions of the $R$ = Sm--Lu compounds in
the $I4/mcm$ phase. Only SmNiO$_2$ is dynamically stable in this
structure.  GdNiO$_2$ exhibits an $X_2^-$ instablity, while the rest
exhibit an additional $M_5^+$ instability.  I used group-theoretical
analysis to identify all twelve isotropy subgroups that are possible
due these two instabilities and generated distorted structures
corresponding to their order parameters. After full structural
relaxations minimizing both the atomic forces and lattice stresses,
only five isotropy subgroups corresponding to the $Cmca$ $X_2^-(0,a)$,
$P4/ncc$ $X_2^-(a,a)$, $Cmca$ $M_5^+(a,0)$, $Pbcn$ $M_5^+(a,a)$, and
$Pbca$ $X_2^-(0,a) + M_5^+(b,0)$ order parameters could be stabilized.

From the non-spin-polarized calculations, the $M_5^+(a,a)$ phase has
the lowest energy for the smallest rare-earth compound LuNiO$_2$
relative to its $I4/mcm$ phase.  Its $X_2^-(0,a) + M_5^+(b,0)$ phase
is higher in energy than the $M_5^+(a,a)$ phase by 4.9 meV/atom.  The
energy difference between these two phases decreases as the rare-earth ion
size is increased, reaching a value of less than 2.0 meV/atom for
compounds with $R$ = Ho or smaller. There is a qualitative change in
the energetics for TbNiO$_2$, which now has the $X_2^-(0,a) +
M_5^+(b,0)$ phase lower in energy than the $M_5^+ (a,a)$ phase, albeit
only by 0.1 meV/atom. The energetics shuffle again for GdNiO$_2$. Its
lowest-energy phase is $X_2^- (0,a)$, and it is only 0.4 and 0.7 meV/atom
higher in energy than the $X_2^- (0,a) + M_5^+ (b,0)$ and $X_2^-
(a,a)$ phases, respectively.

The near degeneracy of the phases corresponding to distinct structural
order parameters and the tunability of the relative energies of these
competing phases both suggest that any structural transition that
might occur due to the phonon instabilities present in the $I4/mcm$
phase of these materials can be suppressed to 0 K.  I find the
inclusion of the nearest-neighbor antiferromagnetic order only lowers
the relative energy of the low-symmetry phases by few meV/atom.  In
particular, the qualitative and quantitative aspects of the
competition between different order parameters remains.  If the
infinite-layer nickelates with smaller rare-earth ions do not exhibit
magnetic ordering like those with the larger ones, this implies that
the proximity to the structural quantum criticality occurs in the
presence of magnetic fluctuations.

Recent experiments on (Ca$_x$Sr$_{1-x}$)$_3$Rh$_4$Sn$_{13}$,
(Ca$_x$Sr$_{1-x}$)$_3$Ir$_4$Sn$_{13}$, and LaCu$_{6-x}$Au$_x$ show
these materials to host a structural quantum critical point, which
manifests as a softening of the associated phonon modes at $T
\rightarrow 0$ K
\cite{goh2015,yu2015,hu2017,cheung2018,poudel2016}. The measured heat
capacity is strongly enhanced near the critical point due to the
ensuing preponderance of the phonon density of states.  Similar
softening of the $X_2^-$ and $M_5^+$ phonon modes and enhancement
of the heat capacity will be the experimental consequences of the
structural quantum criticality proposed here for the infinite-layer
nickelates.  Moreover, superconducting $T_c$ is found to peak near the
quantum critical point in (Ca$_x$Sr$_{1-x}$)$_3$Ir$_4$Sn$_{13}$
\cite{biswas2015}, and the infinite-layer nickelates should exhibit
similar enhancement if phonons play an important role in their
superconductivity.

\section{acknowledgements}
This work was supported by GENCI-TGCC under grant no.\ A0110913028. 


\begin{thebibliography}{103}%
\makeatletter
\providecommand \@ifxundefined [1]{%
 \@ifx{#1\undefined}
}%
\providecommand \@ifnum [1]{%
 \ifnum #1\expandafter \@firstoftwo
 \else \expandafter \@secondoftwo
 \fi
}%
\providecommand \@ifx [1]{%
 \ifx #1\expandafter \@firstoftwo
 \else \expandafter \@secondoftwo
 \fi
}%
\providecommand \natexlab [1]{#1}%
\providecommand \enquote  [1]{``#1''}%
\providecommand \bibnamefont  [1]{#1}%
\providecommand \bibfnamefont [1]{#1}%
\providecommand \citenamefont [1]{#1}%
\providecommand \href@noop [0]{\@secondoftwo}%
\providecommand \href [0]{\begingroup \@sanitize@url \@href}%
\providecommand \@href[1]{\@@startlink{#1}\@@href}%
\providecommand \@@href[1]{\endgroup#1\@@endlink}%
\providecommand \@sanitize@url [0]{\catcode `\\12\catcode `\$12\catcode
  `\&12\catcode `\#12\catcode `\^12\catcode `\_12\catcode `\%12\relax}%
\providecommand \@@startlink[1]{}%
\providecommand \@@endlink[0]{}%
\providecommand \url  [0]{\begingroup\@sanitize@url \@url }%
\providecommand \@url [1]{\endgroup\@href {#1}{\urlprefix }}%
\providecommand \urlprefix  [0]{URL }%
\providecommand \Eprint [0]{\href }%
\providecommand \doibase [0]{https://doi.org/}%
\providecommand \selectlanguage [0]{\@gobble}%
\providecommand \bibinfo  [0]{\@secondoftwo}%
\providecommand \bibfield  [0]{\@secondoftwo}%
\providecommand \translation [1]{[#1]}%
\providecommand \BibitemOpen [0]{}%
\providecommand \bibitemStop [0]{}%
\providecommand \bibitemNoStop [0]{.\EOS\space}%
\providecommand \EOS [0]{\spacefactor3000\relax}%
\providecommand \BibitemShut  [1]{\csname bibitem#1\endcsname}%
\let\auto@bib@innerbib\@empty
\bibitem [{\citenamefont {Li}\ \emph {et~al.}(2019)\citenamefont {Li},
  \citenamefont {Lee}, \citenamefont {Wang}, \citenamefont {Osada},
  \citenamefont {Crossley}, \citenamefont {Lee}, \citenamefont {Cui},
  \citenamefont {Hikita},\ and\ \citenamefont {Hwang}}]{li2019}%
  \BibitemOpen
  \bibfield  {author} {\bibinfo {author} {\bibfnamefont {D.}~\bibnamefont
  {Li}}, \bibinfo {author} {\bibfnamefont {K.}~\bibnamefont {Lee}}, \bibinfo
  {author} {\bibfnamefont {B.~Y.}\ \bibnamefont {Wang}}, \bibinfo {author}
  {\bibfnamefont {M.}~\bibnamefont {Osada}}, \bibinfo {author} {\bibfnamefont
  {S.}~\bibnamefont {Crossley}}, \bibinfo {author} {\bibfnamefont {H.~R.}\
  \bibnamefont {Lee}}, \bibinfo {author} {\bibfnamefont {Y.}~\bibnamefont
  {Cui}}, \bibinfo {author} {\bibfnamefont {Y.}~\bibnamefont {Hikita}},\ and\
  \bibinfo {author} {\bibfnamefont {H.~Y.}\ \bibnamefont {Hwang}},\ }\bibfield
  {title} {\bibinfo {title} {Superconductivity in an infinite-layer
  nickelate},\ }\href@noop {} {\bibfield  {journal} {\bibinfo  {journal}
  {Nature}\ }\textbf {\bibinfo {volume} {572}},\ \bibinfo {pages} {624}
  (\bibinfo {year} {2019})}\BibitemShut {NoStop}%
\bibitem [{\citenamefont {Osada}\ \emph
  {et~al.}(2020{\natexlab{a}})\citenamefont {Osada}, \citenamefont {Wang},
  \citenamefont {Goodge}, \citenamefont {Lee}, \citenamefont {Yoon},
  \citenamefont {Sakuma}, \citenamefont {Li}, \citenamefont {Miura},
  \citenamefont {Kourkoutis},\ and\ \citenamefont {Hwang}}]{osada2020nl}%
  \BibitemOpen
  \bibfield  {author} {\bibinfo {author} {\bibfnamefont {M.}~\bibnamefont
  {Osada}}, \bibinfo {author} {\bibfnamefont {B.~Y.}\ \bibnamefont {Wang}},
  \bibinfo {author} {\bibfnamefont {B.~H.}\ \bibnamefont {Goodge}}, \bibinfo
  {author} {\bibfnamefont {K.}~\bibnamefont {Lee}}, \bibinfo {author}
  {\bibfnamefont {H.}~\bibnamefont {Yoon}}, \bibinfo {author} {\bibfnamefont
  {K.}~\bibnamefont {Sakuma}}, \bibinfo {author} {\bibfnamefont
  {D.}~\bibnamefont {Li}}, \bibinfo {author} {\bibfnamefont {M.}~\bibnamefont
  {Miura}}, \bibinfo {author} {\bibfnamefont {L.~F.}\ \bibnamefont
  {Kourkoutis}},\ and\ \bibinfo {author} {\bibfnamefont {H.~Y.}\ \bibnamefont
  {Hwang}},\ }\bibfield  {title} {\bibinfo {title} {A superconducting
  praseodymium nickelate with infinite layer structure},\ }\href@noop {}
  {\bibfield  {journal} {\bibinfo  {journal} {Nano Lett.}\ }\textbf {\bibinfo
  {volume} {20}},\ \bibinfo {pages} {5735} (\bibinfo {year}
  {2020}{\natexlab{a}})}\BibitemShut {NoStop}%
\bibitem [{\citenamefont {Osada}\ \emph {et~al.}(2021)\citenamefont {Osada},
  \citenamefont {Wang}, \citenamefont {Goodge}, \citenamefont {Harvey},
  \citenamefont {Lee}, \citenamefont {Li}, \citenamefont {Kourkoutis},\ and\
  \citenamefont {Hwang}}]{osada2021}%
  \BibitemOpen
  \bibfield  {author} {\bibinfo {author} {\bibfnamefont {M.}~\bibnamefont
  {Osada}}, \bibinfo {author} {\bibfnamefont {B.~Y.}\ \bibnamefont {Wang}},
  \bibinfo {author} {\bibfnamefont {B.~H.}\ \bibnamefont {Goodge}}, \bibinfo
  {author} {\bibfnamefont {S.~P.}\ \bibnamefont {Harvey}}, \bibinfo {author}
  {\bibfnamefont {K.}~\bibnamefont {Lee}}, \bibinfo {author} {\bibfnamefont
  {D.}~\bibnamefont {Li}}, \bibinfo {author} {\bibfnamefont {L.~F.}\
  \bibnamefont {Kourkoutis}},\ and\ \bibinfo {author} {\bibfnamefont {H.~Y.}\
  \bibnamefont {Hwang}},\ }\bibfield  {title} {\bibinfo {title} {Nickelate
  superconductivity without rare-earth magnetism: {(La,Sr)NiO$_2$}},\
  }\href@noop {} {\bibfield  {journal} {\bibinfo  {journal} {Adv. Mater.}\
  }\textbf {\bibinfo {volume} {33}},\ \bibinfo {pages} {2104083} (\bibinfo
  {year} {2021})}\BibitemShut {NoStop}%
\bibitem [{\citenamefont {Zeng}\ \emph
  {et~al.}(2022{\natexlab{a}})\citenamefont {Zeng}, \citenamefont {Li},
  \citenamefont {Chow}, \citenamefont {Cao}, \citenamefont {Zhang},
  \citenamefont {Tang}, \citenamefont {Yin}, \citenamefont {Lim}, \citenamefont
  {Hu}, \citenamefont {Yang} \emph {et~al.}}]{zeng2022sa}%
  \BibitemOpen
  \bibfield  {author} {\bibinfo {author} {\bibfnamefont {S.}~\bibnamefont
  {Zeng}}, \bibinfo {author} {\bibfnamefont {C.}~\bibnamefont {Li}}, \bibinfo
  {author} {\bibfnamefont {L.~E.}\ \bibnamefont {Chow}}, \bibinfo {author}
  {\bibfnamefont {Y.}~\bibnamefont {Cao}}, \bibinfo {author} {\bibfnamefont
  {Z.}~\bibnamefont {Zhang}}, \bibinfo {author} {\bibfnamefont {C.~S.}\
  \bibnamefont {Tang}}, \bibinfo {author} {\bibfnamefont {X.}~\bibnamefont
  {Yin}}, \bibinfo {author} {\bibfnamefont {Z.~S.}\ \bibnamefont {Lim}},
  \bibinfo {author} {\bibfnamefont {J.}~\bibnamefont {Hu}}, \bibinfo {author}
  {\bibfnamefont {P.}~\bibnamefont {Yang}}, \emph {et~al.},\ }\bibfield
  {title} {\bibinfo {title} {Superconductivity in infinite-layer nickelate
  {La$_{1-x}$Ca$_x$NiO$_2$} thin films},\ }\href@noop {} {\bibfield  {journal}
  {\bibinfo  {journal} {Sci. Adv.}\ }\textbf {\bibinfo {volume} {8}},\ \bibinfo
  {pages} {eabl9927} (\bibinfo {year} {2022}{\natexlab{a}})}\BibitemShut
  {NoStop}%
\bibitem [{\citenamefont {Hepting}\ \emph {et~al.}(2020)\citenamefont
  {Hepting}, \citenamefont {Li}, \citenamefont {Jia}, \citenamefont {Lu},
  \citenamefont {Paris}, \citenamefont {Tseng}, \citenamefont {Feng},
  \citenamefont {Osada}, \citenamefont {Been}, \citenamefont {Hikita} \emph
  {et~al.}}]{hepting2020}%
  \BibitemOpen
  \bibfield  {author} {\bibinfo {author} {\bibfnamefont {M.}~\bibnamefont
  {Hepting}}, \bibinfo {author} {\bibfnamefont {D.}~\bibnamefont {Li}},
  \bibinfo {author} {\bibfnamefont {C.}~\bibnamefont {Jia}}, \bibinfo {author}
  {\bibfnamefont {H.}~\bibnamefont {Lu}}, \bibinfo {author} {\bibfnamefont
  {E.}~\bibnamefont {Paris}}, \bibinfo {author} {\bibfnamefont
  {Y.}~\bibnamefont {Tseng}}, \bibinfo {author} {\bibfnamefont
  {X.}~\bibnamefont {Feng}}, \bibinfo {author} {\bibfnamefont {M.}~\bibnamefont
  {Osada}}, \bibinfo {author} {\bibfnamefont {E.}~\bibnamefont {Been}},
  \bibinfo {author} {\bibfnamefont {Y.}~\bibnamefont {Hikita}}, \emph
  {et~al.},\ }\bibfield  {title} {\bibinfo {title} {Electronic structure of the
  parent compound of superconducting infinite-layer nickelates},\ }\href@noop
  {} {\bibfield  {journal} {\bibinfo  {journal} {Nat. Mater.}\ }\textbf
  {\bibinfo {volume} {19}},\ \bibinfo {pages} {381} (\bibinfo {year}
  {2020})}\BibitemShut {NoStop}%
\bibitem [{\citenamefont {Li}\ \emph {et~al.}(2020{\natexlab{a}})\citenamefont
  {Li}, \citenamefont {He}, \citenamefont {Si}, \citenamefont {Zhu},
  \citenamefont {Zhang},\ and\ \citenamefont {Wen}}]{li2020absence}%
  \BibitemOpen
  \bibfield  {author} {\bibinfo {author} {\bibfnamefont {Q.}~\bibnamefont
  {Li}}, \bibinfo {author} {\bibfnamefont {C.}~\bibnamefont {He}}, \bibinfo
  {author} {\bibfnamefont {J.}~\bibnamefont {Si}}, \bibinfo {author}
  {\bibfnamefont {X.}~\bibnamefont {Zhu}}, \bibinfo {author} {\bibfnamefont
  {Y.}~\bibnamefont {Zhang}},\ and\ \bibinfo {author} {\bibfnamefont {H.-H.}\
  \bibnamefont {Wen}},\ }\bibfield  {title} {\bibinfo {title} {Absence of
  superconductivity in bulk nd1- xsrxnio2},\ }\href@noop {} {\bibfield
  {journal} {\bibinfo  {journal} {Communications Materials}\ }\textbf {\bibinfo
  {volume} {1}},\ \bibinfo {pages} {1} (\bibinfo {year}
  {2020}{\natexlab{a}})}\BibitemShut {NoStop}%
\bibitem [{\citenamefont {Fu}\ \emph {et~al.}(2019)\citenamefont {Fu},
  \citenamefont {Wang}, \citenamefont {Cheng}, \citenamefont {Pei},
  \citenamefont {Zhou}, \citenamefont {Chen}, \citenamefont {Wang},
  \citenamefont {Zhao}, \citenamefont {Jiang}, \citenamefont {Liu} \emph
  {et~al.}}]{fu2019}%
  \BibitemOpen
  \bibfield  {author} {\bibinfo {author} {\bibfnamefont {Y.}~\bibnamefont
  {Fu}}, \bibinfo {author} {\bibfnamefont {L.}~\bibnamefont {Wang}}, \bibinfo
  {author} {\bibfnamefont {H.}~\bibnamefont {Cheng}}, \bibinfo {author}
  {\bibfnamefont {S.}~\bibnamefont {Pei}}, \bibinfo {author} {\bibfnamefont
  {X.}~\bibnamefont {Zhou}}, \bibinfo {author} {\bibfnamefont {J.}~\bibnamefont
  {Chen}}, \bibinfo {author} {\bibfnamefont {S.}~\bibnamefont {Wang}}, \bibinfo
  {author} {\bibfnamefont {R.}~\bibnamefont {Zhao}}, \bibinfo {author}
  {\bibfnamefont {W.}~\bibnamefont {Jiang}}, \bibinfo {author} {\bibfnamefont
  {C.}~\bibnamefont {Liu}}, \emph {et~al.},\ }\bibfield  {title} {\bibinfo
  {title} {Core-level x-ray photoemission and {R}aman spectroscopy studies on
  electronic structures in {M}ott-{H}ubbard type nickelate oxide {NdNiO$_2$}},\
  }\href@noop {} {\bibfield  {journal} {\bibinfo  {journal} {arXiv preprint
  arXiv:1911.03177}\ } (\bibinfo {year} {2019})}\BibitemShut {NoStop}%
\bibitem [{\citenamefont {Lee}\ \emph {et~al.}(2020)\citenamefont {Lee},
  \citenamefont {Goodge}, \citenamefont {Li}, \citenamefont {Osada},
  \citenamefont {Wang}, \citenamefont {Cui}, \citenamefont {Kourkoutis},\ and\
  \citenamefont {Hwang}}]{lee2020ap}%
  \BibitemOpen
  \bibfield  {author} {\bibinfo {author} {\bibfnamefont {K.}~\bibnamefont
  {Lee}}, \bibinfo {author} {\bibfnamefont {B.~H.}\ \bibnamefont {Goodge}},
  \bibinfo {author} {\bibfnamefont {D.}~\bibnamefont {Li}}, \bibinfo {author}
  {\bibfnamefont {M.}~\bibnamefont {Osada}}, \bibinfo {author} {\bibfnamefont
  {B.~Y.}\ \bibnamefont {Wang}}, \bibinfo {author} {\bibfnamefont
  {Y.}~\bibnamefont {Cui}}, \bibinfo {author} {\bibfnamefont {L.~F.}\
  \bibnamefont {Kourkoutis}},\ and\ \bibinfo {author} {\bibfnamefont {H.~Y.}\
  \bibnamefont {Hwang}},\ }\bibfield  {title} {\bibinfo {title} {Aspects of the
  synthesis of thin film superconducting infinite-layer nickelates},\
  }\href@noop {} {\bibfield  {journal} {\bibinfo  {journal} {Apl Mater.}\
  }\textbf {\bibinfo {volume} {8}},\ \bibinfo {pages} {041107} (\bibinfo {year}
  {2020})}\BibitemShut {NoStop}%
\bibitem [{\citenamefont {Li}\ \emph {et~al.}(2020{\natexlab{b}})\citenamefont
  {Li}, \citenamefont {Wang}, \citenamefont {Lee}, \citenamefont {Harvey},
  \citenamefont {Osada}, \citenamefont {Goodge}, \citenamefont {Kourkoutis},\
  and\ \citenamefont {Hwang}}]{li2020}%
  \BibitemOpen
  \bibfield  {author} {\bibinfo {author} {\bibfnamefont {D.}~\bibnamefont
  {Li}}, \bibinfo {author} {\bibfnamefont {B.~Y.}\ \bibnamefont {Wang}},
  \bibinfo {author} {\bibfnamefont {K.}~\bibnamefont {Lee}}, \bibinfo {author}
  {\bibfnamefont {S.~P.}\ \bibnamefont {Harvey}}, \bibinfo {author}
  {\bibfnamefont {M.}~\bibnamefont {Osada}}, \bibinfo {author} {\bibfnamefont
  {B.~H.}\ \bibnamefont {Goodge}}, \bibinfo {author} {\bibfnamefont {L.~F.}\
  \bibnamefont {Kourkoutis}},\ and\ \bibinfo {author} {\bibfnamefont {H.~Y.}\
  \bibnamefont {Hwang}},\ }\bibfield  {title} {\bibinfo {title}
  {Superconducting dome in {Nd$_{1-x}$Sr$_x$NiO$_2$} infinite layer films},\
  }\href@noop {} {\bibfield  {journal} {\bibinfo  {journal} {Phys. Rev. Lett.}\
  }\textbf {\bibinfo {volume} {125}},\ \bibinfo {pages} {027001} (\bibinfo
  {year} {2020}{\natexlab{b}})}\BibitemShut {NoStop}%
\bibitem [{\citenamefont {Zeng}\ \emph {et~al.}(2020)\citenamefont {Zeng},
  \citenamefont {Tang}, \citenamefont {Yin}, \citenamefont {Li}, \citenamefont
  {Li}, \citenamefont {Huang}, \citenamefont {Hu}, \citenamefont {Liu},
  \citenamefont {Omar}, \citenamefont {Jani} \emph {et~al.}}]{zeng2020prl}%
  \BibitemOpen
  \bibfield  {author} {\bibinfo {author} {\bibfnamefont {S.}~\bibnamefont
  {Zeng}}, \bibinfo {author} {\bibfnamefont {C.~S.}\ \bibnamefont {Tang}},
  \bibinfo {author} {\bibfnamefont {X.}~\bibnamefont {Yin}}, \bibinfo {author}
  {\bibfnamefont {C.}~\bibnamefont {Li}}, \bibinfo {author} {\bibfnamefont
  {M.}~\bibnamefont {Li}}, \bibinfo {author} {\bibfnamefont {Z.}~\bibnamefont
  {Huang}}, \bibinfo {author} {\bibfnamefont {J.}~\bibnamefont {Hu}}, \bibinfo
  {author} {\bibfnamefont {W.}~\bibnamefont {Liu}}, \bibinfo {author}
  {\bibfnamefont {G.~J.}\ \bibnamefont {Omar}}, \bibinfo {author}
  {\bibfnamefont {H.}~\bibnamefont {Jani}}, \emph {et~al.},\ }\bibfield
  {title} {\bibinfo {title} {Phase diagram and superconducting dome of
  infinite-layer {Nd$_{1-x}$Sr$_x$NiO$_2$} thin films},\ }\href@noop {}
  {\bibfield  {journal} {\bibinfo  {journal} {Phys. Rev. Lett.}\ }\textbf
  {\bibinfo {volume} {125}},\ \bibinfo {pages} {147003} (\bibinfo {year}
  {2020})}\BibitemShut {NoStop}%
\bibitem [{\citenamefont {Goodge}\ \emph {et~al.}(2021)\citenamefont {Goodge},
  \citenamefont {Li}, \citenamefont {Lee}, \citenamefont {Osada}, \citenamefont
  {Wang}, \citenamefont {Sawatzky}, \citenamefont {Hwang},\ and\ \citenamefont
  {Kourkoutis}}]{goodge2021}%
  \BibitemOpen
  \bibfield  {author} {\bibinfo {author} {\bibfnamefont {B.~H.}\ \bibnamefont
  {Goodge}}, \bibinfo {author} {\bibfnamefont {D.}~\bibnamefont {Li}}, \bibinfo
  {author} {\bibfnamefont {K.}~\bibnamefont {Lee}}, \bibinfo {author}
  {\bibfnamefont {M.}~\bibnamefont {Osada}}, \bibinfo {author} {\bibfnamefont
  {B.~Y.}\ \bibnamefont {Wang}}, \bibinfo {author} {\bibfnamefont {G.~A.}\
  \bibnamefont {Sawatzky}}, \bibinfo {author} {\bibfnamefont {H.~Y.}\
  \bibnamefont {Hwang}},\ and\ \bibinfo {author} {\bibfnamefont {L.~F.}\
  \bibnamefont {Kourkoutis}},\ }\bibfield  {title} {\bibinfo {title} {Doping
  evolution of the {M}ott--{H}ubbard landscape in infinite-layer nickelates},\
  }\href@noop {} {\bibfield  {journal} {\bibinfo  {journal} {Proceedings of the
  National Academy of Sciences}\ }\textbf {\bibinfo {volume} {118}} (\bibinfo
  {year} {2021})}\BibitemShut {NoStop}%
\bibitem [{\citenamefont {Wang}\ \emph
  {et~al.}(2020{\natexlab{a}})\citenamefont {Wang}, \citenamefont {Zheng},
  \citenamefont {Krivyakina}, \citenamefont {Chmaissem}, \citenamefont {Lopes},
  \citenamefont {Lynn}, \citenamefont {Gallington}, \citenamefont {Ren},
  \citenamefont {Rosenkranz}, \citenamefont {Mitchell} \emph
  {et~al.}}]{wang2020synthesis}%
  \BibitemOpen
  \bibfield  {author} {\bibinfo {author} {\bibfnamefont {B.-X.}\ \bibnamefont
  {Wang}}, \bibinfo {author} {\bibfnamefont {H.}~\bibnamefont {Zheng}},
  \bibinfo {author} {\bibfnamefont {E.}~\bibnamefont {Krivyakina}}, \bibinfo
  {author} {\bibfnamefont {O.}~\bibnamefont {Chmaissem}}, \bibinfo {author}
  {\bibfnamefont {P.~P.}\ \bibnamefont {Lopes}}, \bibinfo {author}
  {\bibfnamefont {J.~W.}\ \bibnamefont {Lynn}}, \bibinfo {author}
  {\bibfnamefont {L.~C.}\ \bibnamefont {Gallington}}, \bibinfo {author}
  {\bibfnamefont {Y.}~\bibnamefont {Ren}}, \bibinfo {author} {\bibfnamefont
  {S.}~\bibnamefont {Rosenkranz}}, \bibinfo {author} {\bibfnamefont
  {J.}~\bibnamefont {Mitchell}}, \emph {et~al.},\ }\bibfield  {title} {\bibinfo
  {title} {Synthesis and characterization of bulk {Nd$_{1-x}$Sr$_x$NiO$_2$} and
  {Nd$_{1-x}$Sr$_x$NiO$_3$}},\ }\href@noop {} {\bibfield  {journal} {\bibinfo
  {journal} {Physical review materials}\ }\textbf {\bibinfo {volume} {4}},\
  \bibinfo {pages} {084409} (\bibinfo {year} {2020}{\natexlab{a}})}\BibitemShut
  {NoStop}%
\bibitem [{\citenamefont {Gu}\ \emph {et~al.}(2020{\natexlab{a}})\citenamefont
  {Gu}, \citenamefont {Li}, \citenamefont {Wan}, \citenamefont {Li},
  \citenamefont {Guo}, \citenamefont {Yang}, \citenamefont {Li}, \citenamefont
  {Zhu}, \citenamefont {Pan}, \citenamefont {Nie} \emph {et~al.}}]{gu2020}%
  \BibitemOpen
  \bibfield  {author} {\bibinfo {author} {\bibfnamefont {Q.}~\bibnamefont
  {Gu}}, \bibinfo {author} {\bibfnamefont {Y.}~\bibnamefont {Li}}, \bibinfo
  {author} {\bibfnamefont {S.}~\bibnamefont {Wan}}, \bibinfo {author}
  {\bibfnamefont {H.}~\bibnamefont {Li}}, \bibinfo {author} {\bibfnamefont
  {W.}~\bibnamefont {Guo}}, \bibinfo {author} {\bibfnamefont {H.}~\bibnamefont
  {Yang}}, \bibinfo {author} {\bibfnamefont {Q.}~\bibnamefont {Li}}, \bibinfo
  {author} {\bibfnamefont {X.}~\bibnamefont {Zhu}}, \bibinfo {author}
  {\bibfnamefont {X.}~\bibnamefont {Pan}}, \bibinfo {author} {\bibfnamefont
  {Y.}~\bibnamefont {Nie}}, \emph {et~al.},\ }\bibfield  {title} {\bibinfo
  {title} {Single particle tunneling spectrum of superconducting
  {Nd$_{1-x}$Sr$_x$NiO$_2$} thin films},\ }\href@noop {} {\bibfield  {journal}
  {\bibinfo  {journal} {Nat. Commun.}\ }\textbf {\bibinfo {volume} {11}},\
  \bibinfo {pages} {1} (\bibinfo {year} {2020}{\natexlab{a}})}\BibitemShut
  {NoStop}%
\bibitem [{\citenamefont {Wang}\ \emph
  {et~al.}(2021{\natexlab{a}})\citenamefont {Wang}, \citenamefont {Li},
  \citenamefont {Goodge}, \citenamefont {Lee}, \citenamefont {Osada},
  \citenamefont {Harvey}, \citenamefont {Kourkoutis}, \citenamefont {Beasley},\
  and\ \citenamefont {Hwang}}]{wang2021}%
  \BibitemOpen
  \bibfield  {author} {\bibinfo {author} {\bibfnamefont {B.~Y.}\ \bibnamefont
  {Wang}}, \bibinfo {author} {\bibfnamefont {D.}~\bibnamefont {Li}}, \bibinfo
  {author} {\bibfnamefont {B.~H.}\ \bibnamefont {Goodge}}, \bibinfo {author}
  {\bibfnamefont {K.}~\bibnamefont {Lee}}, \bibinfo {author} {\bibfnamefont
  {M.}~\bibnamefont {Osada}}, \bibinfo {author} {\bibfnamefont {S.~P.}\
  \bibnamefont {Harvey}}, \bibinfo {author} {\bibfnamefont {L.~F.}\
  \bibnamefont {Kourkoutis}}, \bibinfo {author} {\bibfnamefont {M.~R.}\
  \bibnamefont {Beasley}},\ and\ \bibinfo {author} {\bibfnamefont {H.~Y.}\
  \bibnamefont {Hwang}},\ }\bibfield  {title} {\bibinfo {title} {Isotropic
  {P}auli-limited superconductivity in the infinite-layer nickelate
  {Nd$_{0.775}$Sr$_{0.225}$NiO$_2$}},\ }\href@noop {} {\bibfield  {journal}
  {\bibinfo  {journal} {Nat. Phys.}\ }\textbf {\bibinfo {volume} {17}},\
  \bibinfo {pages} {473} (\bibinfo {year} {2021}{\natexlab{a}})}\BibitemShut
  {NoStop}%
\bibitem [{\citenamefont {Xiang}\ \emph {et~al.}(2021)\citenamefont {Xiang},
  \citenamefont {Li}, \citenamefont {Li}, \citenamefont {Yang}, \citenamefont
  {Nie},\ and\ \citenamefont {Wen}}]{xiang2021}%
  \BibitemOpen
  \bibfield  {author} {\bibinfo {author} {\bibfnamefont {Y.}~\bibnamefont
  {Xiang}}, \bibinfo {author} {\bibfnamefont {Q.}~\bibnamefont {Li}}, \bibinfo
  {author} {\bibfnamefont {Y.}~\bibnamefont {Li}}, \bibinfo {author}
  {\bibfnamefont {H.}~\bibnamefont {Yang}}, \bibinfo {author} {\bibfnamefont
  {Y.}~\bibnamefont {Nie}},\ and\ \bibinfo {author} {\bibfnamefont {H.-H.}\
  \bibnamefont {Wen}},\ }\bibfield  {title} {\bibinfo {title} {Physical
  properties revealed by transport measurements for superconducting
  {Nd$_{0.8}$Sr$_{0.2}$NiO$_2$} thin films},\ }\href@noop {} {\bibfield
  {journal} {\bibinfo  {journal} {Chinese Physics Letters}\ }\textbf {\bibinfo
  {volume} {38}},\ \bibinfo {pages} {047401} (\bibinfo {year}
  {2021})}\BibitemShut {NoStop}%
\bibitem [{\citenamefont {Osada}\ \emph
  {et~al.}(2020{\natexlab{b}})\citenamefont {Osada}, \citenamefont {Wang},
  \citenamefont {Lee}, \citenamefont {Li},\ and\ \citenamefont
  {Hwang}}]{osada2020prm}%
  \BibitemOpen
  \bibfield  {author} {\bibinfo {author} {\bibfnamefont {M.}~\bibnamefont
  {Osada}}, \bibinfo {author} {\bibfnamefont {B.~Y.}\ \bibnamefont {Wang}},
  \bibinfo {author} {\bibfnamefont {K.}~\bibnamefont {Lee}}, \bibinfo {author}
  {\bibfnamefont {D.}~\bibnamefont {Li}},\ and\ \bibinfo {author}
  {\bibfnamefont {H.~Y.}\ \bibnamefont {Hwang}},\ }\bibfield  {title} {\bibinfo
  {title} {Phase diagram of infinite layer praseodymium nickelate
  {Pr$_{1-x}$Sr$_x$NiO$_2$} thin films},\ }\href@noop {} {\bibfield  {journal}
  {\bibinfo  {journal} {Phys. Rev. Materials}\ }\textbf {\bibinfo {volume}
  {4}},\ \bibinfo {pages} {121801} (\bibinfo {year}
  {2020}{\natexlab{b}})}\BibitemShut {NoStop}%
\bibitem [{\citenamefont {Rossi}\ \emph
  {et~al.}(2021{\natexlab{a}})\citenamefont {Rossi}, \citenamefont {Lu},
  \citenamefont {Nag}, \citenamefont {Li}, \citenamefont {Osada}, \citenamefont
  {Lee}, \citenamefont {Wang}, \citenamefont {Agrestini}, \citenamefont
  {Garcia-Fernandez}, \citenamefont {Kas} \emph {et~al.}}]{rossi2021orbital}%
  \BibitemOpen
  \bibfield  {author} {\bibinfo {author} {\bibfnamefont {M.}~\bibnamefont
  {Rossi}}, \bibinfo {author} {\bibfnamefont {H.}~\bibnamefont {Lu}}, \bibinfo
  {author} {\bibfnamefont {A.}~\bibnamefont {Nag}}, \bibinfo {author}
  {\bibfnamefont {D.}~\bibnamefont {Li}}, \bibinfo {author} {\bibfnamefont
  {M.}~\bibnamefont {Osada}}, \bibinfo {author} {\bibfnamefont
  {K.}~\bibnamefont {Lee}}, \bibinfo {author} {\bibfnamefont {B.~Y.}\
  \bibnamefont {Wang}}, \bibinfo {author} {\bibfnamefont {S.}~\bibnamefont
  {Agrestini}}, \bibinfo {author} {\bibfnamefont {M.}~\bibnamefont
  {Garcia-Fernandez}}, \bibinfo {author} {\bibfnamefont {J.}~\bibnamefont
  {Kas}}, \emph {et~al.},\ }\bibfield  {title} {\bibinfo {title} {Orbital and
  spin character of doped carriers in infinite-layer nickelates},\ }\href@noop
  {} {\bibfield  {journal} {\bibinfo  {journal} {Physical Review B}\ }\textbf
  {\bibinfo {volume} {104}},\ \bibinfo {pages} {L220505} (\bibinfo {year}
  {2021}{\natexlab{a}})}\BibitemShut {NoStop}%
\bibitem [{\citenamefont {Cui}\ \emph {et~al.}(2021)\citenamefont {Cui},
  \citenamefont {Li}, \citenamefont {Li}, \citenamefont {Zhu}, \citenamefont
  {Hu}, \citenamefont {Yang}, \citenamefont {Zhang}, \citenamefont {Yu},
  \citenamefont {Wen},\ and\ \citenamefont {Yu}}]{cui2021nmr}%
  \BibitemOpen
  \bibfield  {author} {\bibinfo {author} {\bibfnamefont {Y.}~\bibnamefont
  {Cui}}, \bibinfo {author} {\bibfnamefont {C.}~\bibnamefont {Li}}, \bibinfo
  {author} {\bibfnamefont {Q.}~\bibnamefont {Li}}, \bibinfo {author}
  {\bibfnamefont {X.}~\bibnamefont {Zhu}}, \bibinfo {author} {\bibfnamefont
  {Z.}~\bibnamefont {Hu}}, \bibinfo {author} {\bibfnamefont {Y.-f.}\
  \bibnamefont {Yang}}, \bibinfo {author} {\bibfnamefont {J.}~\bibnamefont
  {Zhang}}, \bibinfo {author} {\bibfnamefont {R.}~\bibnamefont {Yu}}, \bibinfo
  {author} {\bibfnamefont {H.-H.}\ \bibnamefont {Wen}},\ and\ \bibinfo {author}
  {\bibfnamefont {W.}~\bibnamefont {Yu}},\ }\bibfield  {title} {\bibinfo
  {title} {{NMR} evidence of antiferromagnetic spin fluctuations in nd0. 85sr0.
  15nio2},\ }\href@noop {} {\bibfield  {journal} {\bibinfo  {journal} {Chinese
  Physics Letters}\ }\textbf {\bibinfo {volume} {38}},\ \bibinfo {pages}
  {067401} (\bibinfo {year} {2021})}\BibitemShut {NoStop}%
\bibitem [{\citenamefont {He}\ \emph {et~al.}(2021)\citenamefont {He},
  \citenamefont {Ming}, \citenamefont {Li}, \citenamefont {Zhu}, \citenamefont
  {Si},\ and\ \citenamefont {Wen}}]{he2021}%
  \BibitemOpen
  \bibfield  {author} {\bibinfo {author} {\bibfnamefont {C.}~\bibnamefont
  {He}}, \bibinfo {author} {\bibfnamefont {X.}~\bibnamefont {Ming}}, \bibinfo
  {author} {\bibfnamefont {Q.}~\bibnamefont {Li}}, \bibinfo {author}
  {\bibfnamefont {X.}~\bibnamefont {Zhu}}, \bibinfo {author} {\bibfnamefont
  {J.}~\bibnamefont {Si}},\ and\ \bibinfo {author} {\bibfnamefont {H.-H.}\
  \bibnamefont {Wen}},\ }\bibfield  {title} {\bibinfo {title} {Synthesis and
  physical properties of perovskite sm1-xsrx{NiO}3 (x = 0, 0.2) and
  infinite-layer sm0.8sr0.2nio2 nickelates},\ }\href
  {https://doi.org/10.1088/1361-648x/abfb90} {\bibfield  {journal} {\bibinfo
  {journal} {Journal of Physics: Condensed Matter}\ }\textbf {\bibinfo {volume}
  {33}},\ \bibinfo {pages} {265701} (\bibinfo {year} {2021})}\BibitemShut
  {NoStop}%
\bibitem [{\citenamefont {Ortiz}\ \emph
  {et~al.}(2021{\natexlab{a}})\citenamefont {Ortiz}, \citenamefont {Menke},
  \citenamefont {Misj{\'a}k}, \citenamefont {Mantadakis}, \citenamefont
  {F{\"u}rsich}, \citenamefont {Schierle}, \citenamefont {Logvenov},
  \citenamefont {Kaiser}, \citenamefont {Keimer}, \citenamefont {Hansmann}
  \emph {et~al.}}]{ortiz2021prb}%
  \BibitemOpen
  \bibfield  {author} {\bibinfo {author} {\bibfnamefont {R.}~\bibnamefont
  {Ortiz}}, \bibinfo {author} {\bibfnamefont {H.}~\bibnamefont {Menke}},
  \bibinfo {author} {\bibfnamefont {F.}~\bibnamefont {Misj{\'a}k}}, \bibinfo
  {author} {\bibfnamefont {D.}~\bibnamefont {Mantadakis}}, \bibinfo {author}
  {\bibfnamefont {K.}~\bibnamefont {F{\"u}rsich}}, \bibinfo {author}
  {\bibfnamefont {E.}~\bibnamefont {Schierle}}, \bibinfo {author}
  {\bibfnamefont {G.}~\bibnamefont {Logvenov}}, \bibinfo {author}
  {\bibfnamefont {U.}~\bibnamefont {Kaiser}}, \bibinfo {author} {\bibfnamefont
  {B.}~\bibnamefont {Keimer}}, \bibinfo {author} {\bibfnamefont
  {P.}~\bibnamefont {Hansmann}}, \emph {et~al.},\ }\bibfield  {title} {\bibinfo
  {title} {Superlattice approach to doping infinite-layer nickelates},\
  }\href@noop {} {\bibfield  {journal} {\bibinfo  {journal} {Physical Review
  B}\ }\textbf {\bibinfo {volume} {104}},\ \bibinfo {pages} {165137} (\bibinfo
  {year} {2021}{\natexlab{a}})}\BibitemShut {NoStop}%
\bibitem [{\citenamefont {Gao}\ \emph {et~al.}(2021)\citenamefont {Gao},
  \citenamefont {Zhao}, \citenamefont {Zhou},\ and\ \citenamefont
  {Zhu}}]{gao2021}%
  \BibitemOpen
  \bibfield  {author} {\bibinfo {author} {\bibfnamefont {Q.}~\bibnamefont
  {Gao}}, \bibinfo {author} {\bibfnamefont {Y.}~\bibnamefont {Zhao}}, \bibinfo
  {author} {\bibfnamefont {X.-J.}\ \bibnamefont {Zhou}},\ and\ \bibinfo
  {author} {\bibfnamefont {Z.}~\bibnamefont {Zhu}},\ }\bibfield  {title}
  {\bibinfo {title} {Preparation of {S}uperconducting {T}hin {F}ilms of
  {I}nfinite-{L}ayer {N}ickelate {Nd$_{0.8}$Sr$_{0.2}$NiO$_2$}},\ }\href@noop
  {} {\bibfield  {journal} {\bibinfo  {journal} {Chinese Physics Letters}\
  }\textbf {\bibinfo {volume} {38}},\ \bibinfo {pages} {077401} (\bibinfo
  {year} {2021})}\BibitemShut {NoStop}%
\bibitem [{\citenamefont {Zeng}\ \emph
  {et~al.}(2022{\natexlab{b}})\citenamefont {Zeng}, \citenamefont {Yin},
  \citenamefont {Li}, \citenamefont {Chow}, \citenamefont {Tang}, \citenamefont
  {Han}, \citenamefont {Huang}, \citenamefont {Cao}, \citenamefont {Wan},
  \citenamefont {Zhang} \emph {et~al.}}]{zeng2022nc}%
  \BibitemOpen
  \bibfield  {author} {\bibinfo {author} {\bibfnamefont {S.}~\bibnamefont
  {Zeng}}, \bibinfo {author} {\bibfnamefont {X.}~\bibnamefont {Yin}}, \bibinfo
  {author} {\bibfnamefont {C.}~\bibnamefont {Li}}, \bibinfo {author}
  {\bibfnamefont {L.}~\bibnamefont {Chow}}, \bibinfo {author} {\bibfnamefont
  {C.}~\bibnamefont {Tang}}, \bibinfo {author} {\bibfnamefont {K.}~\bibnamefont
  {Han}}, \bibinfo {author} {\bibfnamefont {Z.}~\bibnamefont {Huang}}, \bibinfo
  {author} {\bibfnamefont {Y.}~\bibnamefont {Cao}}, \bibinfo {author}
  {\bibfnamefont {D.}~\bibnamefont {Wan}}, \bibinfo {author} {\bibfnamefont
  {Z.}~\bibnamefont {Zhang}}, \emph {et~al.},\ }\bibfield  {title} {\bibinfo
  {title} {Observation of perfect diamagnetism and interfacial effect on the
  electronic structures in infinite layer {Nd$_{0.8}$Sr$_{0.2}$NiO$_2$}
  superconductors},\ }\href@noop {} {\bibfield  {journal} {\bibinfo  {journal}
  {Nat. Commun.}\ }\textbf {\bibinfo {volume} {13}},\ \bibinfo {pages} {1}
  (\bibinfo {year} {2022}{\natexlab{b}})}\BibitemShut {NoStop}%
\bibitem [{\citenamefont {Zhou}\ \emph {et~al.}(2021)\citenamefont {Zhou},
  \citenamefont {Feng}, \citenamefont {Qin}, \citenamefont {Yan}, \citenamefont
  {Wang}, \citenamefont {Nie}, \citenamefont {Wu}, \citenamefont {Zhang},
  \citenamefont {Chen}, \citenamefont {Meng} \emph {et~al.}}]{zhou2021}%
  \BibitemOpen
  \bibfield  {author} {\bibinfo {author} {\bibfnamefont {X.-R.}\ \bibnamefont
  {Zhou}}, \bibinfo {author} {\bibfnamefont {Z.-X.}\ \bibnamefont {Feng}},
  \bibinfo {author} {\bibfnamefont {P.-X.}\ \bibnamefont {Qin}}, \bibinfo
  {author} {\bibfnamefont {H.}~\bibnamefont {Yan}}, \bibinfo {author}
  {\bibfnamefont {X.-N.}\ \bibnamefont {Wang}}, \bibinfo {author}
  {\bibfnamefont {P.}~\bibnamefont {Nie}}, \bibinfo {author} {\bibfnamefont
  {H.-J.}\ \bibnamefont {Wu}}, \bibinfo {author} {\bibfnamefont
  {X.}~\bibnamefont {Zhang}}, \bibinfo {author} {\bibfnamefont {H.-Y.}\
  \bibnamefont {Chen}}, \bibinfo {author} {\bibfnamefont {Z.-A.}\ \bibnamefont
  {Meng}}, \emph {et~al.},\ }\bibfield  {title} {\bibinfo {title} {Negligible
  oxygen vacancies, low critical current density, electric-field modulation,
  in-plane anisotropic and high-field transport of a superconducting
  {Nd$_{0.8}$Sr$_{0.2}$NiO$_2$/SrTiO$_3$} heterostructure},\ }\href@noop {}
  {\bibfield  {journal} {\bibinfo  {journal} {Rare Metals}\ }\textbf {\bibinfo
  {volume} {40}},\ \bibinfo {pages} {2847} (\bibinfo {year}
  {2021})}\BibitemShut {NoStop}%
\bibitem [{\citenamefont {Zhao}\ \emph {et~al.}(2021)\citenamefont {Zhao},
  \citenamefont {Zhou}, \citenamefont {Fu}, \citenamefont {Wang}, \citenamefont
  {Zhou}, \citenamefont {Cheng}, \citenamefont {Li}, \citenamefont {Song},
  \citenamefont {Li}, \citenamefont {Kang} \emph {et~al.}}]{zhao2021}%
  \BibitemOpen
  \bibfield  {author} {\bibinfo {author} {\bibfnamefont {D.}~\bibnamefont
  {Zhao}}, \bibinfo {author} {\bibfnamefont {Y.}~\bibnamefont {Zhou}}, \bibinfo
  {author} {\bibfnamefont {Y.}~\bibnamefont {Fu}}, \bibinfo {author}
  {\bibfnamefont {L.}~\bibnamefont {Wang}}, \bibinfo {author} {\bibfnamefont
  {X.}~\bibnamefont {Zhou}}, \bibinfo {author} {\bibfnamefont {H.}~\bibnamefont
  {Cheng}}, \bibinfo {author} {\bibfnamefont {J.}~\bibnamefont {Li}}, \bibinfo
  {author} {\bibfnamefont {D.}~\bibnamefont {Song}}, \bibinfo {author}
  {\bibfnamefont {S.}~\bibnamefont {Li}}, \bibinfo {author} {\bibfnamefont
  {B.}~\bibnamefont {Kang}}, \emph {et~al.},\ }\bibfield  {title} {\bibinfo
  {title} {Intrinsic {S}pin {S}usceptibility and {P}seudogaplike {B}ehavior in
  {I}nfinite-{L}ayer {LaNiO$_2$}},\ }\href@noop {} {\bibfield  {journal}
  {\bibinfo  {journal} {Physical Review Letters}\ }\textbf {\bibinfo {volume}
  {126}},\ \bibinfo {pages} {197001} (\bibinfo {year} {2021})}\BibitemShut
  {NoStop}%
\bibitem [{\citenamefont {Lin}\ \emph {et~al.}(2022)\citenamefont {Lin},
  \citenamefont {Gawryluk}, \citenamefont {Klein}, \citenamefont {Huangfu},
  \citenamefont {Pomjakushina}, \citenamefont {von Rohr},\ and\ \citenamefont
  {Schilling}}]{lin2022}%
  \BibitemOpen
  \bibfield  {author} {\bibinfo {author} {\bibfnamefont {H.}~\bibnamefont
  {Lin}}, \bibinfo {author} {\bibfnamefont {D.~J.}\ \bibnamefont {Gawryluk}},
  \bibinfo {author} {\bibfnamefont {Y.~M.}\ \bibnamefont {Klein}}, \bibinfo
  {author} {\bibfnamefont {S.}~\bibnamefont {Huangfu}}, \bibinfo {author}
  {\bibfnamefont {E.}~\bibnamefont {Pomjakushina}}, \bibinfo {author}
  {\bibfnamefont {F.}~\bibnamefont {von Rohr}},\ and\ \bibinfo {author}
  {\bibfnamefont {A.}~\bibnamefont {Schilling}},\ }\bibfield  {title} {\bibinfo
  {title} {Universal spin-glass behaviour in bulk {LaNiO$_2$}, {PrNiO$_2$} and
  {NdNiO$_2$}},\ }\href@noop {} {\bibfield  {journal} {\bibinfo  {journal} {New
  Journal of Physics}\ }\textbf {\bibinfo {volume} {24}},\ \bibinfo {pages}
  {013022} (\bibinfo {year} {2022})}\BibitemShut {NoStop}%
\bibitem [{\citenamefont {Hsu}\ \emph {et~al.}(2021)\citenamefont {Hsu},
  \citenamefont {Wang}, \citenamefont {Berben}, \citenamefont {Li},
  \citenamefont {Lee}, \citenamefont {Duffy}, \citenamefont {Ottenbros},
  \citenamefont {Kim}, \citenamefont {Osada}, \citenamefont {Wiedmann} \emph
  {et~al.}}]{hsu2021}%
  \BibitemOpen
  \bibfield  {author} {\bibinfo {author} {\bibfnamefont {Y.-T.}\ \bibnamefont
  {Hsu}}, \bibinfo {author} {\bibfnamefont {B.~Y.}\ \bibnamefont {Wang}},
  \bibinfo {author} {\bibfnamefont {M.}~\bibnamefont {Berben}}, \bibinfo
  {author} {\bibfnamefont {D.}~\bibnamefont {Li}}, \bibinfo {author}
  {\bibfnamefont {K.}~\bibnamefont {Lee}}, \bibinfo {author} {\bibfnamefont
  {C.}~\bibnamefont {Duffy}}, \bibinfo {author} {\bibfnamefont
  {T.}~\bibnamefont {Ottenbros}}, \bibinfo {author} {\bibfnamefont {W.~J.}\
  \bibnamefont {Kim}}, \bibinfo {author} {\bibfnamefont {M.}~\bibnamefont
  {Osada}}, \bibinfo {author} {\bibfnamefont {S.}~\bibnamefont {Wiedmann}},
  \emph {et~al.},\ }\bibfield  {title} {\bibinfo {title} {Insulator-to-metal
  crossover near the edge of the superconducting dome in
  {Nd$_{1-x}$Sr$_x$NiO$_2$}},\ }\href@noop {} {\bibfield  {journal} {\bibinfo
  {journal} {Physical Review Research}\ }\textbf {\bibinfo {volume} {3}},\
  \bibinfo {pages} {L042015} (\bibinfo {year} {2021})}\BibitemShut {NoStop}%
\bibitem [{\citenamefont {Lu}\ \emph {et~al.}(2021)\citenamefont {Lu},
  \citenamefont {Rossi}, \citenamefont {Nag}, \citenamefont {Osada},
  \citenamefont {Li}, \citenamefont {Lee}, \citenamefont {Wang}, \citenamefont
  {Garcia-Fernandez}, \citenamefont {Agrestini}, \citenamefont {Shen} \emph
  {et~al.}}]{lu2021}%
  \BibitemOpen
  \bibfield  {author} {\bibinfo {author} {\bibfnamefont {H.}~\bibnamefont
  {Lu}}, \bibinfo {author} {\bibfnamefont {M.}~\bibnamefont {Rossi}}, \bibinfo
  {author} {\bibfnamefont {A.}~\bibnamefont {Nag}}, \bibinfo {author}
  {\bibfnamefont {M.}~\bibnamefont {Osada}}, \bibinfo {author} {\bibfnamefont
  {D.}~\bibnamefont {Li}}, \bibinfo {author} {\bibfnamefont {K.}~\bibnamefont
  {Lee}}, \bibinfo {author} {\bibfnamefont {B.}~\bibnamefont {Wang}}, \bibinfo
  {author} {\bibfnamefont {M.}~\bibnamefont {Garcia-Fernandez}}, \bibinfo
  {author} {\bibfnamefont {S.}~\bibnamefont {Agrestini}}, \bibinfo {author}
  {\bibfnamefont {Z.}~\bibnamefont {Shen}}, \emph {et~al.},\ }\bibfield
  {title} {\bibinfo {title} {Magnetic excitations in infinite-layer
  nickelates},\ }\href@noop {} {\bibfield  {journal} {\bibinfo  {journal}
  {Science}\ }\textbf {\bibinfo {volume} {373}},\ \bibinfo {pages} {213}
  (\bibinfo {year} {2021})}\BibitemShut {NoStop}%
\bibitem [{\citenamefont {Li}\ \emph {et~al.}(2021)\citenamefont {Li},
  \citenamefont {Sun}, \citenamefont {Yang}, \citenamefont {Cai}, \citenamefont
  {Guo}, \citenamefont {Gu}, \citenamefont {Zhu},\ and\ \citenamefont
  {Nie}}]{li2021}%
  \BibitemOpen
  \bibfield  {author} {\bibinfo {author} {\bibfnamefont {Y.}~\bibnamefont
  {Li}}, \bibinfo {author} {\bibfnamefont {W.}~\bibnamefont {Sun}}, \bibinfo
  {author} {\bibfnamefont {J.}~\bibnamefont {Yang}}, \bibinfo {author}
  {\bibfnamefont {X.}~\bibnamefont {Cai}}, \bibinfo {author} {\bibfnamefont
  {W.}~\bibnamefont {Guo}}, \bibinfo {author} {\bibfnamefont {Z.}~\bibnamefont
  {Gu}}, \bibinfo {author} {\bibfnamefont {Y.}~\bibnamefont {Zhu}},\ and\
  \bibinfo {author} {\bibfnamefont {Y.}~\bibnamefont {Nie}},\ }\bibfield
  {title} {\bibinfo {title} {Impact of cation stoichiometry on the crystalline
  structure and superconductivity in nickelates},\ }\href@noop {} {\bibfield
  {journal} {\bibinfo  {journal} {arXiv preprint arXiv:2106.02485}\ } (\bibinfo
  {year} {2021})}\BibitemShut {NoStop}%
\bibitem [{\citenamefont {Puphal}\ \emph {et~al.}(2021)\citenamefont {Puphal},
  \citenamefont {Wu}, \citenamefont {F{\"u}rsich}, \citenamefont {Lee},
  \citenamefont {Pakdaman}, \citenamefont {Bruin}, \citenamefont {Nuss},
  \citenamefont {Suyolcu}, \citenamefont {van Aken}, \citenamefont {Keimer}
  \emph {et~al.}}]{puphal2021}%
  \BibitemOpen
  \bibfield  {author} {\bibinfo {author} {\bibfnamefont {P.}~\bibnamefont
  {Puphal}}, \bibinfo {author} {\bibfnamefont {Y.-M.}\ \bibnamefont {Wu}},
  \bibinfo {author} {\bibfnamefont {K.}~\bibnamefont {F{\"u}rsich}}, \bibinfo
  {author} {\bibfnamefont {H.}~\bibnamefont {Lee}}, \bibinfo {author}
  {\bibfnamefont {M.}~\bibnamefont {Pakdaman}}, \bibinfo {author}
  {\bibfnamefont {J.~A.}\ \bibnamefont {Bruin}}, \bibinfo {author}
  {\bibfnamefont {J.}~\bibnamefont {Nuss}}, \bibinfo {author} {\bibfnamefont
  {Y.~E.}\ \bibnamefont {Suyolcu}}, \bibinfo {author} {\bibfnamefont {P.~A.}\
  \bibnamefont {van Aken}}, \bibinfo {author} {\bibfnamefont {B.}~\bibnamefont
  {Keimer}}, \emph {et~al.},\ }\bibfield  {title} {\bibinfo {title} {Topotactic
  transformation of single crystals: From perovskite to infinite-layer
  nickelates},\ }\href@noop {} {\bibfield  {journal} {\bibinfo  {journal}
  {Science advances}\ }\textbf {\bibinfo {volume} {7}},\ \bibinfo {pages}
  {eabl8091} (\bibinfo {year} {2021})}\BibitemShut {NoStop}%
\bibitem [{\citenamefont {Chen}\ \emph {et~al.}(2022)\citenamefont {Chen},
  \citenamefont {Osada}, \citenamefont {Li}, \citenamefont {Been},
  \citenamefont {Chen}, \citenamefont {Hashimoto}, \citenamefont {Lu},
  \citenamefont {Mo}, \citenamefont {Lee}, \citenamefont {Wang} \emph
  {et~al.}}]{chen2022}%
  \BibitemOpen
  \bibfield  {author} {\bibinfo {author} {\bibfnamefont {Z.}~\bibnamefont
  {Chen}}, \bibinfo {author} {\bibfnamefont {M.}~\bibnamefont {Osada}},
  \bibinfo {author} {\bibfnamefont {D.}~\bibnamefont {Li}}, \bibinfo {author}
  {\bibfnamefont {E.~M.}\ \bibnamefont {Been}}, \bibinfo {author}
  {\bibfnamefont {S.-D.}\ \bibnamefont {Chen}}, \bibinfo {author}
  {\bibfnamefont {M.}~\bibnamefont {Hashimoto}}, \bibinfo {author}
  {\bibfnamefont {D.}~\bibnamefont {Lu}}, \bibinfo {author} {\bibfnamefont
  {S.-K.}\ \bibnamefont {Mo}}, \bibinfo {author} {\bibfnamefont
  {K.}~\bibnamefont {Lee}}, \bibinfo {author} {\bibfnamefont {B.~Y.}\
  \bibnamefont {Wang}}, \emph {et~al.},\ }\bibfield  {title} {\bibinfo {title}
  {Electronic structure of superconducting nickelates probed by resonant
  photoemission spectroscopy},\ }\href@noop {} {\bibfield  {journal} {\bibinfo
  {journal} {Matter}\ } (\bibinfo {year} {2022})}\BibitemShut {NoStop}%
\bibitem [{\citenamefont {Wang}\ \emph
  {et~al.}(2021{\natexlab{b}})\citenamefont {Wang}, \citenamefont {Yang},
  \citenamefont {Chen}, \citenamefont {Yang}, \citenamefont {Zhang},
  \citenamefont {Zhu}, \citenamefont {Uwatoko}, \citenamefont {Dong},
  \citenamefont {Jin}, \citenamefont {Sun} \emph {et~al.}}]{wang2021pressure}%
  \BibitemOpen
  \bibfield  {author} {\bibinfo {author} {\bibfnamefont {N.~N.}\ \bibnamefont
  {Wang}}, \bibinfo {author} {\bibfnamefont {M.~W.}\ \bibnamefont {Yang}},
  \bibinfo {author} {\bibfnamefont {K.~Y.}\ \bibnamefont {Chen}}, \bibinfo
  {author} {\bibfnamefont {Z.}~\bibnamefont {Yang}}, \bibinfo {author}
  {\bibfnamefont {H.}~\bibnamefont {Zhang}}, \bibinfo {author} {\bibfnamefont
  {Z.~H.}\ \bibnamefont {Zhu}}, \bibinfo {author} {\bibfnamefont
  {Y.}~\bibnamefont {Uwatoko}}, \bibinfo {author} {\bibfnamefont {X.~L.}\
  \bibnamefont {Dong}}, \bibinfo {author} {\bibfnamefont {K.~J.}\ \bibnamefont
  {Jin}}, \bibinfo {author} {\bibfnamefont {J.~P.}\ \bibnamefont {Sun}}, \emph
  {et~al.},\ }\bibfield  {title} {\bibinfo {title} {Pressure-induced monotonic
  enhancement of {$T_c$} to over 30 k in the superconducting
  {Pr$_{0.82}$Sr$_{0.18}$NiO$_2$} thin films},\ }\href@noop {} {\bibfield
  {journal} {\bibinfo  {journal} {arXiv preprint arXiv:2109.12811}\ } (\bibinfo
  {year} {2021}{\natexlab{b}})}\BibitemShut {NoStop}%
\bibitem [{\citenamefont {Zhou}\ \emph {et~al.}(2022)\citenamefont {Zhou},
  \citenamefont {Zhang}, \citenamefont {Yi}, \citenamefont {Qin}, \citenamefont
  {Feng}, \citenamefont {Jiang}, \citenamefont {Zhong}, \citenamefont {Yan},
  \citenamefont {Wang}, \citenamefont {Chen} \emph {et~al.}}]{zhou2022}%
  \BibitemOpen
  \bibfield  {author} {\bibinfo {author} {\bibfnamefont {X.}~\bibnamefont
  {Zhou}}, \bibinfo {author} {\bibfnamefont {X.}~\bibnamefont {Zhang}},
  \bibinfo {author} {\bibfnamefont {J.}~\bibnamefont {Yi}}, \bibinfo {author}
  {\bibfnamefont {P.}~\bibnamefont {Qin}}, \bibinfo {author} {\bibfnamefont
  {Z.}~\bibnamefont {Feng}}, \bibinfo {author} {\bibfnamefont {P.}~\bibnamefont
  {Jiang}}, \bibinfo {author} {\bibfnamefont {Z.}~\bibnamefont {Zhong}},
  \bibinfo {author} {\bibfnamefont {H.}~\bibnamefont {Yan}}, \bibinfo {author}
  {\bibfnamefont {X.}~\bibnamefont {Wang}}, \bibinfo {author} {\bibfnamefont
  {H.}~\bibnamefont {Chen}}, \emph {et~al.},\ }\bibfield  {title} {\bibinfo
  {title} {Antiferromagnetism in ni-based superconductors},\ }\href@noop {}
  {\bibfield  {journal} {\bibinfo  {journal} {Advanced Materials}\ }\textbf
  {\bibinfo {volume} {34}},\ \bibinfo {pages} {2106117} (\bibinfo {year}
  {2022})}\BibitemShut {NoStop}%
\bibitem [{\citenamefont {Ortiz}\ \emph
  {et~al.}(2021{\natexlab{b}})\citenamefont {Ortiz}, \citenamefont {Puphal},
  \citenamefont {Klett}, \citenamefont {Hotz}, \citenamefont {Kremer},
  \citenamefont {Trepka}, \citenamefont {Hemmida}, \citenamefont {von Nidda},
  \citenamefont {Isobe}, \citenamefont {Khasanov} \emph {et~al.}}]{ortiz2021}%
  \BibitemOpen
  \bibfield  {author} {\bibinfo {author} {\bibfnamefont {R.}~\bibnamefont
  {Ortiz}}, \bibinfo {author} {\bibfnamefont {P.}~\bibnamefont {Puphal}},
  \bibinfo {author} {\bibfnamefont {M.}~\bibnamefont {Klett}}, \bibinfo
  {author} {\bibfnamefont {F.}~\bibnamefont {Hotz}}, \bibinfo {author}
  {\bibfnamefont {R.}~\bibnamefont {Kremer}}, \bibinfo {author} {\bibfnamefont
  {H.}~\bibnamefont {Trepka}}, \bibinfo {author} {\bibfnamefont
  {M.}~\bibnamefont {Hemmida}}, \bibinfo {author} {\bibfnamefont {H.-A.~K.}\
  \bibnamefont {von Nidda}}, \bibinfo {author} {\bibfnamefont {M.}~\bibnamefont
  {Isobe}}, \bibinfo {author} {\bibfnamefont {R.}~\bibnamefont {Khasanov}},
  \emph {et~al.},\ }\bibfield  {title} {\bibinfo {title} {Magnetic correlations
  in infinite-layer nickelates: an experimental and theoretical multi-method
  study},\ }\href@noop {} {\bibfield  {journal} {\bibinfo  {journal} {arXiv
  preprint arXiv:2111.13668}\ } (\bibinfo {year}
  {2021}{\natexlab{b}})}\BibitemShut {NoStop}%
\bibitem [{\citenamefont {Rossi}\ \emph
  {et~al.}(2021{\natexlab{b}})\citenamefont {Rossi}, \citenamefont {Osada},
  \citenamefont {Choi}, \citenamefont {Agrestini}, \citenamefont {Jost},
  \citenamefont {Lee}, \citenamefont {Lu}, \citenamefont {Wang}, \citenamefont
  {Lee}, \citenamefont {Nag} \emph {et~al.}}]{rossi2021}%
  \BibitemOpen
  \bibfield  {author} {\bibinfo {author} {\bibfnamefont {M.}~\bibnamefont
  {Rossi}}, \bibinfo {author} {\bibfnamefont {M.}~\bibnamefont {Osada}},
  \bibinfo {author} {\bibfnamefont {J.}~\bibnamefont {Choi}}, \bibinfo {author}
  {\bibfnamefont {S.}~\bibnamefont {Agrestini}}, \bibinfo {author}
  {\bibfnamefont {D.}~\bibnamefont {Jost}}, \bibinfo {author} {\bibfnamefont
  {Y.}~\bibnamefont {Lee}}, \bibinfo {author} {\bibfnamefont {H.}~\bibnamefont
  {Lu}}, \bibinfo {author} {\bibfnamefont {B.~Y.}\ \bibnamefont {Wang}},
  \bibinfo {author} {\bibfnamefont {K.}~\bibnamefont {Lee}}, \bibinfo {author}
  {\bibfnamefont {A.}~\bibnamefont {Nag}}, \emph {et~al.},\ }\bibfield  {title}
  {\bibinfo {title} {A broken translational symmetry state in an infinite-layer
  nickelate},\ }\href@noop {} {\bibfield  {journal} {\bibinfo  {journal} {arXiv
  preprint arXiv:2112.02484}\ } (\bibinfo {year}
  {2021}{\natexlab{b}})}\BibitemShut {NoStop}%
\bibitem [{\citenamefont {Krieger}\ \emph {et~al.}(2021)\citenamefont
  {Krieger}, \citenamefont {Martinelli}, \citenamefont {Zeng}, \citenamefont
  {Chow}, \citenamefont {Kummer}, \citenamefont {Arpaia}, \citenamefont {Sala},
  \citenamefont {Brookes}, \citenamefont {Ariando}, \citenamefont {Viart} \emph
  {et~al.}}]{krieger2021}%
  \BibitemOpen
  \bibfield  {author} {\bibinfo {author} {\bibfnamefont {G.}~\bibnamefont
  {Krieger}}, \bibinfo {author} {\bibfnamefont {L.}~\bibnamefont {Martinelli}},
  \bibinfo {author} {\bibfnamefont {S.}~\bibnamefont {Zeng}}, \bibinfo {author}
  {\bibfnamefont {L.}~\bibnamefont {Chow}}, \bibinfo {author} {\bibfnamefont
  {K.}~\bibnamefont {Kummer}}, \bibinfo {author} {\bibfnamefont
  {R.}~\bibnamefont {Arpaia}}, \bibinfo {author} {\bibfnamefont {M.~M.}\
  \bibnamefont {Sala}}, \bibinfo {author} {\bibfnamefont {N.}~\bibnamefont
  {Brookes}}, \bibinfo {author} {\bibfnamefont {A.}~\bibnamefont {Ariando}},
  \bibinfo {author} {\bibfnamefont {N.}~\bibnamefont {Viart}}, \emph {et~al.},\
  }\bibfield  {title} {\bibinfo {title} {Charge and spin order dichotomy in
  ndnio $ \_2 $ driven by srtio $ \_3 $ capping layer},\ }\href@noop {}
  {\bibfield  {journal} {\bibinfo  {journal} {arXiv preprint arXiv:2112.03341}\
  } (\bibinfo {year} {2021})}\BibitemShut {NoStop}%
\bibitem [{\citenamefont {Tam}\ \emph {et~al.}(2021)\citenamefont {Tam},
  \citenamefont {Choi}, \citenamefont {Ding}, \citenamefont {Agrestini},
  \citenamefont {Nag}, \citenamefont {Huang}, \citenamefont {Luo},
  \citenamefont {Garc{\'\i}a-Fern{\'a}ndez}, \citenamefont {Qiao},\ and\
  \citenamefont {Zhou}}]{tam2021}%
  \BibitemOpen
  \bibfield  {author} {\bibinfo {author} {\bibfnamefont {C.~C.}\ \bibnamefont
  {Tam}}, \bibinfo {author} {\bibfnamefont {J.}~\bibnamefont {Choi}}, \bibinfo
  {author} {\bibfnamefont {X.}~\bibnamefont {Ding}}, \bibinfo {author}
  {\bibfnamefont {S.}~\bibnamefont {Agrestini}}, \bibinfo {author}
  {\bibfnamefont {A.}~\bibnamefont {Nag}}, \bibinfo {author} {\bibfnamefont
  {B.}~\bibnamefont {Huang}}, \bibinfo {author} {\bibfnamefont
  {H.}~\bibnamefont {Luo}}, \bibinfo {author} {\bibfnamefont {M.}~\bibnamefont
  {Garc{\'\i}a-Fern{\'a}ndez}}, \bibinfo {author} {\bibfnamefont
  {L.}~\bibnamefont {Qiao}},\ and\ \bibinfo {author} {\bibfnamefont {K.-J.}\
  \bibnamefont {Zhou}},\ }\bibfield  {title} {\bibinfo {title} {Charge density
  waves in infinite-layer {NdNiO$_2$} nickelates},\ }\href@noop {} {\bibfield
  {journal} {\bibinfo  {journal} {arXiv preprint arXiv:2112.04440}\ } (\bibinfo
  {year} {2021})}\BibitemShut {NoStop}%
\bibitem [{\citenamefont {Goodge}\ \emph {et~al.}(2022)\citenamefont {Goodge},
  \citenamefont {Geisler}, \citenamefont {Lee}, \citenamefont {Osada},
  \citenamefont {Wang}, \citenamefont {Li}, \citenamefont {Hwang},
  \citenamefont {Pentcheva},\ and\ \citenamefont {Kourkoutis}}]{goodge2022}%
  \BibitemOpen
  \bibfield  {author} {\bibinfo {author} {\bibfnamefont {B.~H.}\ \bibnamefont
  {Goodge}}, \bibinfo {author} {\bibfnamefont {B.}~\bibnamefont {Geisler}},
  \bibinfo {author} {\bibfnamefont {K.}~\bibnamefont {Lee}}, \bibinfo {author}
  {\bibfnamefont {M.}~\bibnamefont {Osada}}, \bibinfo {author} {\bibfnamefont
  {B.~Y.}\ \bibnamefont {Wang}}, \bibinfo {author} {\bibfnamefont
  {D.}~\bibnamefont {Li}}, \bibinfo {author} {\bibfnamefont {H.~Y.}\
  \bibnamefont {Hwang}}, \bibinfo {author} {\bibfnamefont {R.}~\bibnamefont
  {Pentcheva}},\ and\ \bibinfo {author} {\bibfnamefont {L.~F.}\ \bibnamefont
  {Kourkoutis}},\ }\bibfield  {title} {\bibinfo {title} {Reconstructing the
  polar interface of infinite-layer nickelate thin films},\ }\href@noop {}
  {\bibfield  {journal} {\bibinfo  {journal} {arXiv preprint arXiv:2201.03613}\
  } (\bibinfo {year} {2022})}\BibitemShut {NoStop}%
\bibitem [{\citenamefont {Chow}\ \emph {et~al.}(2022)\citenamefont {Chow},
  \citenamefont {Sudheesh}, \citenamefont {Nandi}, \citenamefont {Zeng},
  \citenamefont {Zhang}, \citenamefont {Du}, \citenamefont {Lim}, \citenamefont
  {Chia},\ and\ \citenamefont {Ariando}}]{chow2022}%
  \BibitemOpen
  \bibfield  {author} {\bibinfo {author} {\bibfnamefont {L.~E.}\ \bibnamefont
  {Chow}}, \bibinfo {author} {\bibfnamefont {S.~K.}\ \bibnamefont {Sudheesh}},
  \bibinfo {author} {\bibfnamefont {P.}~\bibnamefont {Nandi}}, \bibinfo
  {author} {\bibfnamefont {S.}~\bibnamefont {Zeng}}, \bibinfo {author}
  {\bibfnamefont {Z.}~\bibnamefont {Zhang}}, \bibinfo {author} {\bibfnamefont
  {X.}~\bibnamefont {Du}}, \bibinfo {author} {\bibfnamefont {Z.~S.}\
  \bibnamefont {Lim}}, \bibinfo {author} {\bibfnamefont {E.~E.}\ \bibnamefont
  {Chia}},\ and\ \bibinfo {author} {\bibfnamefont {A.}~\bibnamefont
  {Ariando}},\ }\bibfield  {title} {\bibinfo {title} {Pairing symmetry in
  infinite-layer nickelate superconductor},\ }\href@noop {} {\bibfield
  {journal} {\bibinfo  {journal} {arXiv preprint arXiv:2201.10038}\ } (\bibinfo
  {year} {2022})}\BibitemShut {NoStop}%
\bibitem [{\citenamefont {Fowlie}\ \emph {et~al.}(2022)\citenamefont {Fowlie},
  \citenamefont {Hadjimichael}, \citenamefont {Martins}, \citenamefont {Li},
  \citenamefont {Osada}, \citenamefont {Wang}, \citenamefont {Lee},
  \citenamefont {Lee}, \citenamefont {Salman}, \citenamefont {Prokscha} \emph
  {et~al.}}]{fowlie2022}%
  \BibitemOpen
  \bibfield  {author} {\bibinfo {author} {\bibfnamefont {J.}~\bibnamefont
  {Fowlie}}, \bibinfo {author} {\bibfnamefont {M.}~\bibnamefont
  {Hadjimichael}}, \bibinfo {author} {\bibfnamefont {M.~M.}\ \bibnamefont
  {Martins}}, \bibinfo {author} {\bibfnamefont {D.}~\bibnamefont {Li}},
  \bibinfo {author} {\bibfnamefont {M.}~\bibnamefont {Osada}}, \bibinfo
  {author} {\bibfnamefont {B.~Y.}\ \bibnamefont {Wang}}, \bibinfo {author}
  {\bibfnamefont {K.}~\bibnamefont {Lee}}, \bibinfo {author} {\bibfnamefont
  {Y.}~\bibnamefont {Lee}}, \bibinfo {author} {\bibfnamefont {Z.}~\bibnamefont
  {Salman}}, \bibinfo {author} {\bibfnamefont {T.}~\bibnamefont {Prokscha}},
  \emph {et~al.},\ }\bibfield  {title} {\bibinfo {title} {Intrinsic magnetism
  in superconducting infinite-layer nickelates},\ }\href@noop {} {\bibfield
  {journal} {\bibinfo  {journal} {arXiv preprint arXiv:2201.11943}\ } (\bibinfo
  {year} {2022})}\BibitemShut {NoStop}%
\bibitem [{\citenamefont {Harvey}\ \emph {et~al.}(2022)\citenamefont {Harvey},
  \citenamefont {Wang}, \citenamefont {Fowlie}, \citenamefont {Osada},
  \citenamefont {Lee}, \citenamefont {Lee}, \citenamefont {Li},\ and\
  \citenamefont {Hwang}}]{harvey2022}%
  \BibitemOpen
  \bibfield  {author} {\bibinfo {author} {\bibfnamefont {S.~P.}\ \bibnamefont
  {Harvey}}, \bibinfo {author} {\bibfnamefont {B.~Y.}\ \bibnamefont {Wang}},
  \bibinfo {author} {\bibfnamefont {J.}~\bibnamefont {Fowlie}}, \bibinfo
  {author} {\bibfnamefont {M.}~\bibnamefont {Osada}}, \bibinfo {author}
  {\bibfnamefont {K.}~\bibnamefont {Lee}}, \bibinfo {author} {\bibfnamefont
  {Y.}~\bibnamefont {Lee}}, \bibinfo {author} {\bibfnamefont {D.}~\bibnamefont
  {Li}},\ and\ \bibinfo {author} {\bibfnamefont {H.~Y.}\ \bibnamefont
  {Hwang}},\ }\bibfield  {title} {\bibinfo {title} {Evidence for nodal
  superconductivity in infinite-layer nickelates},\ }\href@noop {} {\bibfield
  {journal} {\bibinfo  {journal} {arXiv preprint arXiv:2201.12971}\ } (\bibinfo
  {year} {2022})}\BibitemShut {NoStop}%
\bibitem [{\citenamefont {Ding}\ \emph {et~al.}(2022)\citenamefont {Ding},
  \citenamefont {Shen}, \citenamefont {Leng}, \citenamefont {Xu}, \citenamefont
  {Zhao}, \citenamefont {Zhao}, \citenamefont {Sui}, \citenamefont {Wu},
  \citenamefont {Xiao}, \citenamefont {Zu} \emph {et~al.}}]{ding2022}%
  \BibitemOpen
  \bibfield  {author} {\bibinfo {author} {\bibfnamefont {X.}~\bibnamefont
  {Ding}}, \bibinfo {author} {\bibfnamefont {S.}~\bibnamefont {Shen}}, \bibinfo
  {author} {\bibfnamefont {H.}~\bibnamefont {Leng}}, \bibinfo {author}
  {\bibfnamefont {M.}~\bibnamefont {Xu}}, \bibinfo {author} {\bibfnamefont
  {Y.}~\bibnamefont {Zhao}}, \bibinfo {author} {\bibfnamefont {J.}~\bibnamefont
  {Zhao}}, \bibinfo {author} {\bibfnamefont {X.}~\bibnamefont {Sui}}, \bibinfo
  {author} {\bibfnamefont {X.}~\bibnamefont {Wu}}, \bibinfo {author}
  {\bibfnamefont {H.}~\bibnamefont {Xiao}}, \bibinfo {author} {\bibfnamefont
  {X.}~\bibnamefont {Zu}}, \emph {et~al.},\ }\bibfield  {title} {\bibinfo
  {title} {Stability of {S}uperconducting {Nd$_{0.8}$Sr$_{0.2}$NiO$_2$} {T}hin
  {F}ilms},\ }\href@noop {} {\bibfield  {journal} {\bibinfo  {journal} {arXiv
  preprint arXiv:2201.13032}\ } (\bibinfo {year} {2022})}\BibitemShut {NoStop}%
\bibitem [{\citenamefont {Botana}\ and\ \citenamefont
  {Norman}(2020)}]{botana2020}%
  \BibitemOpen
  \bibfield  {author} {\bibinfo {author} {\bibfnamefont {A.~S.}\ \bibnamefont
  {Botana}}\ and\ \bibinfo {author} {\bibfnamefont {M.~R.}\ \bibnamefont
  {Norman}},\ }\bibfield  {title} {\bibinfo {title} {Similarities and
  differences between {LaNiO$_2$} and {CaCuO$_2$} and implications for
  superconductivity},\ }\href@noop {} {\bibfield  {journal} {\bibinfo
  {journal} {Phys. Rev. X}\ }\textbf {\bibinfo {volume} {10}},\ \bibinfo
  {pages} {011024} (\bibinfo {year} {2020})}\BibitemShut {NoStop}%
\bibitem [{\citenamefont {Sakakibara}\ \emph {et~al.}(2020)\citenamefont
  {Sakakibara}, \citenamefont {Usui}, \citenamefont {Suzuki}, \citenamefont
  {Kotani}, \citenamefont {Aoki},\ and\ \citenamefont
  {Kuroki}}]{sakakibara2020}%
  \BibitemOpen
  \bibfield  {author} {\bibinfo {author} {\bibfnamefont {H.}~\bibnamefont
  {Sakakibara}}, \bibinfo {author} {\bibfnamefont {H.}~\bibnamefont {Usui}},
  \bibinfo {author} {\bibfnamefont {K.}~\bibnamefont {Suzuki}}, \bibinfo
  {author} {\bibfnamefont {T.}~\bibnamefont {Kotani}}, \bibinfo {author}
  {\bibfnamefont {H.}~\bibnamefont {Aoki}},\ and\ \bibinfo {author}
  {\bibfnamefont {K.}~\bibnamefont {Kuroki}},\ }\bibfield  {title} {\bibinfo
  {title} {Model construction and a possibility of cupratelike pairing in a new
  d 9 nickelate superconductor (nd, sr) nio 2},\ }\href@noop {} {\bibfield
  {journal} {\bibinfo  {journal} {Physical Review Letters}\ }\textbf {\bibinfo
  {volume} {125}},\ \bibinfo {pages} {077003} (\bibinfo {year}
  {2020})}\BibitemShut {NoStop}%
\bibitem [{\citenamefont {Jiang}\ \emph {et~al.}(2020)\citenamefont {Jiang},
  \citenamefont {Berciu},\ and\ \citenamefont {Sawatzky}}]{jiang2020}%
  \BibitemOpen
  \bibfield  {author} {\bibinfo {author} {\bibfnamefont {M.}~\bibnamefont
  {Jiang}}, \bibinfo {author} {\bibfnamefont {M.}~\bibnamefont {Berciu}},\ and\
  \bibinfo {author} {\bibfnamefont {G.~A.}\ \bibnamefont {Sawatzky}},\
  }\bibfield  {title} {\bibinfo {title} {Critical nature of the ni spin state
  in doped {NdNiO$_2$}},\ }\href@noop {} {\bibfield  {journal} {\bibinfo
  {journal} {Phys. Rev. Lett.}\ }\textbf {\bibinfo {volume} {124}},\ \bibinfo
  {pages} {207004} (\bibinfo {year} {2020})}\BibitemShut {NoStop}%
\bibitem [{\citenamefont {Wu}\ \emph {et~al.}(2020)\citenamefont {Wu},
  \citenamefont {Di~Sante}, \citenamefont {Schwemmer}, \citenamefont {Hanke},
  \citenamefont {Hwang}, \citenamefont {Raghu},\ and\ \citenamefont
  {Thomale}}]{wu2020}%
  \BibitemOpen
  \bibfield  {author} {\bibinfo {author} {\bibfnamefont {X.}~\bibnamefont
  {Wu}}, \bibinfo {author} {\bibfnamefont {D.}~\bibnamefont {Di~Sante}},
  \bibinfo {author} {\bibfnamefont {T.}~\bibnamefont {Schwemmer}}, \bibinfo
  {author} {\bibfnamefont {W.}~\bibnamefont {Hanke}}, \bibinfo {author}
  {\bibfnamefont {H.~Y.}\ \bibnamefont {Hwang}}, \bibinfo {author}
  {\bibfnamefont {S.}~\bibnamefont {Raghu}},\ and\ \bibinfo {author}
  {\bibfnamefont {R.}~\bibnamefont {Thomale}},\ }\bibfield  {title} {\bibinfo
  {title} {Robust {$d_{x^2 - y^2}$}-wave superconductivity of infinite-layer
  nickelates},\ }\href@noop {} {\bibfield  {journal} {\bibinfo  {journal}
  {Phys. Rev. B}\ }\textbf {\bibinfo {volume} {101}},\ \bibinfo {pages}
  {060504} (\bibinfo {year} {2020})}\BibitemShut {NoStop}%
\bibitem [{\citenamefont {Nomura}\ \emph {et~al.}(2019)\citenamefont {Nomura},
  \citenamefont {Hirayama}, \citenamefont {Tadano}, \citenamefont {Yoshimoto},
  \citenamefont {Nakamura},\ and\ \citenamefont {Arita}}]{nomura2019}%
  \BibitemOpen
  \bibfield  {author} {\bibinfo {author} {\bibfnamefont {Y.}~\bibnamefont
  {Nomura}}, \bibinfo {author} {\bibfnamefont {M.}~\bibnamefont {Hirayama}},
  \bibinfo {author} {\bibfnamefont {T.}~\bibnamefont {Tadano}}, \bibinfo
  {author} {\bibfnamefont {Y.}~\bibnamefont {Yoshimoto}}, \bibinfo {author}
  {\bibfnamefont {K.}~\bibnamefont {Nakamura}},\ and\ \bibinfo {author}
  {\bibfnamefont {R.}~\bibnamefont {Arita}},\ }\bibfield  {title} {\bibinfo
  {title} {Formation of a two-dimensional single-component correlated electron
  system and band engineering in the nickelate superconductor ndnio 2},\
  }\href@noop {} {\bibfield  {journal} {\bibinfo  {journal} {Physical Review
  B}\ }\textbf {\bibinfo {volume} {100}},\ \bibinfo {pages} {205138} (\bibinfo
  {year} {2019})}\BibitemShut {NoStop}%
\bibitem [{\citenamefont {Ryee}\ \emph {et~al.}(2020)\citenamefont {Ryee},
  \citenamefont {Yoon}, \citenamefont {Kim}, \citenamefont {Jeong},\ and\
  \citenamefont {Han}}]{ryee2020}%
  \BibitemOpen
  \bibfield  {author} {\bibinfo {author} {\bibfnamefont {S.}~\bibnamefont
  {Ryee}}, \bibinfo {author} {\bibfnamefont {H.}~\bibnamefont {Yoon}}, \bibinfo
  {author} {\bibfnamefont {T.~J.}\ \bibnamefont {Kim}}, \bibinfo {author}
  {\bibfnamefont {M.~Y.}\ \bibnamefont {Jeong}},\ and\ \bibinfo {author}
  {\bibfnamefont {M.~J.}\ \bibnamefont {Han}},\ }\bibfield  {title} {\bibinfo
  {title} {Induced magnetic two-dimensionality by hole doping in the
  superconducting infinite-layer nickelate {Nd$_{1-x}$Sr$_x$NiO$_2$}},\
  }\href@noop {} {\bibfield  {journal} {\bibinfo  {journal} {Physical Review
  B}\ }\textbf {\bibinfo {volume} {101}},\ \bibinfo {pages} {064513} (\bibinfo
  {year} {2020})}\BibitemShut {NoStop}%
\bibitem [{\citenamefont {Zhang}\ \emph
  {et~al.}(2020{\natexlab{a}})\citenamefont {Zhang}, \citenamefont {Jin},
  \citenamefont {Wang}, \citenamefont {Xi}, \citenamefont {Shi}, \citenamefont
  {Ye},\ and\ \citenamefont {Mei}}]{zhang2020prr}%
  \BibitemOpen
  \bibfield  {author} {\bibinfo {author} {\bibfnamefont {H.}~\bibnamefont
  {Zhang}}, \bibinfo {author} {\bibfnamefont {L.}~\bibnamefont {Jin}}, \bibinfo
  {author} {\bibfnamefont {S.}~\bibnamefont {Wang}}, \bibinfo {author}
  {\bibfnamefont {B.}~\bibnamefont {Xi}}, \bibinfo {author} {\bibfnamefont
  {X.}~\bibnamefont {Shi}}, \bibinfo {author} {\bibfnamefont {F.}~\bibnamefont
  {Ye}},\ and\ \bibinfo {author} {\bibfnamefont {J.-W.}\ \bibnamefont {Mei}},\
  }\bibfield  {title} {\bibinfo {title} {Effective hamiltonian for nickelate
  oxides {Nd$_{1-x}$Sr$_x$NiO$_2$}},\ }\href@noop {} {\bibfield  {journal}
  {\bibinfo  {journal} {Physical Review Research}\ }\textbf {\bibinfo {volume}
  {2}},\ \bibinfo {pages} {013214} (\bibinfo {year}
  {2020}{\natexlab{a}})}\BibitemShut {NoStop}%
\bibitem [{\citenamefont {Zhang}\ \emph
  {et~al.}(2020{\natexlab{b}})\citenamefont {Zhang}, \citenamefont {Yang},\
  and\ \citenamefont {Zhang}}]{zhang2020}%
  \BibitemOpen
  \bibfield  {author} {\bibinfo {author} {\bibfnamefont {G.-M.}\ \bibnamefont
  {Zhang}}, \bibinfo {author} {\bibfnamefont {Y.-F.}\ \bibnamefont {Yang}},\
  and\ \bibinfo {author} {\bibfnamefont {F.-C.}\ \bibnamefont {Zhang}},\
  }\bibfield  {title} {\bibinfo {title} {Self-doped {M}ott insulator for parent
  compounds of nickelate superconductors},\ }\href@noop {} {\bibfield
  {journal} {\bibinfo  {journal} {Physical Review B}\ }\textbf {\bibinfo
  {volume} {101}},\ \bibinfo {pages} {020501(R)} (\bibinfo {year}
  {2020}{\natexlab{b}})}\BibitemShut {NoStop}%
\bibitem [{\citenamefont {Zhang}\ and\ \citenamefont
  {Vishwanath}(2020)}]{zhang2020type}%
  \BibitemOpen
  \bibfield  {author} {\bibinfo {author} {\bibfnamefont {Y.-H.}\ \bibnamefont
  {Zhang}}\ and\ \bibinfo {author} {\bibfnamefont {A.}~\bibnamefont
  {Vishwanath}},\ }\bibfield  {title} {\bibinfo {title} {Type-{II} {$t-J$}
  model in superconducting nickelate {Nd$_{1-x}$Sr$_x$NiO$_2$}},\ }\href@noop
  {} {\bibfield  {journal} {\bibinfo  {journal} {Physical Review Research}\
  }\textbf {\bibinfo {volume} {2}},\ \bibinfo {pages} {023112} (\bibinfo {year}
  {2020})}\BibitemShut {NoStop}%
\bibitem [{\citenamefont {Jiang}\ \emph {et~al.}(2019)\citenamefont {Jiang},
  \citenamefont {Si}, \citenamefont {Liao},\ and\ \citenamefont
  {Zhong}}]{jiang2019}%
  \BibitemOpen
  \bibfield  {author} {\bibinfo {author} {\bibfnamefont {P.}~\bibnamefont
  {Jiang}}, \bibinfo {author} {\bibfnamefont {L.}~\bibnamefont {Si}}, \bibinfo
  {author} {\bibfnamefont {Z.}~\bibnamefont {Liao}},\ and\ \bibinfo {author}
  {\bibfnamefont {Z.}~\bibnamefont {Zhong}},\ }\bibfield  {title} {\bibinfo
  {title} {Electronic structure of rare-earth infinite-layer {$R$NiO$_2$ (R=
  La, Nd)}},\ }\href@noop {} {\bibfield  {journal} {\bibinfo  {journal}
  {Physical Review B}\ }\textbf {\bibinfo {volume} {100}},\ \bibinfo {pages}
  {201106} (\bibinfo {year} {2019})}\BibitemShut {NoStop}%
\bibitem [{\citenamefont {Werner}\ and\ \citenamefont
  {Hoshino}(2020)}]{werner2020}%
  \BibitemOpen
  \bibfield  {author} {\bibinfo {author} {\bibfnamefont {P.}~\bibnamefont
  {Werner}}\ and\ \bibinfo {author} {\bibfnamefont {S.}~\bibnamefont
  {Hoshino}},\ }\bibfield  {title} {\bibinfo {title} {Nickelate
  superconductors: Multiorbital nature and spin freezing},\ }\href@noop {}
  {\bibfield  {journal} {\bibinfo  {journal} {Physical Review B}\ }\textbf
  {\bibinfo {volume} {101}},\ \bibinfo {pages} {041104} (\bibinfo {year}
  {2020})}\BibitemShut {NoStop}%
\bibitem [{\citenamefont {Hu}\ and\ \citenamefont {Wu}(2019)}]{hu2019}%
  \BibitemOpen
  \bibfield  {author} {\bibinfo {author} {\bibfnamefont {L.-H.}\ \bibnamefont
  {Hu}}\ and\ \bibinfo {author} {\bibfnamefont {C.}~\bibnamefont {Wu}},\
  }\bibfield  {title} {\bibinfo {title} {Two-band model for magnetism and
  superconductivity in nickelates},\ }\href@noop {} {\bibfield  {journal}
  {\bibinfo  {journal} {Physical Review Research}\ }\textbf {\bibinfo {volume}
  {1}},\ \bibinfo {pages} {032046} (\bibinfo {year} {2019})}\BibitemShut
  {NoStop}%
\bibitem [{\citenamefont {Zhou}\ \emph {et~al.}(2020)\citenamefont {Zhou},
  \citenamefont {Gao},\ and\ \citenamefont {Wang}}]{zhou2020spin}%
  \BibitemOpen
  \bibfield  {author} {\bibinfo {author} {\bibfnamefont {T.}~\bibnamefont
  {Zhou}}, \bibinfo {author} {\bibfnamefont {Y.}~\bibnamefont {Gao}},\ and\
  \bibinfo {author} {\bibfnamefont {Z.}~\bibnamefont {Wang}},\ }\bibfield
  {title} {\bibinfo {title} {Spin excitations in nickelate superconductors},\
  }\href@noop {} {\bibfield  {journal} {\bibinfo  {journal} {Science China
  Physics, Mechanics \& Astronomy}\ }\textbf {\bibinfo {volume} {63}},\
  \bibinfo {pages} {1} (\bibinfo {year} {2020})}\BibitemShut {NoStop}%
\bibitem [{\citenamefont {Bernardini}\ \emph {et~al.}(2020)\citenamefont
  {Bernardini}, \citenamefont {Olevano},\ and\ \citenamefont
  {Cano}}]{bernardini2020}%
  \BibitemOpen
  \bibfield  {author} {\bibinfo {author} {\bibfnamefont {F.}~\bibnamefont
  {Bernardini}}, \bibinfo {author} {\bibfnamefont {V.}~\bibnamefont
  {Olevano}},\ and\ \bibinfo {author} {\bibfnamefont {A.}~\bibnamefont
  {Cano}},\ }\bibfield  {title} {\bibinfo {title} {Magnetic penetration depth
  and {$T_c$} in superconducting nickelates},\ }\href@noop {} {\bibfield
  {journal} {\bibinfo  {journal} {Physical Review Research}\ }\textbf {\bibinfo
  {volume} {2}},\ \bibinfo {pages} {013219} (\bibinfo {year}
  {2020})}\BibitemShut {NoStop}%
\bibitem [{\citenamefont {Gu}\ \emph {et~al.}(2020{\natexlab{b}})\citenamefont
  {Gu}, \citenamefont {Zhu}, \citenamefont {Wang}, \citenamefont {Hu},\ and\
  \citenamefont {Chen}}]{gu2020cp}%
  \BibitemOpen
  \bibfield  {author} {\bibinfo {author} {\bibfnamefont {Y.}~\bibnamefont
  {Gu}}, \bibinfo {author} {\bibfnamefont {S.}~\bibnamefont {Zhu}}, \bibinfo
  {author} {\bibfnamefont {X.}~\bibnamefont {Wang}}, \bibinfo {author}
  {\bibfnamefont {J.}~\bibnamefont {Hu}},\ and\ \bibinfo {author}
  {\bibfnamefont {H.}~\bibnamefont {Chen}},\ }\bibfield  {title} {\bibinfo
  {title} {A substantial hybridization between correlated {Ni-$d$} orbital and
  itinerant electrons in infinite-layer nickelates},\ }\href@noop {} {\bibfield
   {journal} {\bibinfo  {journal} {Communications Physics}\ }\textbf {\bibinfo
  {volume} {3}},\ \bibinfo {pages} {1} (\bibinfo {year}
  {2020}{\natexlab{b}})}\BibitemShut {NoStop}%
\bibitem [{\citenamefont {Choi}\ \emph
  {et~al.}(2020{\natexlab{a}})\citenamefont {Choi}, \citenamefont {Lee},\ and\
  \citenamefont {Pickett}}]{choi2020prb}%
  \BibitemOpen
  \bibfield  {author} {\bibinfo {author} {\bibfnamefont {M.-Y.}\ \bibnamefont
  {Choi}}, \bibinfo {author} {\bibfnamefont {K.-W.}\ \bibnamefont {Lee}},\ and\
  \bibinfo {author} {\bibfnamefont {W.~E.}\ \bibnamefont {Pickett}},\
  }\bibfield  {title} {\bibinfo {title} {Role of {$4f$} states in
  infinite-layer {NdNiO$_2$}},\ }\href@noop {} {\bibfield  {journal} {\bibinfo
  {journal} {Physical Review B}\ }\textbf {\bibinfo {volume} {101}},\ \bibinfo
  {pages} {020503} (\bibinfo {year} {2020}{\natexlab{a}})}\BibitemShut
  {NoStop}%
\bibitem [{\citenamefont {Lechermann}(2020{\natexlab{a}})}]{lechermann2020prb}%
  \BibitemOpen
  \bibfield  {author} {\bibinfo {author} {\bibfnamefont {F.}~\bibnamefont
  {Lechermann}},\ }\bibfield  {title} {\bibinfo {title} {Late transition metal
  oxides with infinite-layer structure: Nickelates versus cuprates},\
  }\href@noop {} {\bibfield  {journal} {\bibinfo  {journal} {Physical Review
  B}\ }\textbf {\bibinfo {volume} {101}},\ \bibinfo {pages} {081110} (\bibinfo
  {year} {2020}{\natexlab{a}})}\BibitemShut {NoStop}%
\bibitem [{\citenamefont {Chang}\ \emph {et~al.}(2020)\citenamefont {Chang},
  \citenamefont {Zhao},\ and\ \citenamefont {Ding}}]{chang2020hund}%
  \BibitemOpen
  \bibfield  {author} {\bibinfo {author} {\bibfnamefont {J.}~\bibnamefont
  {Chang}}, \bibinfo {author} {\bibfnamefont {J.}~\bibnamefont {Zhao}},\ and\
  \bibinfo {author} {\bibfnamefont {Y.}~\bibnamefont {Ding}},\ }\bibfield
  {title} {\bibinfo {title} {Hund-heisenberg model in superconducting
  infinite-layer nickelates},\ }\href@noop {} {\bibfield  {journal} {\bibinfo
  {journal} {The European Physical Journal B}\ }\textbf {\bibinfo {volume}
  {93}},\ \bibinfo {pages} {1} (\bibinfo {year} {2020})}\BibitemShut {NoStop}%
\bibitem [{\citenamefont {Liu}\ \emph {et~al.}(2020)\citenamefont {Liu},
  \citenamefont {Ren}, \citenamefont {Zhu}, \citenamefont {Wang},\ and\
  \citenamefont {Yang}}]{liu2020}%
  \BibitemOpen
  \bibfield  {author} {\bibinfo {author} {\bibfnamefont {Z.}~\bibnamefont
  {Liu}}, \bibinfo {author} {\bibfnamefont {Z.}~\bibnamefont {Ren}}, \bibinfo
  {author} {\bibfnamefont {W.}~\bibnamefont {Zhu}}, \bibinfo {author}
  {\bibfnamefont {Z.}~\bibnamefont {Wang}},\ and\ \bibinfo {author}
  {\bibfnamefont {J.}~\bibnamefont {Yang}},\ }\bibfield  {title} {\bibinfo
  {title} {Electronic and magnetic structure of infinite-layer {NdNiO$_2$}:
  trace of antiferromagnetic metal},\ }\href@noop {} {\bibfield  {journal}
  {\bibinfo  {journal} {npj Quantum Materials}\ }\textbf {\bibinfo {volume}
  {5}},\ \bibinfo {pages} {1} (\bibinfo {year} {2020})}\BibitemShut {NoStop}%
\bibitem [{\citenamefont {Karp}\ \emph {et~al.}(2020)\citenamefont {Karp},
  \citenamefont {Botana}, \citenamefont {Norman}, \citenamefont {Park},
  \citenamefont {Zingl},\ and\ \citenamefont {Millis}}]{karp2020}%
  \BibitemOpen
  \bibfield  {author} {\bibinfo {author} {\bibfnamefont {J.}~\bibnamefont
  {Karp}}, \bibinfo {author} {\bibfnamefont {A.~S.}\ \bibnamefont {Botana}},
  \bibinfo {author} {\bibfnamefont {M.~R.}\ \bibnamefont {Norman}}, \bibinfo
  {author} {\bibfnamefont {H.}~\bibnamefont {Park}}, \bibinfo {author}
  {\bibfnamefont {M.}~\bibnamefont {Zingl}},\ and\ \bibinfo {author}
  {\bibfnamefont {A.}~\bibnamefont {Millis}},\ }\bibfield  {title} {\bibinfo
  {title} {Many-body electronic structure of {NdNiO$_2$} and {CaCuO$_2$}},\
  }\href@noop {} {\bibfield  {journal} {\bibinfo  {journal} {Physical Review
  X}\ }\textbf {\bibinfo {volume} {10}},\ \bibinfo {pages} {021061} (\bibinfo
  {year} {2020})}\BibitemShut {NoStop}%
\bibitem [{\citenamefont {Kitatani}\ \emph {et~al.}(2020)\citenamefont
  {Kitatani}, \citenamefont {Si}, \citenamefont {Janson}, \citenamefont
  {Arita}, \citenamefont {Zhong},\ and\ \citenamefont {Held}}]{kitatani2020}%
  \BibitemOpen
  \bibfield  {author} {\bibinfo {author} {\bibfnamefont {M.}~\bibnamefont
  {Kitatani}}, \bibinfo {author} {\bibfnamefont {L.}~\bibnamefont {Si}},
  \bibinfo {author} {\bibfnamefont {O.}~\bibnamefont {Janson}}, \bibinfo
  {author} {\bibfnamefont {R.}~\bibnamefont {Arita}}, \bibinfo {author}
  {\bibfnamefont {Z.}~\bibnamefont {Zhong}},\ and\ \bibinfo {author}
  {\bibfnamefont {K.}~\bibnamefont {Held}},\ }\bibfield  {title} {\bibinfo
  {title} {Nickelate superconductors—a renaissance of the one-band hubbard
  model},\ }\href@noop {} {\bibfield  {journal} {\bibinfo  {journal} {npj
  Quantum Materials}\ }\textbf {\bibinfo {volume} {5}},\ \bibinfo {pages} {1}
  (\bibinfo {year} {2020})}\BibitemShut {NoStop}%
\bibitem [{\citenamefont {Been}\ \emph {et~al.}(2021)\citenamefont {Been},
  \citenamefont {Lee}, \citenamefont {Hwang}, \citenamefont {Cui},
  \citenamefont {Zaanen}, \citenamefont {Devereaux}, \citenamefont {Moritz},\
  and\ \citenamefont {Jia}}]{been2021}%
  \BibitemOpen
  \bibfield  {author} {\bibinfo {author} {\bibfnamefont {E.}~\bibnamefont
  {Been}}, \bibinfo {author} {\bibfnamefont {W.-S.}\ \bibnamefont {Lee}},
  \bibinfo {author} {\bibfnamefont {H.~Y.}\ \bibnamefont {Hwang}}, \bibinfo
  {author} {\bibfnamefont {Y.}~\bibnamefont {Cui}}, \bibinfo {author}
  {\bibfnamefont {J.}~\bibnamefont {Zaanen}}, \bibinfo {author} {\bibfnamefont
  {T.}~\bibnamefont {Devereaux}}, \bibinfo {author} {\bibfnamefont
  {B.}~\bibnamefont {Moritz}},\ and\ \bibinfo {author} {\bibfnamefont
  {C.}~\bibnamefont {Jia}},\ }\bibfield  {title} {\bibinfo {title} {Electronic
  structure trends across the rare-earth series in superconducting
  infinite-layer nickelates},\ }\href@noop {} {\bibfield  {journal} {\bibinfo
  {journal} {Phys. Rev. X}\ }\textbf {\bibinfo {volume} {11}},\ \bibinfo
  {pages} {011050} (\bibinfo {year} {2021})}\BibitemShut {NoStop}%
\bibitem [{\citenamefont {Leonov}\ \emph {et~al.}(2020)\citenamefont {Leonov},
  \citenamefont {Skornyakov},\ and\ \citenamefont {Savrasov}}]{leonov2020}%
  \BibitemOpen
  \bibfield  {author} {\bibinfo {author} {\bibfnamefont {I.}~\bibnamefont
  {Leonov}}, \bibinfo {author} {\bibfnamefont {S.}~\bibnamefont {Skornyakov}},\
  and\ \bibinfo {author} {\bibfnamefont {S.}~\bibnamefont {Savrasov}},\
  }\bibfield  {title} {\bibinfo {title} {Lifshitz transition and frustration of
  magnetic moments in infinite-layer {NdNiO$_2$} upon hole doping},\
  }\href@noop {} {\bibfield  {journal} {\bibinfo  {journal} {Physical Review
  B}\ }\textbf {\bibinfo {volume} {101}},\ \bibinfo {pages} {241108} (\bibinfo
  {year} {2020})}\BibitemShut {NoStop}%
\bibitem [{\citenamefont {Lechermann}(2020{\natexlab{b}})}]{lechermann2020}%
  \BibitemOpen
  \bibfield  {author} {\bibinfo {author} {\bibfnamefont {F.}~\bibnamefont
  {Lechermann}},\ }\bibfield  {title} {\bibinfo {title} {Multiorbital processes
  rule the {Nd$_{1-x}$Sr$_x$NiO$_2$} normal state},\ }\href@noop {} {\bibfield
  {journal} {\bibinfo  {journal} {Physical Review X}\ }\textbf {\bibinfo
  {volume} {10}},\ \bibinfo {pages} {041002} (\bibinfo {year}
  {2020}{\natexlab{b}})}\BibitemShut {NoStop}%
\bibitem [{\citenamefont {Choi}\ \emph
  {et~al.}(2020{\natexlab{b}})\citenamefont {Choi}, \citenamefont {Pickett},\
  and\ \citenamefont {Lee}}]{choi2020}%
  \BibitemOpen
  \bibfield  {author} {\bibinfo {author} {\bibfnamefont {M.-Y.}\ \bibnamefont
  {Choi}}, \bibinfo {author} {\bibfnamefont {W.~E.}\ \bibnamefont {Pickett}},\
  and\ \bibinfo {author} {\bibfnamefont {K.-W.}\ \bibnamefont {Lee}},\
  }\bibfield  {title} {\bibinfo {title} {Fluctuation-frustrated flat band
  instabilities in {NdNiO$_2$}},\ }\href@noop {} {\bibfield  {journal}
  {\bibinfo  {journal} {Physical Review Research}\ }\textbf {\bibinfo {volume}
  {2}},\ \bibinfo {pages} {033445} (\bibinfo {year}
  {2020}{\natexlab{b}})}\BibitemShut {NoStop}%
\bibitem [{\citenamefont {Wang}\ \emph
  {et~al.}(2020{\natexlab{b}})\citenamefont {Wang}, \citenamefont {Kang},
  \citenamefont {Miao},\ and\ \citenamefont {Kotliar}}]{wang2020}%
  \BibitemOpen
  \bibfield  {author} {\bibinfo {author} {\bibfnamefont {Y.}~\bibnamefont
  {Wang}}, \bibinfo {author} {\bibfnamefont {C.-J.}\ \bibnamefont {Kang}},
  \bibinfo {author} {\bibfnamefont {H.}~\bibnamefont {Miao}},\ and\ \bibinfo
  {author} {\bibfnamefont {G.}~\bibnamefont {Kotliar}},\ }\bibfield  {title}
  {\bibinfo {title} {Hund's metal physics: {F}rom {SrNiO$_2$} to {LaNiO$_2$}},\
  }\href@noop {} {\bibfield  {journal} {\bibinfo  {journal} {Physical Review
  B}\ }\textbf {\bibinfo {volume} {102}},\ \bibinfo {pages} {161118} (\bibinfo
  {year} {2020}{\natexlab{b}})}\BibitemShut {NoStop}%
\bibitem [{\citenamefont {Kang}\ \emph {et~al.}(2020)\citenamefont {Kang},
  \citenamefont {Melnick}, \citenamefont {Semon}, \citenamefont {Ryee},
  \citenamefont {Han}, \citenamefont {Kotliar},\ and\ \citenamefont
  {Choi}}]{kang2020}%
  \BibitemOpen
  \bibfield  {author} {\bibinfo {author} {\bibfnamefont {B.}~\bibnamefont
  {Kang}}, \bibinfo {author} {\bibfnamefont {C.}~\bibnamefont {Melnick}},
  \bibinfo {author} {\bibfnamefont {P.}~\bibnamefont {Semon}}, \bibinfo
  {author} {\bibfnamefont {S.}~\bibnamefont {Ryee}}, \bibinfo {author}
  {\bibfnamefont {M.~J.}\ \bibnamefont {Han}}, \bibinfo {author} {\bibfnamefont
  {G.}~\bibnamefont {Kotliar}},\ and\ \bibinfo {author} {\bibfnamefont
  {S.}~\bibnamefont {Choi}},\ }\bibfield  {title} {\bibinfo {title}
  {Infinite-layer nickelates as ni-eg hund's metals},\ }\href@noop {}
  {\bibfield  {journal} {\bibinfo  {journal} {arXiv preprint arXiv:2007.14610}\
  } (\bibinfo {year} {2020})}\BibitemShut {NoStop}%
\bibitem [{\citenamefont {Kapeghian}\ and\ \citenamefont
  {Botana}(2020)}]{kapeghian2020}%
  \BibitemOpen
  \bibfield  {author} {\bibinfo {author} {\bibfnamefont {J.}~\bibnamefont
  {Kapeghian}}\ and\ \bibinfo {author} {\bibfnamefont {A.~S.}\ \bibnamefont
  {Botana}},\ }\bibfield  {title} {\bibinfo {title} {Electronic structure and
  magnetism in infinite-layer nickelates {$R$NiO$_2$ ($R$ = La--Lu)}},\
  }\href@noop {} {\bibfield  {journal} {\bibinfo  {journal} {Physical Review
  B}\ }\textbf {\bibinfo {volume} {102}},\ \bibinfo {pages} {205130} (\bibinfo
  {year} {2020})}\BibitemShut {NoStop}%
\bibitem [{\citenamefont {Kang}\ and\ \citenamefont
  {Kotliar}(2021)}]{kang2021}%
  \BibitemOpen
  \bibfield  {author} {\bibinfo {author} {\bibfnamefont {C.-J.}\ \bibnamefont
  {Kang}}\ and\ \bibinfo {author} {\bibfnamefont {G.}~\bibnamefont {Kotliar}},\
  }\bibfield  {title} {\bibinfo {title} {Optical properties of the
  infinite-layer {La$_{1-x}$Sr$_x$NiO$_2$} and hidden {H}und's physics},\
  }\href@noop {} {\bibfield  {journal} {\bibinfo  {journal} {Physical review
  letters}\ }\textbf {\bibinfo {volume} {126}},\ \bibinfo {pages} {127401}
  (\bibinfo {year} {2021})}\BibitemShut {NoStop}%
\bibitem [{\citenamefont {Zhang}\ \emph {et~al.}(2021)\citenamefont {Zhang},
  \citenamefont {Lane}, \citenamefont {Singh}, \citenamefont {Nokelainen},
  \citenamefont {Barbiellini}, \citenamefont {Markiewicz}, \citenamefont
  {Bansil},\ and\ \citenamefont {Sun}}]{zhang2021magnetic}%
  \BibitemOpen
  \bibfield  {author} {\bibinfo {author} {\bibfnamefont {R.}~\bibnamefont
  {Zhang}}, \bibinfo {author} {\bibfnamefont {C.}~\bibnamefont {Lane}},
  \bibinfo {author} {\bibfnamefont {B.}~\bibnamefont {Singh}}, \bibinfo
  {author} {\bibfnamefont {J.}~\bibnamefont {Nokelainen}}, \bibinfo {author}
  {\bibfnamefont {B.}~\bibnamefont {Barbiellini}}, \bibinfo {author}
  {\bibfnamefont {R.~S.}\ \bibnamefont {Markiewicz}}, \bibinfo {author}
  {\bibfnamefont {A.}~\bibnamefont {Bansil}},\ and\ \bibinfo {author}
  {\bibfnamefont {J.}~\bibnamefont {Sun}},\ }\bibfield  {title} {\bibinfo
  {title} {Magnetic and {$f$}-electron effects in {LaNiO$_2$} and {NdNiO$_2$}
  nickelates with cuprate-like {$3d_{x^2-y^2}$} band},\ }\href@noop {}
  {\bibfield  {journal} {\bibinfo  {journal} {Communications Physics}\ }\textbf
  {\bibinfo {volume} {4}},\ \bibinfo {pages} {1} (\bibinfo {year}
  {2021})}\BibitemShut {NoStop}%
\bibitem [{\citenamefont {Lechermann}(2021)}]{lechermann2021}%
  \BibitemOpen
  \bibfield  {author} {\bibinfo {author} {\bibfnamefont {F.}~\bibnamefont
  {Lechermann}},\ }\bibfield  {title} {\bibinfo {title} {Doping-dependent
  character and possible magnetic ordering of {NdNiO$_2$}},\ }\href@noop {}
  {\bibfield  {journal} {\bibinfo  {journal} {Physical Review Materials}\
  }\textbf {\bibinfo {volume} {5}},\ \bibinfo {pages} {044803} (\bibinfo {year}
  {2021})}\BibitemShut {NoStop}%
\bibitem [{\citenamefont {Katukuri}\ \emph {et~al.}(2020)\citenamefont
  {Katukuri}, \citenamefont {Bogdanov}, \citenamefont {Weser}, \citenamefont
  {Van~den Brink},\ and\ \citenamefont {Alavi}}]{katukuri2020}%
  \BibitemOpen
  \bibfield  {author} {\bibinfo {author} {\bibfnamefont {V.~M.}\ \bibnamefont
  {Katukuri}}, \bibinfo {author} {\bibfnamefont {N.~A.}\ \bibnamefont
  {Bogdanov}}, \bibinfo {author} {\bibfnamefont {O.}~\bibnamefont {Weser}},
  \bibinfo {author} {\bibfnamefont {J.}~\bibnamefont {Van~den Brink}},\ and\
  \bibinfo {author} {\bibfnamefont {A.}~\bibnamefont {Alavi}},\ }\bibfield
  {title} {\bibinfo {title} {Electronic correlations and magnetic interactions
  in infinite-layer ndnio 2},\ }\href@noop {} {\bibfield  {journal} {\bibinfo
  {journal} {Physical Review B}\ }\textbf {\bibinfo {volume} {102}},\ \bibinfo
  {pages} {241112(R)} (\bibinfo {year} {2020})}\BibitemShut {NoStop}%
\bibitem [{\citenamefont {Malyi}\ \emph {et~al.}(2022)\citenamefont {Malyi},
  \citenamefont {Varignon},\ and\ \citenamefont {Zunger}}]{malyi2022}%
  \BibitemOpen
  \bibfield  {author} {\bibinfo {author} {\bibfnamefont {O.~I.}\ \bibnamefont
  {Malyi}}, \bibinfo {author} {\bibfnamefont {J.}~\bibnamefont {Varignon}},\
  and\ \bibinfo {author} {\bibfnamefont {A.}~\bibnamefont {Zunger}},\
  }\bibfield  {title} {\bibinfo {title} {Bulk {NdNiO$_2$} is thermodynamically
  unstable with respect to decomposition while hydrogenation reduces the
  instability and transforms it from metal to insulator},\ }\href@noop {}
  {\bibfield  {journal} {\bibinfo  {journal} {Physical Review B}\ }\textbf
  {\bibinfo {volume} {105}},\ \bibinfo {pages} {014106} (\bibinfo {year}
  {2022})}\BibitemShut {NoStop}%
\bibitem [{\citenamefont {Plienbumrung}\ \emph {et~al.}(2021)\citenamefont
  {Plienbumrung}, \citenamefont {Schmid}, \citenamefont {Daghofer},\ and\
  \citenamefont {Ole{\'s}}}]{plienbumrung2021}%
  \BibitemOpen
  \bibfield  {author} {\bibinfo {author} {\bibfnamefont {T.}~\bibnamefont
  {Plienbumrung}}, \bibinfo {author} {\bibfnamefont {M.~T.}\ \bibnamefont
  {Schmid}}, \bibinfo {author} {\bibfnamefont {M.}~\bibnamefont {Daghofer}},\
  and\ \bibinfo {author} {\bibfnamefont {A.~M.}\ \bibnamefont {Ole{\'s}}},\
  }\bibfield  {title} {\bibinfo {title} {Character of doped holes in nd1-
  xsrxnio2},\ }\href@noop {} {\bibfield  {journal} {\bibinfo  {journal}
  {Condensed Matter}\ }\textbf {\bibinfo {volume} {6}},\ \bibinfo {pages} {33}
  (\bibinfo {year} {2021})}\BibitemShut {NoStop}%
\bibitem [{\citenamefont {Klett}\ \emph {et~al.}(2022)\citenamefont {Klett},
  \citenamefont {Hansmann},\ and\ \citenamefont {Sch{\"a}fer}}]{klett2022}%
  \BibitemOpen
  \bibfield  {author} {\bibinfo {author} {\bibfnamefont {M.}~\bibnamefont
  {Klett}}, \bibinfo {author} {\bibfnamefont {P.}~\bibnamefont {Hansmann}},\
  and\ \bibinfo {author} {\bibfnamefont {T.}~\bibnamefont {Sch{\"a}fer}},\
  }\bibfield  {title} {\bibinfo {title} {Magnetic properties and pseudogap
  formation in infinite-layer nickelates: insights from the single-band
  {H}ubbard model},\ }\href@noop {} {\bibfield  {journal} {\bibinfo  {journal}
  {Frontiers in Physics}\ ,\ \bibinfo {pages} {45}} (\bibinfo {year}
  {2022})}\BibitemShut {NoStop}%
\bibitem [{\citenamefont {Xia}\ \emph {et~al.}(2021)\citenamefont {Xia},
  \citenamefont {Wu}, \citenamefont {Chen},\ and\ \citenamefont
  {Chen}}]{xia2021}%
  \BibitemOpen
  \bibfield  {author} {\bibinfo {author} {\bibfnamefont {C.}~\bibnamefont
  {Xia}}, \bibinfo {author} {\bibfnamefont {J.}~\bibnamefont {Wu}}, \bibinfo
  {author} {\bibfnamefont {Y.}~\bibnamefont {Chen}},\ and\ \bibinfo {author}
  {\bibfnamefont {H.}~\bibnamefont {Chen}},\ }\bibfield  {title} {\bibinfo
  {title} {Dynamical structural instability and a new crystal-electronic
  structure of infinite-layer nickelates},\ }\href@noop {} {\bibfield
  {journal} {\bibinfo  {journal} {arXiv preprint arXiv:2110.12405}\ } (\bibinfo
  {year} {2021})}\BibitemShut {NoStop}%
\bibitem [{\citenamefont {Bernardini}\ \emph {et~al.}(2021)\citenamefont
  {Bernardini}, \citenamefont {Bosin},\ and\ \citenamefont
  {Cano}}]{bernardini2021}%
  \BibitemOpen
  \bibfield  {author} {\bibinfo {author} {\bibfnamefont {F.}~\bibnamefont
  {Bernardini}}, \bibinfo {author} {\bibfnamefont {A.}~\bibnamefont {Bosin}},\
  and\ \bibinfo {author} {\bibfnamefont {A.}~\bibnamefont {Cano}},\ }\bibfield
  {title} {\bibinfo {title} {Geometric effects in the infinite-layer
  nickelates},\ }\href@noop {} {\bibfield  {journal} {\bibinfo  {journal}
  {arXiv preprint arXiv:2110.13580}\ } (\bibinfo {year} {2021})}\BibitemShut
  {NoStop}%
\bibitem [{\citenamefont {Carrasco~{\'A}lvarez}\ \emph
  {et~al.}(2021)\citenamefont {Carrasco~{\'A}lvarez}, \citenamefont {Petit},
  \citenamefont {Iglesias}, \citenamefont {Prellier}, \citenamefont {Bibes},\
  and\ \citenamefont {Varignon}}]{carrasco2021}%
  \BibitemOpen
  \bibfield  {author} {\bibinfo {author} {\bibfnamefont {{\'A}.~A.}\
  \bibnamefont {Carrasco~{\'A}lvarez}}, \bibinfo {author} {\bibfnamefont
  {S.}~\bibnamefont {Petit}}, \bibinfo {author} {\bibfnamefont
  {L.}~\bibnamefont {Iglesias}}, \bibinfo {author} {\bibfnamefont
  {W.}~\bibnamefont {Prellier}}, \bibinfo {author} {\bibfnamefont
  {M.}~\bibnamefont {Bibes}},\ and\ \bibinfo {author} {\bibfnamefont
  {J.}~\bibnamefont {Varignon}},\ }\bibfield  {title} {\bibinfo {title}
  {Structural instabilities of infinite-layer nickelates from first-principles
  simulations},\ }\href@noop {} {\bibfield  {journal} {\bibinfo  {journal}
  {arXiv preprint arXiv:2112.02642}\ } (\bibinfo {year} {2021})}\BibitemShut
  {NoStop}%
\bibitem [{\citenamefont {Zhang}\ \emph {et~al.}(2022)\citenamefont {Zhang},
  \citenamefont {Zhang}, \citenamefont {He}, \citenamefont {Wang},\ and\
  \citenamefont {Ghosez}}]{zhang2022phase}%
  \BibitemOpen
  \bibfield  {author} {\bibinfo {author} {\bibfnamefont {Y.}~\bibnamefont
  {Zhang}}, \bibinfo {author} {\bibfnamefont {J.}~\bibnamefont {Zhang}},
  \bibinfo {author} {\bibfnamefont {X.}~\bibnamefont {He}}, \bibinfo {author}
  {\bibfnamefont {J.}~\bibnamefont {Wang}},\ and\ \bibinfo {author}
  {\bibfnamefont {P.}~\bibnamefont {Ghosez}},\ }\bibfield  {title} {\bibinfo
  {title} {Phase {D}iagram of {I}nfinite-layer {N}ickelate {C}ompounds from
  {F}irst-and {S}econd-principles {C}alculations},\ }\href@noop {} {\bibfield
  {journal} {\bibinfo  {journal} {arXiv preprint arXiv:2201.00709}\ } (\bibinfo
  {year} {2022})}\BibitemShut {NoStop}%
\bibitem [{\citenamefont {Karp}\ \emph {et~al.}(2022)\citenamefont {Karp},
  \citenamefont {Hampel},\ and\ \citenamefont {Millis}}]{karp2022}%
  \BibitemOpen
  \bibfield  {author} {\bibinfo {author} {\bibfnamefont {J.}~\bibnamefont
  {Karp}}, \bibinfo {author} {\bibfnamefont {A.}~\bibnamefont {Hampel}},\ and\
  \bibinfo {author} {\bibfnamefont {A.~J.}\ \bibnamefont {Millis}},\ }\bibfield
   {title} {\bibinfo {title} {Superconductivity and {A}ntiferromagnetism in
  {NdNiO$_2$} and {CaCuO$_2$}: {A} {C}luster {DMFT} {S}tudy},\ }\href@noop {}
  {\bibfield  {journal} {\bibinfo  {journal} {arXiv preprint arXiv:2201.10481}\
  } (\bibinfo {year} {2022})}\BibitemShut {NoStop}%
\bibitem [{\citenamefont {Jiang}(2022)}]{jiang2022}%
  \BibitemOpen
  \bibfield  {author} {\bibinfo {author} {\bibfnamefont {M.}~\bibnamefont
  {Jiang}},\ }\bibfield  {title} {\bibinfo {title} {Characterizing the
  superconducting instability in a two-orbital {$d$-$s$} model: insights to
  infinite-layer nickelate superconductors},\ }\href@noop {} {\bibfield
  {journal} {\bibinfo  {journal} {arXiv preprint arXiv:2201.12967}\ } (\bibinfo
  {year} {2022})}\BibitemShut {NoStop}%
\bibitem [{\citenamefont {Si}\ \emph {et~al.}(2020)\citenamefont {Si},
  \citenamefont {Xiao}, \citenamefont {Kaufmann}, \citenamefont {Tomczak},
  \citenamefont {Lu}, \citenamefont {Zhong},\ and\ \citenamefont
  {Held}}]{si2020}%
  \BibitemOpen
  \bibfield  {author} {\bibinfo {author} {\bibfnamefont {L.}~\bibnamefont
  {Si}}, \bibinfo {author} {\bibfnamefont {W.}~\bibnamefont {Xiao}}, \bibinfo
  {author} {\bibfnamefont {J.}~\bibnamefont {Kaufmann}}, \bibinfo {author}
  {\bibfnamefont {J.~M.}\ \bibnamefont {Tomczak}}, \bibinfo {author}
  {\bibfnamefont {Y.}~\bibnamefont {Lu}}, \bibinfo {author} {\bibfnamefont
  {Z.}~\bibnamefont {Zhong}},\ and\ \bibinfo {author} {\bibfnamefont
  {K.}~\bibnamefont {Held}},\ }\bibfield  {title} {\bibinfo {title} {Topotactic
  hydrogen in nickelate superconductors and akin infinite-layer oxides a b o
  2},\ }\href@noop {} {\bibfield  {journal} {\bibinfo  {journal} {Physical
  Review Letters}\ }\textbf {\bibinfo {volume} {124}},\ \bibinfo {pages}
  {166402} (\bibinfo {year} {2020})}\BibitemShut {NoStop}%
\bibitem [{\citenamefont {Crespin}\ \emph {et~al.}(1983)\citenamefont
  {Crespin}, \citenamefont {Levitz},\ and\ \citenamefont
  {Gatineau}}]{crespin1983}%
  \BibitemOpen
  \bibfield  {author} {\bibinfo {author} {\bibfnamefont {M.}~\bibnamefont
  {Crespin}}, \bibinfo {author} {\bibfnamefont {P.}~\bibnamefont {Levitz}},\
  and\ \bibinfo {author} {\bibfnamefont {L.}~\bibnamefont {Gatineau}},\
  }\bibfield  {title} {\bibinfo {title} {Reduced forms of lanio 3 perovskite.
  part 1.—evidence for new phases: La 2 ni 2 o 5 and lanio 2},\ }\href@noop
  {} {\bibfield  {journal} {\bibinfo  {journal} {Journal of the Chemical
  Society, Faraday Transactions 2: Molecular and Chemical Physics}\ }\textbf
  {\bibinfo {volume} {79}},\ \bibinfo {pages} {1181} (\bibinfo {year}
  {1983})}\BibitemShut {NoStop}%
\bibitem [{\citenamefont {Levitz}\ \emph {et~al.}(1983)\citenamefont {Levitz},
  \citenamefont {Crespin},\ and\ \citenamefont {Gatineau}}]{levitz1983}%
  \BibitemOpen
  \bibfield  {author} {\bibinfo {author} {\bibfnamefont {P.}~\bibnamefont
  {Levitz}}, \bibinfo {author} {\bibfnamefont {M.}~\bibnamefont {Crespin}},\
  and\ \bibinfo {author} {\bibfnamefont {L.}~\bibnamefont {Gatineau}},\
  }\bibfield  {title} {\bibinfo {title} {Reduced forms of lanio 3 perovskite.
  part 2.—x-ray structure of lanio 2 and extended x-ray absorption fine
  structure study: local environment of monovalent nickel},\ }\href@noop {}
  {\bibfield  {journal} {\bibinfo  {journal} {Journal of the Chemical Society,
  Faraday Transactions 2: Molecular and Chemical Physics}\ }\textbf {\bibinfo
  {volume} {79}},\ \bibinfo {pages} {1195} (\bibinfo {year}
  {1983})}\BibitemShut {NoStop}%
\bibitem [{\citenamefont {Hayward}\ \emph {et~al.}(1999)\citenamefont
  {Hayward}, \citenamefont {Green}, \citenamefont {Rosseinsky},\ and\
  \citenamefont {Sloan}}]{hayward1999}%
  \BibitemOpen
  \bibfield  {author} {\bibinfo {author} {\bibfnamefont {M.}~\bibnamefont
  {Hayward}}, \bibinfo {author} {\bibfnamefont {M.}~\bibnamefont {Green}},
  \bibinfo {author} {\bibfnamefont {M.}~\bibnamefont {Rosseinsky}},\ and\
  \bibinfo {author} {\bibfnamefont {J.}~\bibnamefont {Sloan}},\ }\bibfield
  {title} {\bibinfo {title} {Sodium hydride as a powerful reducing agent for
  topotactic oxide deintercalation: synthesis and characterization of the
  nickel (i) oxide lanio2},\ }\href@noop {} {\bibfield  {journal} {\bibinfo
  {journal} {Journal of the American Chemical Society}\ }\textbf {\bibinfo
  {volume} {121}},\ \bibinfo {pages} {8843} (\bibinfo {year}
  {1999})}\BibitemShut {NoStop}%
\bibitem [{\citenamefont {Hayward}\ and\ \citenamefont
  {Rosseinsky}(2003)}]{hayward2003}%
  \BibitemOpen
  \bibfield  {author} {\bibinfo {author} {\bibfnamefont {M.}~\bibnamefont
  {Hayward}}\ and\ \bibinfo {author} {\bibfnamefont {M.}~\bibnamefont
  {Rosseinsky}},\ }\bibfield  {title} {\bibinfo {title} {Synthesis of the
  infinite layer ni (i) phase ndnio2+ x by low temperature reduction of ndnio3
  with sodium hydride},\ }\href@noop {} {\bibfield  {journal} {\bibinfo
  {journal} {Solid state sciences}\ }\textbf {\bibinfo {volume} {5}},\ \bibinfo
  {pages} {839} (\bibinfo {year} {2003})}\BibitemShut {NoStop}%
\bibitem [{\citenamefont {Baroni}\ \emph {et~al.}(2001)\citenamefont {Baroni},
  \citenamefont {de~Gironcoli}, \citenamefont {Dal~Corso},\ and\ \citenamefont
  {Giannozzi}}]{dfpt}%
  \BibitemOpen
  \bibfield  {author} {\bibinfo {author} {\bibfnamefont {S.}~\bibnamefont
  {Baroni}}, \bibinfo {author} {\bibfnamefont {S.}~\bibnamefont
  {de~Gironcoli}}, \bibinfo {author} {\bibfnamefont {A.}~\bibnamefont
  {Dal~Corso}},\ and\ \bibinfo {author} {\bibfnamefont {P.}~\bibnamefont
  {Giannozzi}},\ }\bibfield  {title} {\bibinfo {title} {Phonons and related
  crystal properties from density-functional perturbation theory},\ }\href
  {https://doi.org/10.1103/RevModPhys.73.515} {\bibfield  {journal} {\bibinfo
  {journal} {Rev. Mod. Phys.}\ }\textbf {\bibinfo {volume} {73}},\ \bibinfo
  {pages} {515} (\bibinfo {year} {2001})}\BibitemShut {NoStop}%
\bibitem [{\citenamefont {Giannozzi}\ \emph {et~al.}(2009)\citenamefont
  {Giannozzi}, \citenamefont {Baroni}, \citenamefont {Bonini}, \citenamefont
  {Calandra}, \citenamefont {Car}, \citenamefont {Cavazzoni}, \citenamefont
  {Ceresoli}, \citenamefont {Chiarotti}, \citenamefont {Cococcioni},
  \citenamefont {Dabo} \emph {et~al.}}]{qe}%
  \BibitemOpen
  \bibfield  {author} {\bibinfo {author} {\bibfnamefont {P.}~\bibnamefont
  {Giannozzi}}, \bibinfo {author} {\bibfnamefont {S.}~\bibnamefont {Baroni}},
  \bibinfo {author} {\bibfnamefont {N.}~\bibnamefont {Bonini}}, \bibinfo
  {author} {\bibfnamefont {M.}~\bibnamefont {Calandra}}, \bibinfo {author}
  {\bibfnamefont {R.}~\bibnamefont {Car}}, \bibinfo {author} {\bibfnamefont
  {C.}~\bibnamefont {Cavazzoni}}, \bibinfo {author} {\bibfnamefont
  {D.}~\bibnamefont {Ceresoli}}, \bibinfo {author} {\bibfnamefont {G.~L.}\
  \bibnamefont {Chiarotti}}, \bibinfo {author} {\bibfnamefont {M.}~\bibnamefont
  {Cococcioni}}, \bibinfo {author} {\bibfnamefont {I.}~\bibnamefont {Dabo}},
  \emph {et~al.},\ }\bibfield  {title} {\bibinfo {title} {{\sc quantum
  espresso}: a modular and open-source software project for quantum simulations
  of materials},\ }\href {https://doi.org/10.1088/0953-8984/21/39/395502}
  {\bibfield  {journal} {\bibinfo  {journal} {J. Phys.: Condens. Matter}\
  }\textbf {\bibinfo {volume} {21}},\ \bibinfo {pages} {395502} (\bibinfo
  {year} {2009})}\BibitemShut {NoStop}%
\bibitem [{\citenamefont {Dal~Corso}(2014)}]{pslib}%
  \BibitemOpen
  \bibfield  {author} {\bibinfo {author} {\bibfnamefont {A.}~\bibnamefont
  {Dal~Corso}},\ }\bibfield  {title} {\bibinfo {title} {Pseudopotentials
  periodic table: From {H} to {Pu}},\ }\href@noop {} {\bibfield  {journal}
  {\bibinfo  {journal} {Computational Materials Science}\ }\textbf {\bibinfo
  {volume} {95}},\ \bibinfo {pages} {337} (\bibinfo {year} {2014})}\BibitemShut
  {NoStop}%
\bibitem [{\citenamefont {Perdew}\ \emph {et~al.}(1996)\citenamefont {Perdew},
  \citenamefont {Burke},\ and\ \citenamefont {Ernzerhof}}]{pbe}%
  \BibitemOpen
  \bibfield  {author} {\bibinfo {author} {\bibfnamefont {J.~P.}\ \bibnamefont
  {Perdew}}, \bibinfo {author} {\bibfnamefont {K.}~\bibnamefont {Burke}},\ and\
  \bibinfo {author} {\bibfnamefont {M.}~\bibnamefont {Ernzerhof}},\ }\bibfield
  {title} {\bibinfo {title} {{G}eneralized {G}radient {A}pproximation {M}ade
  {S}imple},\ }\href {https://doi.org/10.1103/PhysRevLett.77.3865} {\bibfield
  {journal} {\bibinfo  {journal} {Phys. Rev. Lett.}\ }\textbf {\bibinfo
  {volume} {77}},\ \bibinfo {pages} {3865} (\bibinfo {year}
  {1996})}\BibitemShut {NoStop}%
\bibitem [{\citenamefont {Stokes}\ \emph {et~al.}()\citenamefont {Stokes},
  \citenamefont {Campbell},\ and\ \citenamefont {Hatch}}]{isotropy}%
  \BibitemOpen
  \bibfield  {author} {\bibinfo {author} {\bibfnamefont {H.~T.}\ \bibnamefont
  {Stokes}}, \bibinfo {author} {\bibfnamefont {B.~J.}\ \bibnamefont
  {Campbell}},\ and\ \bibinfo {author} {\bibfnamefont {D.~M.}\ \bibnamefont
  {Hatch}},\ }\href@noop {} {\bibinfo {title} {{\sc isotropy} software
  suite}},\ \bibinfo {howpublished} {\url{iso.byu.edu}}\BibitemShut {NoStop}%
\bibitem [{\citenamefont {Kresse}\ and\ \citenamefont
  {Furthm{\"u}ller}(1996)}]{vasp}%
  \BibitemOpen
  \bibfield  {author} {\bibinfo {author} {\bibfnamefont {G.}~\bibnamefont
  {Kresse}}\ and\ \bibinfo {author} {\bibfnamefont {J.}~\bibnamefont
  {Furthm{\"u}ller}},\ }\bibfield  {title} {\bibinfo {title} {Efficient
  iterative schemes for {\emph{ab initio}} total-energy calculations using a
  plane-wave basis set},\ }\href@noop {} {\bibfield  {journal} {\bibinfo
  {journal} {Physical review B}\ }\textbf {\bibinfo {volume} {54}},\ \bibinfo
  {pages} {11169} (\bibinfo {year} {1996})}\BibitemShut {NoStop}%
\bibitem [{\citenamefont {Stokes}\ and\ \citenamefont {Hatch}(2005)}]{findsym}%
  \BibitemOpen
  \bibfield  {author} {\bibinfo {author} {\bibfnamefont {H.~T.}\ \bibnamefont
  {Stokes}}\ and\ \bibinfo {author} {\bibfnamefont {D.~M.}\ \bibnamefont
  {Hatch}},\ }\bibfield  {title} {\bibinfo {title} {{\sc findsym}: program for
  identifying the space-group symmetry of a crystal},\ }\href@noop {}
  {\bibfield  {journal} {\bibinfo  {journal} {Journal of Applied
  Crystallography}\ }\textbf {\bibinfo {volume} {38}},\ \bibinfo {pages} {237}
  (\bibinfo {year} {2005})}\BibitemShut {NoStop}%
\bibitem [{\citenamefont {Togo}\ and\ \citenamefont {Tanaka}(2018)}]{spglib}%
  \BibitemOpen
  \bibfield  {author} {\bibinfo {author} {\bibfnamefont {A.}~\bibnamefont
  {Togo}}\ and\ \bibinfo {author} {\bibfnamefont {I.}~\bibnamefont {Tanaka}},\
  }\bibfield  {title} {\bibinfo {title} {{\sc spglib}: a software library for
  crystal symmetry search},\ }\href@noop {} {\bibfield  {journal} {\bibinfo
  {journal} {arXiv preprint arXiv:1808.01590}\ } (\bibinfo {year}
  {2018})}\BibitemShut {NoStop}%
\bibitem [{\citenamefont {Subedi}(2018)}]{subedi2018lno3}%
  \BibitemOpen
  \bibfield  {author} {\bibinfo {author} {\bibfnamefont {A.}~\bibnamefont
  {Subedi}},\ }\bibfield  {title} {\bibinfo {title} {Breathing distortions in
  the metallic, antiferromagnetic phase of lanio $ \_3$},\ }\href@noop {}
  {\bibfield  {journal} {\bibinfo  {journal} {SciPost Physics}\ }\textbf
  {\bibinfo {volume} {5}},\ \bibinfo {pages} {020} (\bibinfo {year}
  {2018})}\BibitemShut {NoStop}%
\bibitem [{\citenamefont {Lee}\ and\ \citenamefont {Pickett}(2004)}]{lee2004}%
  \BibitemOpen
  \bibfield  {author} {\bibinfo {author} {\bibfnamefont {K.-W.}\ \bibnamefont
  {Lee}}\ and\ \bibinfo {author} {\bibfnamefont {W.~E.}\ \bibnamefont
  {Pickett}},\ }\bibfield  {title} {\bibinfo {title} {Infinite-layer
  {LaNiO$_2$}: {Ni$^{1+}$} is not {Cu$^{2+}$}},\ }\href@noop {} {\bibfield
  {journal} {\bibinfo  {journal} {Physical Review B}\ }\textbf {\bibinfo
  {volume} {70}},\ \bibinfo {pages} {165109} (\bibinfo {year}
  {2004})}\BibitemShut {NoStop}%
\bibitem [{\citenamefont {Goh}\ \emph {et~al.}(2015)\citenamefont {Goh},
  \citenamefont {Tompsett}, \citenamefont {Saines}, \citenamefont {Chang},
  \citenamefont {Matsumoto}, \citenamefont {Imai}, \citenamefont {Yoshimura},\
  and\ \citenamefont {Grosche}}]{goh2015}%
  \BibitemOpen
  \bibfield  {author} {\bibinfo {author} {\bibfnamefont {S.~K.}\ \bibnamefont
  {Goh}}, \bibinfo {author} {\bibfnamefont {D.~A.}\ \bibnamefont {Tompsett}},
  \bibinfo {author} {\bibfnamefont {P.~J.}\ \bibnamefont {Saines}}, \bibinfo
  {author} {\bibfnamefont {H.~C.}\ \bibnamefont {Chang}}, \bibinfo {author}
  {\bibfnamefont {T.}~\bibnamefont {Matsumoto}}, \bibinfo {author}
  {\bibfnamefont {M.}~\bibnamefont {Imai}}, \bibinfo {author} {\bibfnamefont
  {K.}~\bibnamefont {Yoshimura}},\ and\ \bibinfo {author} {\bibfnamefont
  {F.~M.}\ \bibnamefont {Grosche}},\ }\bibfield  {title} {\bibinfo {title}
  {Ambient pressure structural quantum critical point in the phase diagram of
  {(Ca$_x$Sr$_{1-x}$)$_3$Rh$_4$Sn$_{13}$}},\ }\href@noop {} {\bibfield
  {journal} {\bibinfo  {journal} {Physical review letters}\ }\textbf {\bibinfo
  {volume} {114}},\ \bibinfo {pages} {097002} (\bibinfo {year}
  {2015})}\BibitemShut {NoStop}%
\bibitem [{\citenamefont {Yu}\ \emph {et~al.}(2015)\citenamefont {Yu},
  \citenamefont {Cheung}, \citenamefont {Saines}, \citenamefont {Imai},
  \citenamefont {Matsumoto}, \citenamefont {Michioka}, \citenamefont
  {Yoshimura},\ and\ \citenamefont {Goh}}]{yu2015}%
  \BibitemOpen
  \bibfield  {author} {\bibinfo {author} {\bibfnamefont {W.~C.}\ \bibnamefont
  {Yu}}, \bibinfo {author} {\bibfnamefont {Y.~W.}\ \bibnamefont {Cheung}},
  \bibinfo {author} {\bibfnamefont {P.~J.}\ \bibnamefont {Saines}}, \bibinfo
  {author} {\bibfnamefont {M.}~\bibnamefont {Imai}}, \bibinfo {author}
  {\bibfnamefont {T.}~\bibnamefont {Matsumoto}}, \bibinfo {author}
  {\bibfnamefont {C.}~\bibnamefont {Michioka}}, \bibinfo {author}
  {\bibfnamefont {K.}~\bibnamefont {Yoshimura}},\ and\ \bibinfo {author}
  {\bibfnamefont {S.~K.}\ \bibnamefont {Goh}},\ }\bibfield  {title} {\bibinfo
  {title} {{S}trong {C}oupling {S}uperconductivity in the {V}icinity of the
  {S}tructural {Q}uantum {C}ritical {P}oint in
  {(Ca$_x$Sr$_{1-x}$)$_3$Rh$_4$Sn$_{13}$}},\ }\href@noop {} {\bibfield
  {journal} {\bibinfo  {journal} {Physical review letters}\ }\textbf {\bibinfo
  {volume} {115}},\ \bibinfo {pages} {207003} (\bibinfo {year}
  {2015})}\BibitemShut {NoStop}%
\bibitem [{\citenamefont {Hu}\ \emph {et~al.}(2017)\citenamefont {Hu},
  \citenamefont {Cheung}, \citenamefont {Yu}, \citenamefont {Imai},
  \citenamefont {Kanagawa}, \citenamefont {Murakawa}, \citenamefont
  {Yoshimura},\ and\ \citenamefont {Goh}}]{hu2017}%
  \BibitemOpen
  \bibfield  {author} {\bibinfo {author} {\bibfnamefont {Y.}~\bibnamefont
  {Hu}}, \bibinfo {author} {\bibfnamefont {Y.}~\bibnamefont {Cheung}}, \bibinfo
  {author} {\bibfnamefont {W.}~\bibnamefont {Yu}}, \bibinfo {author}
  {\bibfnamefont {M.}~\bibnamefont {Imai}}, \bibinfo {author} {\bibfnamefont
  {H.}~\bibnamefont {Kanagawa}}, \bibinfo {author} {\bibfnamefont
  {J.}~\bibnamefont {Murakawa}}, \bibinfo {author} {\bibfnamefont
  {K.}~\bibnamefont {Yoshimura}},\ and\ \bibinfo {author} {\bibfnamefont
  {S.~K.}\ \bibnamefont {Goh}},\ }\bibfield  {title} {\bibinfo {title} {Soft
  phonon modes in the vicinity of the structural quantum critical point},\
  }\href@noop {} {\bibfield  {journal} {\bibinfo  {journal} {Physical Review
  B}\ }\textbf {\bibinfo {volume} {95}},\ \bibinfo {pages} {155142} (\bibinfo
  {year} {2017})}\BibitemShut {NoStop}%
\bibitem [{\citenamefont {Cheung}\ \emph {et~al.}(2018)\citenamefont {Cheung},
  \citenamefont {Hu}, \citenamefont {Imai}, \citenamefont {Tanioku},
  \citenamefont {Kanagawa}, \citenamefont {Murakawa}, \citenamefont {Moriyama},
  \citenamefont {Zhang}, \citenamefont {Lai}, \citenamefont {Yoshimura} \emph
  {et~al.}}]{cheung2018}%
  \BibitemOpen
  \bibfield  {author} {\bibinfo {author} {\bibfnamefont {Y.~W.}\ \bibnamefont
  {Cheung}}, \bibinfo {author} {\bibfnamefont {Y.~J.}\ \bibnamefont {Hu}},
  \bibinfo {author} {\bibfnamefont {M.}~\bibnamefont {Imai}}, \bibinfo {author}
  {\bibfnamefont {Y.}~\bibnamefont {Tanioku}}, \bibinfo {author} {\bibfnamefont
  {H.}~\bibnamefont {Kanagawa}}, \bibinfo {author} {\bibfnamefont
  {J.}~\bibnamefont {Murakawa}}, \bibinfo {author} {\bibfnamefont
  {K.}~\bibnamefont {Moriyama}}, \bibinfo {author} {\bibfnamefont
  {W.}~\bibnamefont {Zhang}}, \bibinfo {author} {\bibfnamefont {K.~T.}\
  \bibnamefont {Lai}}, \bibinfo {author} {\bibfnamefont {K.}~\bibnamefont
  {Yoshimura}}, \emph {et~al.},\ }\bibfield  {title} {\bibinfo {title}
  {Evidence of a structural quantum critical point in
  {(Ca$_x$Sr$_{1-x}$)$_3$Rh$_4$Sn$_{13}$} from a lattice dynamics study},\
  }\href@noop {} {\bibfield  {journal} {\bibinfo  {journal} {Physical Review
  B}\ }\textbf {\bibinfo {volume} {98}},\ \bibinfo {pages} {161103} (\bibinfo
  {year} {2018})}\BibitemShut {NoStop}%
\bibitem [{\citenamefont {Poudel}\ \emph {et~al.}(2016)\citenamefont {Poudel},
  \citenamefont {May}, \citenamefont {Koehler}, \citenamefont {McGuire},
  \citenamefont {Mukhopadhyay}, \citenamefont {Calder}, \citenamefont
  {Baumbach}, \citenamefont {Mukherjee}, \citenamefont {Sapkota}, \citenamefont
  {de~la Cruz} \emph {et~al.}}]{poudel2016}%
  \BibitemOpen
  \bibfield  {author} {\bibinfo {author} {\bibfnamefont {L.}~\bibnamefont
  {Poudel}}, \bibinfo {author} {\bibfnamefont {A.~F.}\ \bibnamefont {May}},
  \bibinfo {author} {\bibfnamefont {M.~R.}\ \bibnamefont {Koehler}}, \bibinfo
  {author} {\bibfnamefont {M.~A.}\ \bibnamefont {McGuire}}, \bibinfo {author}
  {\bibfnamefont {S.}~\bibnamefont {Mukhopadhyay}}, \bibinfo {author}
  {\bibfnamefont {S.}~\bibnamefont {Calder}}, \bibinfo {author} {\bibfnamefont
  {R.~E.}\ \bibnamefont {Baumbach}}, \bibinfo {author} {\bibfnamefont
  {R.}~\bibnamefont {Mukherjee}}, \bibinfo {author} {\bibfnamefont
  {D.}~\bibnamefont {Sapkota}}, \bibinfo {author} {\bibfnamefont
  {C.}~\bibnamefont {de~la Cruz}}, \emph {et~al.},\ }\bibfield  {title}
  {\bibinfo {title} {{C}andidate {E}lastic {Q}uantum {C}ritical {P}oint in
  {LaCu$_{6-x}$Au$_x$}},\ }\href@noop {} {\bibfield  {journal} {\bibinfo
  {journal} {Physical review letters}\ }\textbf {\bibinfo {volume} {117}},\
  \bibinfo {pages} {235701} (\bibinfo {year} {2016})}\BibitemShut {NoStop}%
\bibitem [{\citenamefont {Biswas}\ \emph {et~al.}(2015)\citenamefont {Biswas},
  \citenamefont {Guguchia}, \citenamefont {Khasanov}, \citenamefont {Chinotti},
  \citenamefont {Li}, \citenamefont {Wang}, \citenamefont {Petrovic},\ and\
  \citenamefont {Morenzoni}}]{biswas2015}%
  \BibitemOpen
  \bibfield  {author} {\bibinfo {author} {\bibfnamefont {P.~K.}\ \bibnamefont
  {Biswas}}, \bibinfo {author} {\bibfnamefont {Z.}~\bibnamefont {Guguchia}},
  \bibinfo {author} {\bibfnamefont {R.}~\bibnamefont {Khasanov}}, \bibinfo
  {author} {\bibfnamefont {M.}~\bibnamefont {Chinotti}}, \bibinfo {author}
  {\bibfnamefont {L.}~\bibnamefont {Li}}, \bibinfo {author} {\bibfnamefont
  {K.}~\bibnamefont {Wang}}, \bibinfo {author} {\bibfnamefont {C.}~\bibnamefont
  {Petrovic}},\ and\ \bibinfo {author} {\bibfnamefont {E.}~\bibnamefont
  {Morenzoni}},\ }\bibfield  {title} {\bibinfo {title} {Strong enhancement of
  {$s$}-wave superconductivity near a quantum critical point of
  {Ca$_3$Ir$_4$Sn$_{13}$}},\ }\href@noop {} {\bibfield  {journal} {\bibinfo
  {journal} {Physical Review B}\ }\textbf {\bibinfo {volume} {92}},\ \bibinfo
  {pages} {195122} (\bibinfo {year} {2015})}\BibitemShut {NoStop}%
\end{thebibliography}%

%

\end{document}